\documentclass[reprint, amsmath, amssymb, aps, prx,floatfix]{revtex4-2}

\usepackage{amsmath}
\allowdisplaybreaks[2]
\usepackage{lineno}
\usepackage{color,soul}
\usepackage{graphicx}
\usepackage{dcolumn}
\usepackage{bm}

\usepackage{verbatim}
\usepackage[dvipsnames]{xcolor}

\newcommand{\ket}[1]{\vert #1 \rangle}
\newcommand{\bra}[1]{\langle #1 \vert}
\newcommand{\expectation}[1]{\langle #1 \rangle}
\newcommand{\coh}{\mathfrak{C}}
\newcommand{\braket}[2]{\langle #1 \vert #2\rangle}
\definecolor{darkgreen}{rgb}{0.0, 0.5, 0.0}
\usepackage[hidelinks,bookmarks=false]{hyperref}

\newtheorem{theorem}{Theorem}[section]

\begin{document}
\global\long\def\bracket#1#2#3{\bigl\langle#1\bigl|#2\bigr|#3\bigr\rangle}%

\global\long\def\braket#1#2{\bigl\langle#1|#2\bigr\rangle}%

\global\long\def\bra#1{\langle#1|}%

\global\long\def\ket#1{|#1\rangle}%

\global\long\def\innerprod#1#2{\bigl|#1\bigr\rangle\bigl\langle#2\bigr|}%

\global\long\def\Bbracket#1#2#3{\left\langle #1\left|#2\vphantom{#1#3}\right|#3\right\rangle }%

\global\long\def\Bbraket#1#2{\left\langle #1\left|\vphantom{#1#2}\right.#2\right\rangle }%

\global\long\def\Bbra#1{\left\langle #1\right|}%

\global\long\def\Bket#1{\left|#1\right\rangle }%

\global\long\def\Binnerprod#1#2{\left|#1\vphantom{#2}\right\rangle \left\langle #2\vphantom{#1}\right|}%

\global\long\def\dd#1#2{\frac{\text{d}#1}{\text{d}#2}}%

\global\long\def\pp#1#2{\frac{\partial#1}{\partial#2}}%

\global\long\def\expectation#1{\bigl\langle#1\bigr\rangle}%

\sloppy

\global\long\def\coh{\mathfrak{C}}%

\title{Collectively pumped superradiant laser with coherence scaling beyond the standard quantum limit}

\author{Ori Somech}
\email{orisomech@gmail.com}
 
\author{Howard M. Wiseman}%
\email{h.wiseman@griffith.edu.au}

\affiliation{%
Centre for Quantum Computation and Communication Technology (Australian
Research Council), Quantum and Advanced Technology Research Institute, Griffith
University, Yuggera Country, Brisbane, QLD 4111, Australia
}%

\date{November 4, 2026}

\begin{abstract}
For more than half a century, the standard quantum limit (SQL) was thought to limit the laser coherence $\coh$ –– the number of photons emitted from the laser into the beam in one coherence time –– to a scaling $\Theta(\mu^2)$, where $\mu$ is the mean number of optical-frequency excitations stored inside the laser. However, recently it has been shown [Baker \textit{et al.}, Nat.~Phys.~\textbf{17}, 179 (2021)] that the Heisenberg limit –– an achievable ultimate limit set by quantum mechanics for the task of producing a beam with the standard properties of a laser beam –– is $\coh=\Theta(\mu^4)$, a quadratic enhancement. So far, proposals to demonstrate beyond-SQL scaling of $\coh$ have been limited to circuit QED, at microwave frequencies (i.e.,~a maser). Here, we introduce a class of generalized far-above-threshold (FAT) laser models, obtained by replacing the ordinary bosonic lowering operator $\hat{a}$ by a general lowering operator $\hat{L}$ in the superoperators appearing in the master equation that governs the FAT laser dynamics.
We then show that an optical-frequency superradiant laser in the bad-cavity limit, with feedback-controlled collective pumping, realizes one such model, in which $\hat{L}$ is identified with the collective superradiant lowering operator $\hat{J}$. This model can surpass the SQL scaling, with the coherence exhibiting a scaling as large as $\coh= \Theta\left(\mu^{8/3}\right)$ in the optimal parameter regime. Here, in the bad-cavity limit, $\mu\approx N/2$, where $N$ is the number of superradiant atoms.
\end{abstract}

\maketitle


\section{Introduction}
\label{sec:Intro}
The widespread use of lasers in precision technology is due to their ability to generate beams with a high degree of coherence. A laser beam may exhibit a wide range of physical properties and, in general, cannot be fully characterized by a single parameter.  Nevertheless, in 1961, Mandel introduced a single dimensionless measure of coherence, the degeneracy parameter, defined as the mean number of photons in the beam that occupy the same quantum state or within a ``coherence volume'' \cite{degeneracyMandel1961,mandel_wolf_1995}.
Recently, a more precise definition was given by Baker \textit{et al.}~in 2020 \cite{HL}, as the mean number of photons in the maximally populated mode of the beam (within some frequency band). In Ref.~\cite{HL}, this dimensionless quantity is denoted by $\coh$, namely coherence. 
For a continuous-wave laser with `ideal' properties (see Sec.~\ref{thesissec:HLtheorem} and Ref.~\cite{HL}), the mode that maximizes $\coh$ is infinitely long, uniform, and matches the centre frequency of the laser. Then, for a laser with a Lorentzian spectrum, $\coh$ can be intuitively expressed in terms of the beam's photon flux (photons per unit time), $\mathcal{N}$, and its linewidth (twice the reciprocal of coherence time), $\ell$, as \cite{HL}
\begin{equation}
\label{eq:C}
\mathfrak{C}=4\mathcal{N}/{\ell}. 
\end{equation}
Equation (\ref{eq:C}) shows the coherence $\coh$, being proportional to the product of flux and coherence time, can be thought of roughly as the number of photons emitted consecutively into the beam with the same phase. 
Equation (\ref{eq:C}), divided by two, is how Mandel originally evaluated the degeneracy parameter of a laser beam \cite{degeneracyMandel1961}.
Lasers remain the only device that serves as a source of light with high coherence $\coh$ \cite{degeneracyMandel1961,mandel_wolf_1995,HL}.
 
Although lasers were first invented as a device that pumps excitations into a mode of a cavity, the medium which stores the coherent excitations is not limited to a cavity. Here, ``coherent excitations'' refer to the excitations that store the laser's phase, e.g., for the standard laser these are the photons in the lasing cavity mode. 
Reference \cite{HL} defines a laser model in a general way, as a device that creates a high degree of coherence from sources that have a negligible amount of coherence. 
While the coherence from a source that is not pumped during emission, such as Q-switched lasers \cite{Qswitchedlaser} or atom lasers \cite{RevModPhysatomlasers}, cannot be larger than the maximal number of excitations initially stored in the source, for a laser that is pumped while it is gradually dumped, the value of $\mathfrak{C}$ can significantly exceed the number of excitations $\mu$ stored within the laser \cite{HL,mandel_wolf_1995}. In the 1960s and 1970s, quantum mechanical models of standard masers and lasers were formulated by several authors, and the linewidth and flux calculated from them \cite{scully_zubairy_1997,sargent1974laser,mandel_wolf_1995,louisell1973quantum,carmichael1999methods}.
In the ideal case, where a standard laser operates far above threshold, and in the absence of technical noise, it is described by the far above threshold (FAT)
laser model \cite{howard1999,howard_atom_laser,howard_stochastic}. 
This model yields a linewidth of $\ell=\kappa/2\mu$ (commonly, but inaccurately, referred to as the Schawlow--Townes linewidth, after the qualitatively similar result in their famous paper \cite{SchawlowTownes}) and a flux of $\mathcal{N}=\kappa\mu$, where $\kappa$ is the `bare' (or `cold') linewidth of the lasing cavity. Here, $\mu$ is the mean number of photons in the laser mode.
Using Eq.~(\ref{eq:C}), this ideal laser model defines the standard quantum limit (SQL) for laser coherence \cite{HL,howard1999}:
\begin{equation}
\label{eq:SQL}
    \mathfrak{C}_{\text{SQL}} = 8 \mu^2.
\end{equation}

The term SQL is appropriate because it follows from a set of ``standard'' assumptions about how a laser must operate.
The most crucial of these is that the coherent excitations are stored in a single linear oscillator mode, whose couplings to external degrees of freedom, including a gain medium, the output beam, and other environmental reservoirs, are linear in the corresponding mode operators.
This restriction applies to the underlying coupling processes, but not necessarily to the effective reduced dynamics, which may be nonlinear.
For example, in the standard laser, the effective gain dynamics are nonlinear because gain saturation arises from the nonlinearity of the atoms in the gain medium. 
The SQL limit was long thought to be an absolute limit on the scaling. Recently, however, it has been realized that the above assumptions reflect current technological paradigms rather than fundamental constraints imposed by quantum mechanics. In particular, proposals for circuit-QED models at microwave frequencies (i.e., masers) have shown that superior scaling can be achieved in principle \cite{HL,circutQED}. The key to achieving beyond-SQL scaling of the coherence $\mathfrak{C}$ is to make both the pumping and the output coupling nonlinear, in a particular way.

Recent work \cite{HL} also established the Heisenberg limit (HL) for $\mathfrak{C}$, i.e., the ultimate achievable limit imposed by quantum mechanics rather than by conventional technologies, under natural conditions (see also Refs.~\cite{LucasPRL,LucasPRA,LucasAnalytical}), which scales as
\begin{equation}
\label{eq:HLscale}
\mathfrak{C}_{\mathrm{HL}} = \Theta(\mu^4).
\end{equation}
Here, $\Theta$ indicates an achievable upper-bound scaling \cite{BigO}. The Heisenberg limit in Eq.~(\ref{eq:HLscale}) is a quadratic enhancement over the SQL, Eq.~(\ref{eq:SQL}). Importantly, in Eq.~(\ref{eq:HLscale}), $\mu$ is the mean number of \textit{total} laser-frequency excitations within the laser, which can generally include photons, atomic excitations, excitons, etc.

Superradiance is the cooperative interaction of the electromagnetic field with quantum emitters, where each emitter interacts with the electromagnetic field generated by itself and by the other emitters. It was introduced by Dicke in 1954 \cite{Dicke,superressay,mandel_wolf_1995} as a transient phenomenon: a dense gas of atoms initially prepared in an excited state relaxes rapidly to the ground state, producing a collective emission burst that then terminates. The atoms spontaneously choose a common phase and become locked to it, creating a large collective dipole that scales linearly with the number of atoms $N$, producing radiation with an intensity proportional to $N^2$. In Dicke’s original setting, the atoms are identically coupled to a single electromagnetic mode; this special case is known as Dicke superradiance \cite{Dicke,superressay,mandel_wolf_1995}. Dicke superradiance remains one of the only few cases of superradiance that is fully tractable analytically in solving the state dynamics for a large number of atoms $N$ and in the nonlinear regime of highly excited (saturated) atoms, due to its SU(2) symmetry \cite{Dicke,superressay,mandel_wolf_1995}.

With the emergence of cavity-QED and waveguide QED, superradiance has been achieved in these platforms \cite{RevModPhysWQEDsuperradiance,supercascaded,Nature2012,doi:10.1126/sciadv.1601231,PhysRevX.8.021036,superlasertheory1}. 
In cavity-QED implementations, superradiance is achieved in the bad-cavity limit, where most excitations reside in the atoms and the cavity mode is only negligibly excited, acting as a mediator between the atoms. Steady state superradiance can be achieved with coherent pumping \cite{ReyPRA,myPRA,myarticle,Carmichael1980} or with incoherent pumping. The latter can be called a superradiant laser, first introduced theoretically in Refs.~\cite{PhysRevLettHolland,PhysRevA2010,intensityfluctuations} and subsequently pursued further theoretically \cite{superradiantlasertheory1,HollandSU(4),crossoverHolland,crossoverMolmer,MolmerMonteCarlo,superradiantlasertheory2,Hollandfullycollective,PhysRevLett.134.193603,PhysRevLett.133.073601,PhysRevLett.127.073603,PhysRevLett.126.123602} and experimentally \cite{Nature2012,superradiantlaser1,doi:10.1126/sciadv.1601231,PhysRevXcrossover,pulsedelay,PhysRevX.8.021036,superradiantlasetexpirement2,superradiantlasetexpirement3,superradiantlaserexperiment4,Thresholdlessuperradiance}. In this model, which we refer to in this paper as the ``standard'' superradiant laser model (to distinguish it from the laser model proposed here), atoms are pumped non-collectively, breaking the SU(2) symmetry of Dicke superradiance; Nevertheless, the atoms are still highly correlated in phase \cite{PhysRevA2010}, producing radiation with intensity proportional to $N^2$ \cite{PhysRevA2010,PhysRevLettHolland,intensityfluctuations}, and a linewidth bounded from below by $\Gamma$, the single atom emission rate through the cavity.
Since the ``standard" superradiant laser operates optimally within the range where $\mu \propto N$, its coherence scales quadratically with the mean number of excitations, i.e., $\coh_{\text{S}}=\Theta(\mu^2)$, like the SQL, Eq.~(\ref{eq:SQL}).

In this paper, we introduce a class of generalized FAT laser models, obtained by replacing the ordinary bosonic lowering operator $\hat{a}$ by a general lowering operator $\hat{L}$ in the superoperators appearing in the master equation that governs the FAT laser dynamics.
We show that an optical-frequency superradiant laser in the bad-cavity limit, with feedback-controlled collective pumping, could realize one such model, in which $\hat{L}$ equals the collective superradiant lowering operator $\hat{J}$. In addition to the collective loss mechanism of the standard superradiant laser, our laser model’s gain mechanism is collective as well, i.e., composed of collective atomic operators. Thus, the model preserves the SU(2) symmetry, and the relevant Hilbert space is composed only of $N+1$ nondegenerate states, where $N$ is the number of atoms.
To collectively pump the atoms, our model uses a highly detuned third level. The atoms are driven identically by light from an independent laser (which need not be narrowband) from the first level to an auxiliary third level, and induced to decay to the second level by coupling identically to a single cavity mode.
The output of this cavity mode is continuously monitored by a single-photon detector, and the detection signal is used for feedback control of the driving laser. We show that our model can obtain a linewidth far narrower than the standard case and surpass the SQL scaling of Eq.~(\ref{eq:SQL}).

The paper is organized as follows. In Sec.~\ref{sec:background}, we provide background material. In Sec.~\ref{sec:model}, we introduce the generalized FAT laser models and the master equation for our superradiant laser model, and motivate its form and why it is expected to surpass the SQL, Eq.~(\ref{eq:SQL}).
In Sec.~\ref{sec:physicalrealzation}, we suggest a physical realization for the master equation introduced in Sec.~\ref{sec:model}. In Sec.~\ref{sec:metastable}, we discuss the metastable regime in which the laser operates optimally, where it remains in a long-lived metastable state.
In Sec.~\ref{sec:poissonian}, we show that the laser beam exhibits ideal Poissonian output photon statistics, like an ideal standard laser. In Sec.~\ref{sec:coherenceprop}, we derive the linewidth and coherence properties of the laser, demonstrating that high coherence $\coh$, beyond the SQL scaling, can be achieved while operating in the metastable regime. In Sec.~\ref{sec:generalFAT}, for completeness, we provide a qualitative analysis for any generalized FAT laser model. We conclude in Sec.~\ref{sec:conclusions}. Additionally,
we provide appendices with detailed calculations to support the claims in the main text. In Appendix \ref{app:superrdiantlasersemi}, we present an analysis of the semiclassical dynamics of the standard superradiant laser model, which, to our knowledge, has not been given before.

\section{Background}
\label{sec:background}
In this section, we present background material.
We first review the standard laser model and its optimal operating regime, far above threshold. We then discuss the Heisenberg-limit theorem of Ref.~\cite{HL}, the conditions assumed in its proof, and the Heisenberg-limited laser model introduced therein. Finally, we discuss Dicke superradiance, the coherently radiating spin states (CRSS) introduced in Ref.~\cite{myarticle}, and the standard superradiant laser model.

\subsection{The standard laser and the far above threshold (FAT) regime}
\label{sec:fat}
Here, we review the standard laser model and the corresponding master equation governing the cavity-mode dynamics. We then discuss its ideal operating regime, namely, far above threshold (FAT), and present the corresponding FAT laser master equation.
The standard laser appears in many references \cite{howard_atom_laser,howard1999,Wiseman1994PhD,scully_zubairy_1997,sargent1974laser,mandel_wolf_1995,louisell1973quantum,carmichael1999methods}. Here, we follow Refs.~\cite{howard_atom_laser,howard1999,Wiseman1994PhD}.

A standard four-level laser consists of an optical cavity that supports a single relevant lasing mode and is resonantly coupled to a large ensemble of atoms that forms the gain medium. The atoms are continuously pumped at a constant rate $P$ from their ground state to their upper lasing level via an additional auxiliary level. The atomic excitations are then either transferred through stimulated emission into the cavity mode, thereby maintaining a steady-state photon population, or lost through spontaneous emission. The lower lasing level subsequently relaxes to the ground state, allowing population inversion to be maintained.
Standard laser operation takes place in the good-cavity limit, in which the cavity decay rate $\kappa$ is slow compared with the atomic dynamics. In this regime, the atomic degrees of freedom can be adiabatically eliminated, yielding the master equation for the cavity mode. Neglecting thermal photons and photon reabsorption \cite{howard_atom_laser,howard1999}, this is
\begin{equation}
\label{thesis:lasermasterequation}
\dot{\rho}=P\mathcal{D}[\hat{a}^{\dagger}](\mathcal{A}[\hat{a}^{\dagger}]+n_{\text{s}})^{-1}\rho+\kappa \mathcal{D}[\hat{a}]\rho.
\end{equation} 
Here, we follow the superoperator notation of Refs.~\cite{howard_atom_laser,howard1999,howard_stochastic}, in which $\mathcal{A}[\hat{A}] \hat{B}:=\frac{1}{2}\{\hat{A}^{\dagger} \hat{A}, \hat{B}\}$, $\mathcal{J} [\hat{A}] \hat{B}:=\hat{A} \hat{B} \hat{A}^{\dagger}$, and  $\mathcal{D}[\hat{A}] := \mathcal{J}[\hat{A}]-\mathcal{A}[\hat{A}]$. 
Here, $P$, $n_{\mathrm{s}}$, and $\kappa$ denote the pumping rate, saturation parameter and the cavity decay rate, respectively.
The first term in Eq.~(\ref{thesis:lasermasterequation}) describes the nonlinear gain dynamics resulting from gain saturation, whereas the second term describes the cavity loss that generates the laser output beam. 

The saturation parameter $n_{\mathrm{s}}$ originates from spontaneous emission on the lasing transition and is proportional to the corresponding emission rate. At low pump rates, spontaneous emission dominates over stimulated emission. As the pump rate approaches the threshold $P_{\mathrm{th}}=\kappa n_{\mathrm{s}}$, stimulated emission becomes significant. For $P>P_{\mathrm{th}}$, the steady-state mean photon number is approximately $\mu\approx P/\kappa-n_{\mathrm{s}}$ \cite{howard_atom_laser}.

We focus on the far-above-threshold (FAT) regime, $P\gg\kappa n_{\mathrm{s}}$, for which $\mu\gg n_{\mathrm{s}}$. In this regime, spontaneous-emission events are rare relative to stimulated-emission events, and their effect on the gain dynamics can be neglected by setting $n_{\mathrm{s}}=0$.
This approximation follows directly from Eq.~(\ref{thesis:lasermasterequation}). For a photon-number distribution concentrated near $n=\mu$, the relevant eigenvalues of $\mathcal{A}[\hat{a}^{\dagger}]$ are of order $\mu+1$. Thus, when $\mu\gg n_{\mathrm{s}}$, we approximate $\bigl(\mathcal{A}[\hat{a}^{\dagger}]+n_{\mathrm{s}}\bigr)^{-1}\approx\mathcal{A}[\hat{a}^{\dagger}]^{-1}$. Under this approximation, Eq.~(\ref{thesis:lasermasterequation}) reduces to the FAT laser master equation
\begin{equation}
\label{eq:FAT}
\dot{\rho}
=
\kappa
\left\{\frac{P}{\kappa} \mathcal{D}[\hat{a}^{\dagger}]
\mathcal{A}[\hat{a}^{\dagger}]^{-1}\rho
+
\mathcal{D}[\hat{a}]\rho
\right\}.
\end{equation}
In this model, $\mu=P/\kappa$ at steady state. 
Since in this ideal model spontaneous emission is neglected, every excited atom sends its excitation into the lasing cavity through stimulated emission. In this case, the gain rate becomes independent of the state of the cavity mode. Due to this, the steady state of the system is given by an incoherent mixture of coherent states of fixed amplitude $\sqrt{\mu}$ over all phases:
\begin{equation}
\label{eq:incoherentmix123}
\rho^{\text{ss}}=\frac{1}{2\pi}\int_0^{2\pi}\ket{\sqrt{\mu}e^{i\theta}}\bra{\sqrt{\mu}e^{i\theta}}\mathrm{d}\theta.
\end{equation}
Using Eq.~(\ref{eq:incoherentmix123}) and input–output theory \cite{Wiseman_2009}, where the non-vacuum part of the output field operator $\hat{b}$ is equal to $\sqrt{\kappa}\hat{a}$, the FAT laser model exhibits ideal Poissonian output photon statistics.
An alternative way to obtain the master equation (\ref{eq:FAT}) without relying on gain saturation is by monitoring and controlling the injected excitations so that they are added to the lasing cavity mode at a constant rate independent of its state \cite{howard_atom_laser,howard1999}.
We will use this latter mechanism for the physical realization of our laser model (see Sec.~\ref{sec:physicalrealzation}).
The photon flux from the lasing cavity, $\expectation{\hat{b}^\dagger\hat{b}}$, equals $\mathcal{N}=\kappa \mu$ in steady state, which equals the gain rate. When $\mu\gg1$, the solution to the FAT laser master equation (\ref{eq:FAT}) can be approximately described by a classical coherent state that undergoes pure phase diffusion (see Sec.~\ref{thesissec:HLtheorem}) \cite{scully_zubairy_1997,howard_stochastic,howard_atom_laser}, and thus has a Lorentzian spectrum \cite{howard_atom_laser,howard1999,howard_stochastic,louisell1973quantum}. In this regime, the linewidth of the FAT laser is $\ell=\kappa/2\mu$ \cite{howard_atom_laser,howard1999,howard_stochastic,louisell1973quantum}. Using Eq.~(\ref{eq:C}), the ideal FAT laser model defines the SQL, Eq.~(\ref{eq:SQL}) \cite{HL,howard1999}.

\subsection{The Heisenberg limit theorem}
In Sec.~\ref{sec:Intro}, we reviewed the definition of laser coherence, denoted by $\coh$, as the mean number of photons in the most populated mode of the laser beam \cite{HL}.
We also discussed the SQL limit for laser coherence, Eq.~(\ref{eq:SQL}), as the coherence of the FAT laser model. Additionally, we briefly discussed the Heisenberg limit (HL) for laser coherence.
In this section, we will further discuss the HL for laser coherence.
First, in Sec.~\ref{thesissec:HLtheorem}, we will review the Heisenberg limit theorem and present the four conditions required to prove the theorem. Then, in Sec.~\ref{thesissec:HLmodel}, we present the model that achieves the Heisenberg limit scaling of Eq.~(\ref{eq:HLscale}). 
\subsubsection{The Heisenberg limit theorem}
\label{thesissec:HLtheorem}
Recent work by Baker \textit{et al.} \cite{HL} established the Heisenberg limit (HL) for $\mathfrak{C}$, i.e., the ultimate achievable limit imposed by quantum mechanics rather than by conventional technologies, under four natural assumptions.
\begin{theorem}
\mbox{}\par
\noindent\textnormal{\textbf{The Heisenberg Limit for Laser Coherence}}
\par\noindent
For a laser satisfying conditions 1–4 stated below, the coherence $\coh^{\mathrm{ideal}}$ of the beam in the limit $\mu\rightarrow\infty$ is bounded from above by
\begin{equation}
    \mathfrak{C}^{\mathrm {ideal }} \lesssim 2.9748 \mu^4.
\end{equation}
Here, $\mu$ is the number of excitations stored within the laser. \end{theorem} 
Here, we use the notation $A\lesssim B$, for two given quantities $A$ and $B$, to mean that $A/B\leq 1+o(1)$ in the limit $\mu\to\infty$ \cite{BigO}.
This form of the Heisenberg-limit theorem gives not only the \(\mu^4\) scaling given in Eq.~(\ref{eq:HLscale}), but also its exact prefactor. However, this prefactor may not be tight.

Additionally, a bound on the coherence in terms of the variance of the population distribution can be obtained from Eq.~(S83) of the Supplemental Material of Ref.~\cite{HL}. This equation implies that the phase-estimation error obtained from the emitted field cannot be smaller than that of an optimal direct phase measurement on the cavity; since the conditioned cavity state has the same population statistics as the steady state, applying the number--phase uncertainty relation \cite{Wiseman_2009} to the latter yields, asymptotically,
\begin{equation}
\label{eq:irfhtcmgohhvnhkjfhfgfh897653958}
    \mathfrak{C}
    \lesssim
    \frac{512}{3}V(\hat n)^2.
\end{equation}

Now, we review the four conditions from Ref.~\cite{HL} that are required on the laser and its beam in the proof of the Heisenberg-limit theorem for laser coherence.
We make no assumptions about the physical medium storing excitations in the laser. We denote the laser stored excitation number operator as $\hat{n}_{\text{c}}$. It can be degenerate.

\emph{1. One-dimensional (1D) beam---} The beam propagates away from the laser in a particular direction, at a constant speed, and has a single transverse mode and a single polarization.
Mathematically, at any time $T\in\mathbb{R}$, the extant beam can be described as a scalar quantum field with a single argument $t\in(-\infty,T]$, which is the time at which that infinitesimal part of the beam, $b_t$, was created by the laser. The quantum state of $b_t$ is thus independent of $T$ (as long as $T>t$). The annihilation operator $\hat{b}(t)$ for the field satisfies
$[\hat{b}(t),\hat{b}^\dagger(s)] = \delta(t-s)$,
so that $\hat{b}^\dagger(t)\hat{b}(t)$ is the operator for photon flux (photons per unit time).

\emph{2. Endogenous phase---}
The coherence of the beam proceeds only from the laser. That is, a phase shift imposed on the laser state at some time $T$ will lead, in the future, to the same phase shift on the beam emitted after time $T$, as well as on the laser state. The phase shift at time $T$ on the laser is described by the unitary transformation $\hat{U}_c^{\zeta}=\exp(i\zeta \hat{n}_{\text{c}})$, where $\hat n_{\text{c}}$ is defined above.
The effect of this, at any time $T'>T$, on the state of the laser plus the beam segment generated in the interval $(T,T']$, is described by the unitary transformation $
\hat{U}_{\text{cb}}^{\zeta}=\exp(i\zeta(\hat{n}_\text{c}+\hat{n}_{\text{b}}))$,
where
$ \hat{n}_\mathrm{b}=\int_T^{T'} \hat{b}^\dagger(s)\hat{b}(s)\,\mathrm{d}s$
is the photon number operator for the generated beam segment.

\emph{3. Stationarity---}
The statistics of the laser and beam have a long-time limit that is unique and invariant under time translation. In particular, $\langle \hat{n}_\text{c}\rangle$ has a unique long time limit, $\mu$.

\emph{4. Ideal Glauber coherence---}
The beam is close (see below) to the ideal FAT laser beam. That is, it can be described as an eigenstate of $\hat b(t)$ (a coherent state) with eigenvalue $\beta(t)=\sqrt{\mathcal{N}} e^{i\phi(t)}$, where $\phi(t):=\sqrt{\ell}\,W(t)$, and $W(t)$ is a Wiener process \cite{mandel_wolf_1995,Wiseman_2009}, so that $\phi(t)$ is a stochastic phase undergoing pure phase diffusion.
By `close' to the ideal FAT laser beam, we mean that the first- and second-order Glauber coherence functions \cite{Glaubercoherence,mandel_wolf_1995} well
approximate \cite{HL} those of the ideal FAT laser beam. The first- and second-order Glauber coherence functions are defined as
\begin{equation}
\label{firstorderGlauber}
\begin{aligned}
G^{(1)}(s,t) &= \langle \hat{b}^\dagger(s)\hat{b}(t)\rangle,\\
G^{(2)}(s,s',t',t) &= \langle \hat{b}^\dagger(s)\hat{b}^\dagger(s')\hat{b}(t')\hat{b}(t)\rangle.
\end{aligned}
\end{equation}
$G^{(1)}$ yields the photon flux $\mathcal{N}=G^{(1)}(t,t)$, and the coherence \cite{HL}
\begin{equation}
\mathfrak{C}=\max_{\omega}\,\biggl|\int_{-\infty}^{\infty} G^{(1)}(s,t)\,e^{-i\omega s}\,ds\biggr|.
\end{equation}
The ideal beam coherence functions are evaluated by letting $\hat{b}(t)\to \beta(t)$. The requirement on $G^{(1)}$ in this condition is that the laser power spectrum is Lorentzian (or close to it \cite{HL}) with linewidth $\ell= 4\mathcal{N}/\mathfrak{C}$, while the requirement on $G^{(2)}$ implies that the photon statistics are Poissonian (or close to it \cite{HL}).

\subsubsection{The laser model that achieves the Heisenberg limit scaling}
\begin{figure}
\centering
\includegraphics[width=\linewidth]{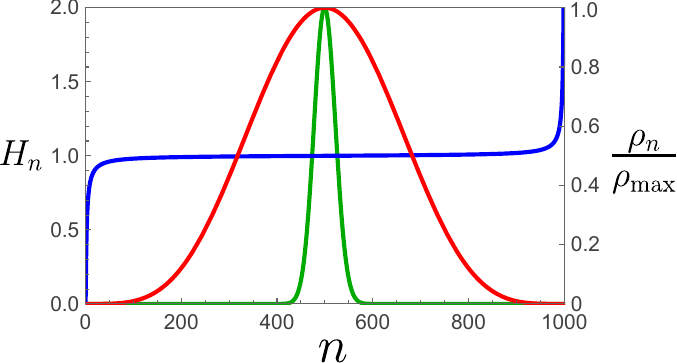}
\caption[Heisenberg-limited laser-model coefficients and population distributions]{ Heisenberg-limited laser model of dimension $D=1001$. The blue curve represents the off-diagonal coefficients given in Eq.~(\ref{ThesisHLcoeff}) and is plotted against the left axis. 
The red curve represents $\rho_n/\rho_{\mathrm{max}}$ for the HL laser, obtained from Eq.~(\ref{thesiseq:HLpop}). 
The green curve represents $\rho_n/\rho_{\mathrm{max}}$ for the Poissonian population distribution of a FAT laser with the same mean excitation number, $\mu=500$. 
The red and green curves are plotted against the right axis.}
\label{fig:HLpop}
\end{figure}
\label{thesissec:HLmodel}
Here, we present the laser model from Ref.~\cite{HL} that achieves the HL scaling for laser coherence of Eq.~(\ref{eq:HLscale}). 
This model is defined on $D$-dimensional Hilbert space, with a non-degenerate number operator $\hat{n}_{\mathrm{c}}=\sum_{n=0}^{D-1} n|n\rangle\langle n|$. The dynamics are described by the master equation 
\begin{equation}
\label{thesiseq:HLmodel}
    \dot{\rho}=\mathcal{D}[\hat{G}]\rho+\mathcal{D}[\hat{H}]\rho.
\end{equation}
Here, $\hat{G}$ and $\hat{H}$ are raising and lowering operators, respectively, with off-diagonal coefficients 
\begin{align}
G_n &:=\bra{n} \hat{G} \ket{n-1}= 1, \qquad \\ \label{ThesisHLcoeff}
H_n &:=  \bra{n-1} \hat{H} \ket{n} = \frac{ \sin^2\left(\pi \frac{n}{D+1}\right)}{ \sin^2\left(\pi \frac{n+1}{D+1}\right)},
\end{align}
for $1 \leq n \leq D-1$. We refer to $\mathcal{D}[\hat{G}]$ as isometric gain because $\hat{G}$ shifts the populations and coherences without changing them. The steady state is a diagonal matrix with the populations given by 
\begin{equation}
\label{thesiseq:HLpop}
    \rho_n \propto \sin ^4\left(\pi \frac{n+1}{D+1}\right).
\end{equation}
The off-diagonal coefficients $H_n$ of Eq.~(\ref{ThesisHLcoeff}) and the normalized populations $\rho_n/\rho_{\max}$ of Eq.~(\ref{thesiseq:HLpop}), where $\rho_{\max}=\max_n\rho_n$, are plotted in Fig.~\ref{fig:HLpop} for $D=1001$. The HL population distribution is compared with the Poissonian distribution of a FAT laser having the same mean excitation number $\mu$. As shown in the figure, the HL distribution has a much larger variance than the FAT distribution. More generally, the variance of the HL distribution scales as $D^2$, whereas that of the Poissonian distribution is $\mu$. Since for both population distributions $\mu$ scales linearly with $D$, the ratio of the two variances increases linearly with $D$. 
The off-diagonal coefficients $H_n$ are highly nonlinear, varying slowly in the central region and rapidly near the boundaries. More generally, as $D$ increases, their profile becomes progressively flatter in the central region and steeper near the boundaries. This slow variation of $H_n$ over the region in which the population is appreciable gives this laser model its high coherence, as will be shown in Sec.~\ref{sec:generalFAT}.

\subsection{Superradiance and superradiant lasers}
\label{subsec:superrandlasers}
\subsubsection{Dicke superradiance}
\label{subsec:Dicke}
Dicke superradiance refers to the collective, dissipative interaction of an ensemble of identical two-level quantum emitters that radiate into a single photonic mode, to which all emitters are coupled identically \cite{Dicke,superressay,mandel_wolf_1995}. 
The collective radiation dipole lowering operator is defined as $\hat{J}:=\sum_{a=1}^N \hat{\sigma}_a$ \cite{superressay,mandel_wolf_1995}, with $\hat{\sigma}_a=\ket{g_a}\bra{e_a}$ being the lowering operator for the two-level quantum emitter $a$, and $N$ is the number of superradiant quantum emitters. The dynamics of Dicke superradiance can be described using only the collective operators $\hat{J}$, $\hat{J}^\dagger$, and $\hat{J}_z=[\hat{J}^\dagger,\hat{J}]/2$ \cite{superressay}. 
These operators satisfy the usual SU(2) commutation relations \cite{superressay,mandel_wolf_1995}. 
Thus, the $2^N$-dimensional Hilbert space of two-level quantum emitters is spanned by the eigenstates of two commuting collective observables \cite{superressay,mandel_wolf_1995} 
\begin{equation}
    \hat{\mathbf{J}}^2 \ket{j,m,\alpha}= j(j+1) \ket{j,m,\alpha},
\end{equation}
\begin{equation}
    \hat{J}_z \ket{j,m,\alpha}= m\ket{j,m,\alpha}.
\end{equation}
Here, the quantum number $j$ is referred to as the total pseudo-angular momentum. When $N$ is even, $j\in\{0,1,\ldots,N/2$\}; when $N$ is odd, $j\in \{1/2,3/2,\ldots,N/2$\}. The quantum number $m$, which is referred to as the pseudo-angular momentum projection on the $z$-axis, ranges from $-j$ to $j$ in unit steps \cite{superressay,mandel_wolf_1995}. Finally, $\alpha$ is an additional quantum number that accounts for the degeneracy.

Any dynamics that can be described using only collective operators (these can include Dicke superradiance dissipative dynamics or the interaction of quantum emitters with coherent light, where each quantum emitter has the same Rabi frequency), conserve the quantum numbers $j$ and $\alpha$. In mathematical terms, this means that the $2^N$-dimensional Hilbert space can be decomposed into a sum of irreducible representations (irreps) for the SU(2) group, each with a dimension of $2j+1$ and represented by the two quantum numbers $j$ and $\alpha$ \cite{jones1998groups,U(3)}. The quantum number $\alpha$ represents the multiplicity of the degenerate and isomorphic irreps with the same quantum number $j$. 
The off-diagonal elements for the lowering operator $\hat{J}$ are given by \cite{superressay,mandel_wolf_1995} 
\begin{equation}
\label{eq:coeff}
\bra{j,m-1,\alpha}\hat{J}\ket{j,m,\alpha'}
=\delta_{\alpha,\alpha'} J_{j,m}.
\end{equation}
Here, $J_{j,m}$ are defined as
\begin{equation}
\label{eq:J_{j,m}}
    J_{j,m}:=
\sqrt{j\left(j+1\right)-m(m-1)}.
\end{equation}
The case we consider in this paper is the non-degenerate irrep $j=N/2$.
This irrep comprises $N+1$ states, $\ket{N/2,m}$, where $m$ is a half-integer ranging from $-N/2$ to $N/2$. Denote by $\ket{n_1, n_2}_\text{S}$ the state of symmetric superposition of $\ket{g}^{\otimes n_1} \otimes \ket{e}^{\otimes n_2}$, where $n_1+n_2=N$ \cite{superressay}. Then, we have $\ket{N/2,m}=\ket{N/2-m,N/2+m}_\text{S}$ \cite{superressay}.
For this irrep, we use the notation 
\begin{equation}
\label{eq:J_m}
J_m:=J_{N/2,m}=\sqrt{N/2\left(N/2+1\right)-m(m-1)}. 
\end{equation}

\subsubsection{Coherently radiating spin states (CRSS)}
\label{sec:CRSS}
Here, we discuss the coherently radiating spin states (CRSS) from Ref.~\cite{myarticle}. Further details on the CRSS can be found there, and in Secs.~\ref{sec:metastable} and \ref{sec:poissonian}. As we will see in Sec.~\ref{sec:metastable}, the metastable state in which our laser model operates has the same excitation number statistics as those of the CRSS.

The CRSS, defined below, are pure states that are approximate eigenstates of the collective lowering operator $\hat{J}$. Consequently, an ensemble of two-level quantum emitters prepared in a CRSS and coupled identically to a common radiation mode through $\hat{J}$ emits coherent light into that mode, as will be shown below.
The CRSS arise as the (approximate) steady-state solution of driven-dissipative Dicke superradiance.
Driven-dissipative Dicke superradiance consists of $N$ two-level atoms coherently driven by resonant laser light with Rabi frequency $\Omega$ and identically coupled to a common dissipative optical mode.
The master equation for the system is given by 
\begin{equation}
\label{thesiseq:DDsuperradiance}
\dot{\rho}=-i[\Omega\hat{J}^{\dagger}+\Omega^{*}\hat{J}
,\rho]+\Gamma\mathcal{D}[\hat{J}]\rho \, .
\end{equation}
Here, $\Gamma$ is the single atom emission rate into the common optical mode. The master equation (\ref{thesiseq:DDsuperradiance}) comprises only collective atomic operators, and thus, the different irreps of the SU(2) group can be analyzed separately (see Sec.~\ref{subsec:Dicke}). When all atoms are initialized in the ground state, corresponding to the state $\ket{N/2 -N/2}$, the system will remain in the Hilbert space with $j=N/2$.

If one assumes that there exist pure states that are eigenstates of the collective atomic lowering operator $\hat J$,
then one of these will be the steady state solution to the master equation (\ref{thesiseq:DDsuperradiance}) (see below).
Strictly, since $\hat {J}^{2N+1}=0$, the operator $\hat J$ does not have any exact eigenstates (see below). Nevertheless, it is useful to consider the approximate eigenstate relation 
\begin{equation}
\label{thesiseq:eig}
 \hat J \ket{\psi} \approx \alpha \ket{\psi}.
\end{equation}
Assuming Eq.~(\ref{thesiseq:eig}) holds exactly, and
substituting Eq.~(\ref{thesiseq:eig}) into Eq.~(\ref{thesiseq:DDsuperradiance}), we find that 
\begin{equation}
\label{thesiseq:alpha}
  \alpha=-2i{\Omega}/{ \Gamma} .
\end{equation}
To find the approximate solution for Eq.~(\ref{thesiseq:eig}), we expand the state in the Dicke basis and restrict the dynamics to a fixed irrep with a $j$-value 
\begin{equation}
\label{eq:fgfgjkgfjuhruaer}
    \ket{\psi}=\sum_{m=-j}^{j} \tilde{a}_m \ket{j,m}.
\end{equation}
Substituting Eq.~(\ref{eq:fgfgjkgfjuhruaer}) into Eq.~(\ref{thesiseq:eig}) as an equality yields the following $2j+1$ conditions
\begin{equation}
\label{eq:dfgjklSDF}
\tilde{a}_m=\frac{\alpha}{J_{j,m}}\tilde{a}_{m-1},
\end{equation}
for $-j<m\leq j$, and 
\begin{equation}
     \tilde{a}_{j}=0.
\end{equation}
Here, the coefficients $J_{j,m}$ are defined in Eq.~(\ref{eq:J_{j,m}}).
These conditions admit only the trivial solution, $\tilde{a}_m=0$ for all $m$.
Now, we identify the states that most closely satisfy Eq.~(\ref{thesiseq:eig}).
This will be done in an informal way below.
Since it is impossible to satisfy all $2j+1$ conditions simultaneously, at least one must be violated. We therefore 
choose a state that satisfies the recursion relation in Eq.~(\ref{eq:dfgjklSDF}) for $m\leq m_{\text{t}}$ and set $\tilde{a}_m=0$ for $m>m_{\text{t}}$; that is, we truncate the state at $m=m_{\text{t}}$. With this construction, only the condition corresponding to $m=m_{\text{t}}+1$ in Eq.~(\ref{eq:dfgjklSDF}) is violated.
We choose the truncation point $m_{\mathrm{t}}$ at the minimum of the solution to the recursion relation in Eq.~(\ref{eq:dfgjklSDF}), thereby minimizing the resulting violation of the conditions above.
Denote by $m_{\pm}$ the solutions to the equation $\alpha=J_{j,m}$ (rounded to integers). These are the minimum and maximum of the solution to the recursion relation (\ref{eq:dfgjklSDF}), respectively. These are given approximately by
\begin{equation}
\label{eq:mpmthesis}
    m_{\pm} \sim \pm j\sqrt{1-r^{2}}.
\end{equation}
Here, $\sim$ denotes the asymptotic limit of a large $j$, and we defined the normalized driving rate
\begin{equation}
    r:=\vert\alpha\vert/j.
\end{equation}
Thus, we define the CRSS as the solution of the recursion relation Eq.~(\ref{eq:dfgjklSDF}) truncated at  $m_{\text{t}}=m_+$: 
\begin{equation}
\label{eq:yofhjksjhdgkfng}
    \ket{j,\alpha}_{\hat J}= \sum_{m=-j}^{m_+}  a_{m}  \ket{j,m}=\frac{1}{S} \sum_{m=-j}^{m_+}  \frac{\alpha^{m+j}}{\prod_{a=-j+1}^{m}J_{j,a}}  \ket{j,m}.
\end{equation}
Here, $a_{m}$ satisfy the recursion relation Eq.~(\ref{eq:dfgjklSDF}), and $S$ is a normalization constant, so that the state in Eq.~(\ref{eq:yofhjksjhdgkfng}) is normalized.
These states are approximate eigenstates of $\hat{J}$ when $a_{m_+}$ is small enough. A rough estimation for $a_{m_+}$ in the limit of large $j$ (see Ref.~\cite{myarticle} and Sec.~\ref{sec:metastable}), is
\begin{equation}
\label{eq:dfgsghzdtygjzdfghzdfhzdfh}
    {a_{m_+}} \sim j^{3 / 4}\left(r^2 \sqrt{1-r^2} / \pi\right)^{1 / 4} e^{-2j g(r)},
\end{equation}
Here, $\sim$ indicates asymptotic scaling in the limit of large $j$, and we have defined the function 
\begin{equation}
g(r):=\text{arctanh}[\sqrt{1-r^2}]-\sqrt{1-r^2}.
\end{equation}
Thus, $a_{m_+}$ decays exponentially with $j$, and $g(r)$ is decreasing with $r$.
To estimate where $a_{m_+}$ is exponentially suppressed, we need to find the values of $r$ for a given large $j$ that the negative exponent in Eq.~(\ref{eq:dfgsghzdtygjzdfghzdfhzdfh}), $2jg(r)$, is large enough. To do that, we
first define $r_j$ as the solution to the equation $2j g(r_j)=1$. Expanding $g(r)$ around $r=1$ we obtain the asymptotic expression for $r_j$ in the limit $j\rightarrow\infty$ 
\begin{equation}
\label{thesiseq:r_j}
r_j \sim \sqrt{1-\left(\frac{3}{2 j}\right)^{2 / 3}}.
\end{equation}
Expanding the negative exponent in Eq.~(\ref{eq:dfgsghzdtygjzdfghzdfhzdfh}), $2jg(r)$, around $r_j$, and requiring it to be a lot larger than 1 yields
\begin{multline}
    2jg(r)=2j(g(r_j)+g'(r_j)(r-r_j)) \\
    \approx1+2\left(\frac{3}{2}\right)^{1/3}j^{2/3}(r_j-r)\gg 1
\end{multline}
Here, we used the asymptotic expression of Eq.~(\ref{thesiseq:r_j}) when evaluating $g'(r_j)$.
Thus, in the asymptotic limit $j\rightarrow \infty$, when 
\begin{equation}
\label{eq:ritjydfgjdfgdrydrgsdfsdgdfhxftgjfg}
    r_j-r\gg j^{-2/3},
\end{equation} 
$a_{m_+}$ is exponentially suppressed, and the CRSS can be treated as approximate eigenstates. 
Equation (\ref{eq:ritjydfgjdfgdrydrgsdfsdgdfhxftgjfg}) does not mean that the ratio of the right-hand side to the left-hand side vanishes as $j\rightarrow\infty$; It just means that the ratio is quite a small number. Note that $r_j\rightarrow1$ as $j\rightarrow\infty$.

To estimate the regime in which the CRSS are approximate steady state solutions of Eq.~(\ref{thesiseq:DDsuperradiance}), one sets $j=N/2$ for atoms initially prepared in the ground state and uses Eqs.~(\ref{thesiseq:alpha}) and (\ref{thesiseq:r_j}).
Outside the regime of Eq.~(\ref{eq:ritjydfgjdfgdrydrgsdfsdgdfhxftgjfg}) in which the CRSS provide approximate solutions of Eq.~(\ref{thesiseq:DDsuperradiance}), the steady-state solution is no longer a pure state \cite{ReyPRA}.
For $j\gg 1 $, and in the regime of Eq.~(\ref{eq:ritjydfgjdfgdrydrgsdfsdgdfhxftgjfg}), the coefficients $\left|a_m\right|^2$ are well described by a Gaussian (see Ref.~\cite{myarticle} and Sec.~\ref{sec:metastable})
\begin{equation}
    \left|a_m\right|^2 \approx \frac{1}{\sqrt{2 \pi w^2}} e^{-{\left(m-m_{-}\right)^2}/{2 w^2}},
\end{equation}
of width $w^2=j r^2 /\left(2 \sqrt{1-r^2}\right)$.

Finally, we show the system of Eq.~(\ref{thesiseq:DDsuperradiance}) described above, operating in a steady state, in the regime of Eq.~(\ref{eq:ritjydfgjdfgdrydrgsdfsdgdfhxftgjfg}) emits coherent light.
Using input output theory \cite{Wiseman_2009}, the radiation dipole of the output mode in the Heisenberg picture satisfies
\begin{equation}
\label{eq:iuhueahrtu3489894tu34tueihgf}
\hat{b}_{\text{out}}=\hat{b}_{\text{in}}+\sqrt{\Gamma}\hat{J}
\end{equation}
Here, $\hat{b}_{\text{in}}$ and $\hat{b}_{\text{out}}$ are the annihilation operators for the input and output fields, respectively. Since the system is in a pure state $\ket{N/2,\alpha}_{\hat{J}}$ at all times, the mode of the output field has to be in a pure state as well. Thus, using Eq.~(\ref{eq:iuhueahrtu3489894tu34tueihgf}), we obtain
\begin{multline}
    \hat{b}_{\text{out}} \ket{N/2,\alpha}_{\hat{J}} \ket{0}_{\text{in}}
    =(\hat{b}_{\text{in}}+\sqrt{\Gamma}\hat{J}) \ket{N/2,\alpha}_{\hat{J}} \ket{0}_{\text{in}} \\
    =\sqrt{\Gamma}\alpha \ket{N/2,\alpha}_{\hat{J}} \ket{0}_{\text{in}}.
\end{multline}
Here, $\ket{0}_{\text{in}}$ is the vacuum state of the output mode before interacting with the system, which satisfies $\hat{b}_{\text{in}}\ket{0}_{\text{in}}=0$.
Thus, the output mode is an eigenstate of its bosonic annihilation operator and therefore it is a coherent state, as stated.

In this paper, we propose a superradiant laser model with ideal Poissonian output photon statistics.
We assume that the system couples to the output field through the collective lowering operator $\hat{J}$, such that the corresponding loss process is described by the superoperator $\Gamma\mathcal{D}[\hat{J}]$. Since two level emitters in a CRSS state emit coherent light through this channel, this suggests that the laser steady state should be an incoherent mixture of the CRSS. Furthermore, since a laser has no absolute or mean phase, we can expect, analogously to the FAT-laser steady state in Eq.~(\ref{eq:incoherentmix123}), an incoherent mixture of uniform phase of the CRSS with fixed amplitude $r$,
\begin{equation}
    \rho^{\mathrm{ms}}
    =
    \frac{1}{2\pi}
    \int_{0}^{2\pi}
    \ket{N/2,re^{i\theta}}
    \bra{N/2,re^{i\theta}}
    \,\mathrm{d}\theta.
\end{equation}
Because the CRSS are approximate, rather than exact, eigenstates of $\hat{J}$, this state may be metastable rather than stationary, as discussed in Sec.~\ref{sec:metastable}.

\subsubsection{The standard superradiant laser model- review}
\label{subsec:superradiantlaserreview}
Here, we review the standard superradiant laser model, including its physical motivation, basic operating principles, and main results. In Appendix \ref{app:superrdiantlasersemi}, we provide an analysis  for the semiclassical dynamics of the standard superradiant laser model. 

Conventional lasers, using an atomic gain medium, operate in the good cavity limit where the resonance bandwidth of the cavity is significantly narrower than the atomic spectral width. The achievable emission bandwidth is limited by intrinsic thermal fluctuations of the cavity materials \cite{thermal1,thermal2,thermal3}, hindering further improvements in the precision of atomic clocks \cite{RevModPhys.87.637}. To overcome these limitations, an alternative laser design, known as the superradiant laser, utilizes alkaline earth atoms with an ultra-narrow clock transition combined with a relatively broad cavity bandwidth.  In this bad cavity regime, steady state superradiance is achievable where the average intracavity photon number can be even below one. The phase is retained in the atoms, resulting in a significant reduction of sensitivity to thermal fluctuations of the cavity and technical noise sources \cite{Nature2012,PhysRevX.8.021036}.

The standard superradiant laser model was first introduced by Meiser \textit{et al.}~\cite{PhysRevLettHolland,PhysRevA2010,intensityfluctuations}.
It consists of $N$ cold atoms trapped inside a bad cavity that supports a single relevant mode with decay rate $\kappa$. The atoms are arranged in transverse layers and are identically coupled to the cavity mode with coupling constant $g$.
The atoms are pumped non-collectively from $\ket{g}$ to $\ket{e}$. This can be achieved by resonantly driving a transition from $\ket{g}$ to a third auxiliary state $\ket{e'}$, which rapidly decays to $\ket{e}$.
Assuming the bad-cavity limit $gN \ll \kappa$, the dynamics are governed by the master equation \cite{PhysRevA2010}
\begin{equation}
\label{eq:basicsuperr}
\dot{\rho}=\nu\Gamma\sum_{n=1}^N \mathcal{D}[\hat{\sigma}^\dagger_n]\rho+\Gamma\mathcal{D}[\hat{J}]\rho.
\end{equation}
Here, $\Gamma=4g^2/\kappa$ is the single-atom emission rate into the output beam, and $\nu$ is a dimensionless pumping rate. The first term in Eq.~(\ref{eq:basicsuperr}) describes non-collective pumping of the atoms, whereas the second term describes collective emission. Despite the fact that the non-collective pumping of the atoms breaks the SU(2) symmetry of Dicke superradiance, the atoms are still highly correlated in phase \cite{PhysRevA2010}, and produce radiation with intensity proportional to $N^2$ \cite{PhysRevA2010,PhysRevLettHolland} (see below).
Equation~(\ref{eq:basicsuperr}) includes only non-collective pumping and collective emission. In a realistic physical system, noise arises from individual spontaneous emission at rate $\gamma$ and dephasing at rate $1/T_2$. These processes can be neglected when the collective decay rate dominates, namely $\Gamma N \gg \gamma, 1/T_2$ \cite{PhysRevLettHolland,PhysRevA2010}.

To leading order in $N$, and in the regime $\nu=\Theta(N)$, i.e.~$\nu$ of order $N$, the mean number of excitations at steady state is given by \cite{PhysRevA2010}
\begin{equation}
\label{eq:mmm6}
    \mu=\begin{cases}
\frac{1}{2}\left(N+\nu\right), & \nu<N\\
N, & \nu\geqslant N
\end{cases},
\end{equation}
and the steady-state flux by \cite{PhysRevA2010}
\begin{equation}
\label{eq:fl}
    \mathcal{N}=\Gamma N^{2}\times\left\{ \begin{array}{ll}
\frac{1}{2}\frac{\nu}{N}\left(1-\frac{\nu}{N}\right), & \nu<N\\
0, & \nu\geqslant N
\end{array}.\right.
\end{equation}
From Eq.~(\ref{eq:fl}) we can extract the maximum flux, $\mathcal{N}_{\max}=\Gamma N^{2}/8$, obtained at $\nu=N/2$ \cite{PhysRevA2010}. The scaling of that flux with $N^{2}$ and indicates the presence of strong correlations between atoms \cite{PhysRevA2010}, and thus the collective nature of the light emission \cite{PhysRevA2010}.
$\mathcal{N}_{\max}$ is only a factor $1/2$ smaller than the maximum possible emission rate that can be achieved by any state of the atoms, $\Gamma\left\langle N/2,0\left|\hat{J}^{\dagger}\hat{J}\right|N/2,0\right\rangle=\Gamma N^2/4$ \cite{PhysRevA2010}.

The linewidth of the superradiant laser in the regime $\nu=\Theta\left(N\right)$, with $\nu<N$,
is given by 
\cite{PhysRevA2010} 
\begin{equation}
\label{eq:line}
    \ell=\Gamma\left(1+\frac{2 \nu}{N}\right).
\end{equation}
Equation (\ref{eq:line}) shows that the linewidth is bounded from below by $\ell_\text{min}=\Gamma$ and is given by $\ell=2\Gamma$ at peak flux ($\nu=N/2$).
Using Eqs.~(\ref{eq:C}), (\ref{eq:mmm6}), (\ref{eq:fl}) and (\ref{eq:line}), we calculate the coherence of the superradiant laser at peak flux, $\nu=N/2$ 
\begin{equation}
\label{eq:standardsuperrcoh}
    \coh=\frac{1}{4}N^2=\frac{4}{9}\mu^2=\Theta(\mu^2).
\end{equation}
The coherence scales quadratically with $\mu$, like the SQL of Eq.~(\ref{eq:SQL}), however with a prefactor that is much lower than that of the SQL, which is 8. 
Additionally, using Eqs.~(\ref{eq:C}), (\ref{eq:mmm6}), (\ref{eq:fl}), (\ref{eq:line}), we find the maximal achievable coherence as a function of the number of excitation in the laser, $\mu$, obtained at $\nu/N=0.23$ which is 
\begin{equation}
\label{eq:coh/mu^2}
 \coh=0.64\mu^2=0.08\coh_{\text{SQL}}.   
\end{equation}
Moreover, using Eqs.~(\ref{eq:C}), (\ref{eq:fl}), (\ref{eq:line}), we find the maximal achievable coherence as a function of the number of atoms, $N$ obtained at $\nu/N=0.366$ which is 
\begin{equation}
\label{eq:coh/N^2}
    \coh=0.27N^2.
\end{equation}

\section{Model}
\label{sec:model}
Here, we present the superradiant laser model analyzed in this paper. We begin in Sec.~\ref{subsec:generalmodelsphasenoise} by introducing the generalized FAT laser models, which form the framework within which our model is defined. Second, we present a generalized FAT laser model that achieves the HL scaling of the laser coherence, Eq.~(\ref{eq:HLscale}). Finally, we introduce our superradiant laser model as a specific generalized FAT laser model. Its physical realization and the derivation of its master equation will be presented in Sec.~\ref{sec:physicalrealzation}.

\subsection{Generalized FAT laser models}
\label{subsec:generalmodelsphasenoise}

Here, we introduce the generalized FAT laser models. These models extend the FAT laser model of Eq.~(\ref{eq:FAT}). An analysis of the generalized FAT laser models will be presented in Sec.~\ref{sec:generalFAT}.

The generalized FAT laser models are defined on a Hilbert space of dimension $D$ (which may be infinite), with a nondegenerate number basis $\{\ket{m}\}_{m=0}^{D-1}$. The master equation is obtained by replacing the \emph{linear} (bosonic) lowering operator $\hat{a}$ with a \emph{non-linear} lowering operator $\hat{L}$ in the FAT laser master equation~(\ref{eq:FAT}) (also, replacing $\kappa$ with $\Gamma$ and $P/\kappa$ with $\nu$):
\begin{equation}
\label{eq:generalmodel}
\dot{\rho}=\Gamma\{\nu\mathcal{D}[\hat{L}^{\dagger}]\mathcal{A}[\hat{L}^{\dagger}]^{-1}\rho+\mathcal{D}[\hat{L}]\rho\}.
\end{equation} 
Here, $\hat{L}:=\sum_{m=1}^{D-1} L_m \ket{m-1}\bra{m}$ is a general lowering operator with off-diagonal coefficients $L_m$, $\Gamma$ has units of frequency, and $\nu$ is a dimensionless positive continuously tunable pumping rate. 
We assume that the off-diagonal matrix elements $L_m$ of the lowering operator $\hat{L}$ are sampled from a slowly varying function of $m$, in a sense that will be made precise later.

\begin{figure}
    \centering
\includegraphics[width=\linewidth]{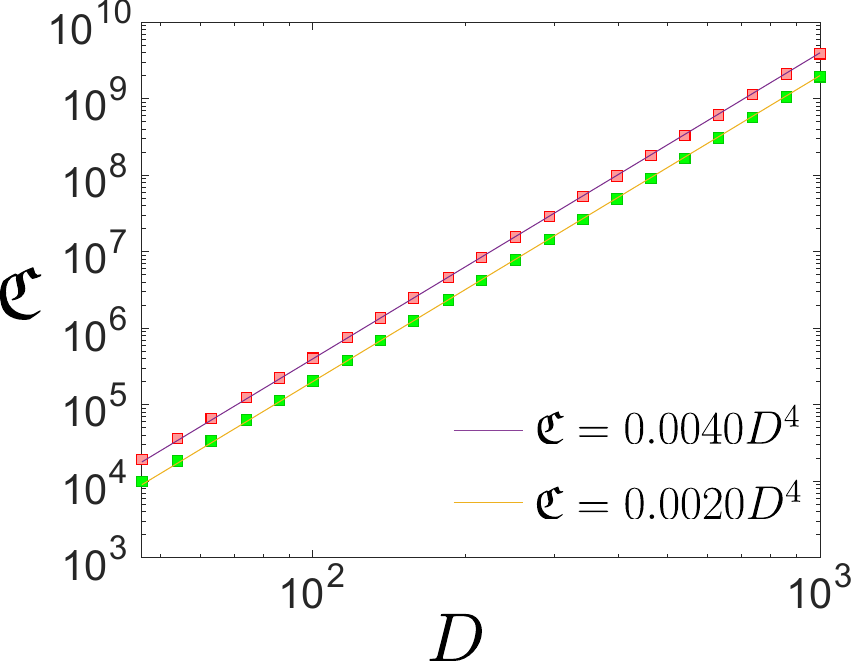}
    \caption[Coherence for generalized FAT Heisenberg limited laser model]{Numerical results of $\mathfrak{C}$ as a function of the dimension of the Hilbert space, $D=2\mu+1$, for the model of Eq.~(\ref{eq:HL}) (green squares) and for the model from Ref.~\cite{HL}, Eq.~(\ref{thesiseq:HLmodel}) (red squares). 
    The lines represent fits to $\coh\propto D^4$, with fitting factors $\coh =0.0040 D^4$ (purple) and $\coh =0.0020 D^4$
  (orange).}
    \label{fig:HL}
\end{figure}

In the following two subsections, we discuss two specific generalized FAT laser models. In Sec.~\ref{subsec:HL}, we consider a model closely related to the HL model of Eq.~(\ref{thesiseq:HLmodel}), which achieves the HL scaling of the laser coherence, Eq.~(\ref{eq:HLscale}). This model is obtained from Eq.~(\ref{eq:generalmodel}) by choosing $\hat{L}=\hat{H}$ and $\nu=1$. Here, $\hat{H}$ is the lowering operator with off-diagonal coefficients given in Eq.~(\ref{ThesisHLcoeff}). In Sec.~\ref{subsec:ourmodel}, we introduce the generalized laser model analyzed throughout this paper. It corresponds to choosing $\hat{L}=\hat{J}$ in Eq.~(\ref{eq:generalmodel}), where $\hat{J}$ is the superradiant lowering operator with off-diagonal coefficients given in Eq.~(\ref{eq:J_m}).

\subsection{Heisenberg-limited laser}
\label{subsec:HL}
Here, we present a generalized FAT laser model that achieves the HL scaling of Eq.~(\ref{eq:HLscale}). This model is obtained by choosing $\hat{L}=\hat{H}$ in Eq.~(\ref{eq:generalmodel})  and setting $\nu=1$
\begin{equation}
\label{eq:HL}
    \dot{\rho}= \Gamma \{\mathcal{D}[\hat{H}^{\dagger}]\mathcal{A}[\hat{H}^{\dagger}]^{-1}\rho+\mathcal{D}[\hat{H}]\rho\}. 
\end{equation}
Here, $\hat{H}$ is the same lowering operator in the HL model of Eq.~(\ref{thesiseq:HLmodel}). 
Here, the system has the same Hilbert space as the model of Eq.~(\ref{thesiseq:HLmodel}), with dimension $D$. The models of Eqs.~(\ref{eq:HL}) and (\ref{thesiseq:HLmodel}) share the same loss superoperator. The two models also obey the same population-dynamics equation \cite{howard1999}, and therefore have the same steady state, and mean excitation number, $\mu=(D-1)/2$.

Figure~\ref{fig:HL} shows the numerical results for the coherence $\coh$ as a function of the Hilbert-space dimension $D=2\mu+1$ for both models. The numerical method employed to obtain the results is similar to the one used in Sec.~\ref{sec:coherenceprop} (see Appendix \ref{subsec:numericalmethod} and Sec.~\ref{sec:generalFAT}).
In Fig.~\ref{fig:HL}, the fitted curves, $\coh\propto D^4$, agree closely with the numerical data, confirming the HL scaling of Eq.~(\ref{eq:HLscale}). Notably, the coherence of the model of Eq.~(\ref{thesiseq:HLmodel}) is twice that of the model of Eq.~(\ref{eq:HL}). This difference arises because the gain superoperator in the model of Eq.~(\ref{thesiseq:HLmodel}), which we referred to as isometric gain, does not introduce phase noise \cite{howard1999}: its off-diagonal coefficients are constant. By contrast, in the model of Eq.~(\ref{eq:HL}), the gain and loss processes contribute equally to the phase noise (see Sec.~\ref{subsec:generalcoherenceporperties}).

\subsection{Our laser model}
\label{subsec:ourmodel}

Here, we present the generalized FAT laser master equation that is the main subject of this paper:
\begin{equation}
\label{eq:model}
\dot{\rho}=\Gamma\{\nu\mathcal{D}[\hat{J}^{\dagger}]\mathcal{A}[\hat{J}^{\dagger}]^{-1}\rho+\mathcal{D}[\hat{J}]\rho\}.
\end{equation}
Here, $\Gamma$ is the single atom emission rate into the output beam, $\nu$ is the dimensionless pumping rate, and $\hat{J}$ is the superradiant lowering operator with the off-diagonal coefficients of Eq.~(\ref{eq:J_m}).
Strictly, the superoperator $\mathcal{A}[\hat{J}^{\dagger}]$ is not invertible; we avoid the superfluous divergence by using the definition of Eq.~(\ref{eq:redefine}) in Sec.~\ref{sec:Markovdynamics}. The second term in Eq.~(\ref{eq:model}), which represents the loss superoperator, is the collective atomic decay and is the same as the standard superradiant laser model. On the other hand, the first term, which is the gain superoperator, is a term of collective pumping. It differs from the standard superradiant laser model, which has a non-collective pumping mechanism (see Sec.~\ref{subsec:superradiantlaserreview}). 
Thus, the total master equation of our model, Eq.~(\ref{eq:model}), is composed only of collective operators and thus does not mix among the different irreps of the SU(2) group.
Our system is naturally initialized in the state where all the atoms are in the ground state, represented by $\ket{N/2, -N/2}$. Thus, the system remains within the SU(2) non-degenerate irrep $j = N/2$, with dimension $D=N+1$.

\begin{figure}
\centering
\includegraphics[width=\linewidth]{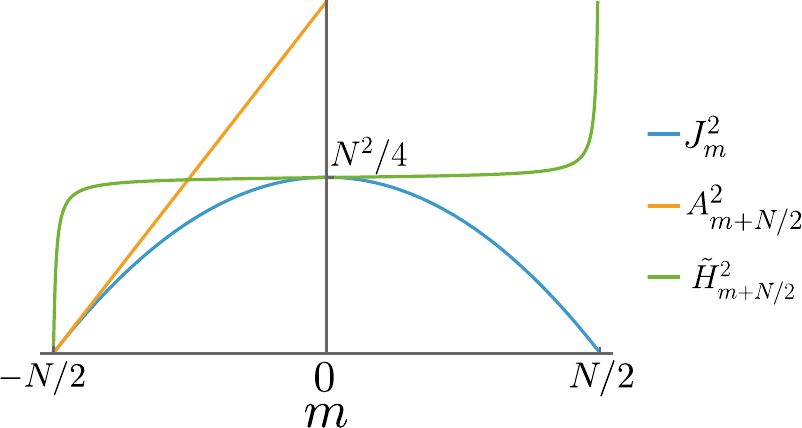}
\caption[Off-diagonal coefficients comparison for the generalized FAT laser models]
{Plot of the off-diagonal elements squared for the lowering operators of the three generalized FAT laser models as a function of $m$, where $m+N/2$ is the number of excitations: for our superradiant laser model $J^2_{m}$ (blue); standard FAT laser $A^2_{m+N/2}:={Na^2_{m+N/2}}$ (orange); and the Heisenberg-limited laser model of Eq.~(\ref{eq:HL}) (with $D=N+1$) $\tilde{H}^2_{m+N/2}=H^2_{m+N/2}N^2/4$ (green). Here $a_n$ are the coefficients of the standard annihilation operator, while $H_n$ is given in Eq.~(\ref{ThesisHLcoeff}). The scaling factors for the elements were chosen to make the relevant comparisons visually apparent. Here, $N=1000$, which can be considered the asymptotic limit of large $N$, so that the $J_m$ curve exhibits a scale-invariant shape (with the $A_m$ curve exhibiting scale invariance for any $N$). The $H_m$ curve sharpens with increasing $N$, flattening in the middle and steepening near the ends.}
\label{fig:coefficients}
\end{figure}

To compare the two generalized FAT laser models introduced in this section, Eqs.~(\ref{eq:HL}) and (\ref{eq:model}), with the standard FAT laser model, Eq.~(\ref{eq:FAT}), we examine the off-diagonal coefficients of their respective lowering operators. These coefficients are plotted in Fig.~\ref{fig:coefficients}. 
If our laser model operates with low pumping, then it will be in the linear regime, where the coefficients $J_m^2$ are linear as function of the excitation number $m+N/2$ (see Fig.~\ref{fig:coefficients}), so the laser will operate at the SQL. However, a superradiant laser with higher excitation levels, i.e.,~a mean population $\mu$ close to $N/2$  (the center in Fig.~\ref{fig:coefficients}), has flatter coupling, and more resembles the Heisenberg-limited laser model. It can thus be expected to exhibit reduced phase noise and possibly surpass the SQL scaling in terms of coherence $\mathfrak{C}$ as a function of the number of excitations $\mu$. In Sec.~\ref{sec:generalFAT}, we explicitly find the linewidth and the coherence of the generalized FAT laser and we verify that flat coupling constants indeed lead to higher coherence. Operating with a mean population $\mu$ close to $N/2$ is clearly not trivial to achieve because of the decline of the superradiant coefficients $J_m$ beyond $m=0$. This implies that a stable steady state with a population near $m=0$ might not exist. This final point will be discussed comprehensively in Sec.~\ref{sec:metastable}. 

\section{Physical realization}
\label{sec:physicalrealzation}
In this section, we present a physical realization for the model of Eq.~(\ref{eq:model}).

\subsection{Physical model}
\subsubsection{Physical implementation}
\label{subsubsec:Implementation}
\begin{figure}
\centering
\includegraphics[width=\linewidth]{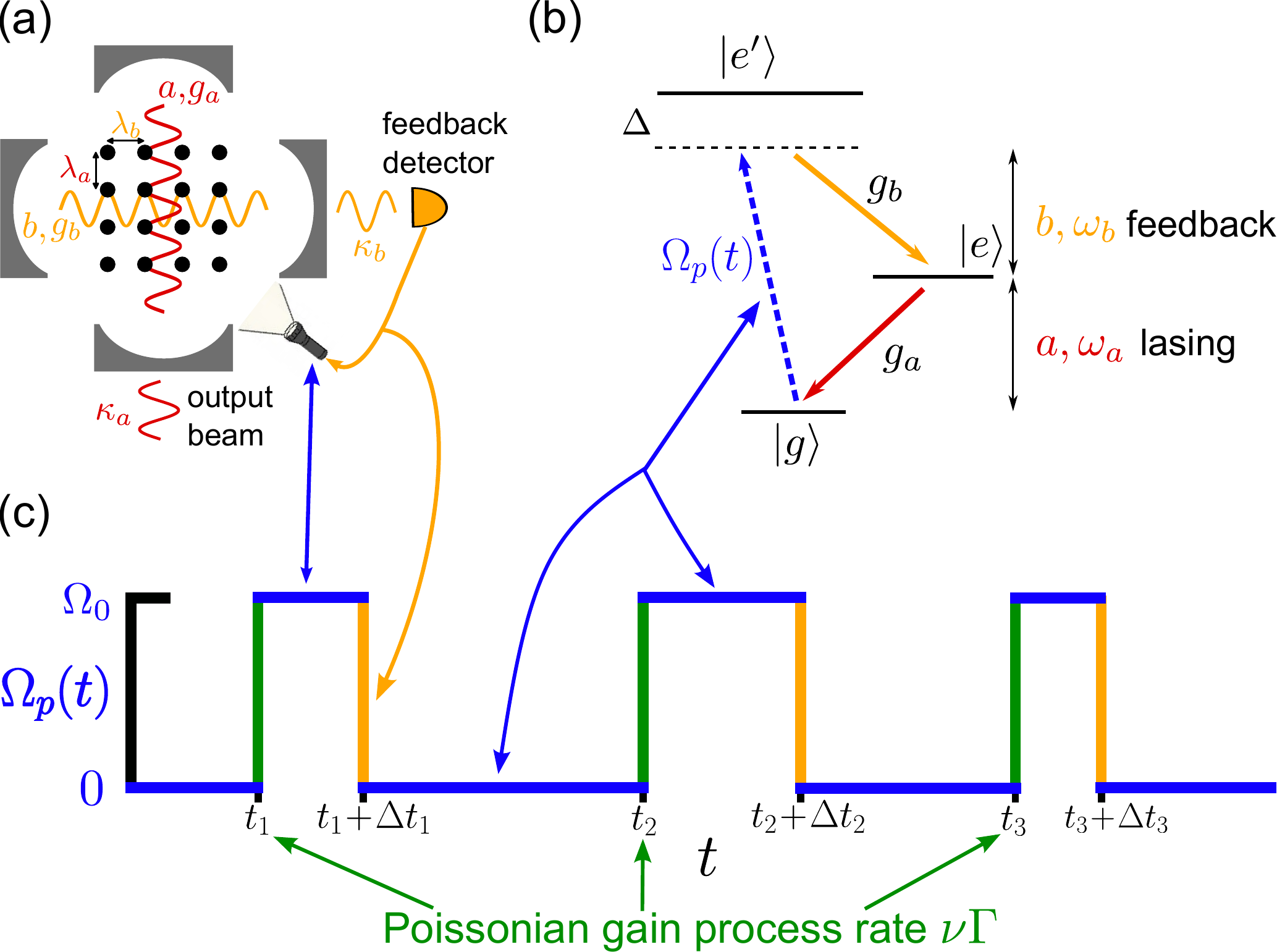}
\caption[Physical realization]{Physical realization for the model of Eq.~(\ref{eq:model}). (a) The physical setup. An array of atoms (black dots) inside two cavities (grey), each supporting one relevant mode (mode $a$ in red and mode $b$ in orange). The array is enlarged for visual appearance; its actual spatial extent is much smaller than the cavity dimensions.
The sinusoidal curves in green and red indicate the spatial dependence of the atom–mode coupling amplitude along the axis of each cavity for the corresponding mode.  The output of mode $a$ yields the coherent beam produced by the laser model. The output from the ``gain'' cavity (mode $b$) is measured continuously by a single-photon detector (orange).  (b) Energy level structure of the atoms. The colors for the transitions match the color of the modes from (a). (c) The drive (Rabi frequency $\Omega(t)$, blue) as a function of time, composed of short ``on'' periods [$\Omega(t)=\Omega_0$] between longer ``off'' periods [$\Omega(t)=0$]. An ``on'' period is initiated by a Poisson process in time (green) and is terminated upon detection of a photon (orange) in the detector.}
    \label{fig:physicalrealization}
\end{figure}
The physical setup consists of an array of trapped three-level atoms, with levels $\ket{g}$ and $\ket{e}$ as in Sec.~\ref{subsec:superrandlasers}, and a third auxiliary level $\ket{e'}$. The transition 
$\left|g\right\rangle \leftrightarrow|e'\rangle$ is driven using a laser beam (frequency $\omega_{\text{L}}$) with a real Rabi frequency $\Omega(t)\in\{0,\Omega_0\}$ (Figs.~\ref{fig:physicalrealization}b and \ref{fig:physicalrealization}c). 
We set the detuning of the laser $\Delta$ away from the transition $\ket{g}\leftrightarrow\ket{e'}$, i.e.~$\omega_{\text{L}}=\omega_{e'g}-\Delta$. The atomic array is held inside two cavities, each supporting one relevant mode, as shown in Fig.~\ref{fig:physicalrealization}a.
The first cavity (mode $a$, represented in red in Fig.~\ref{fig:physicalrealization}a), which we refer to as the ``output'' cavity, is on resonance with the transition $\left|g\right\rangle \leftrightarrow|e\rangle$ (coupling strength $g_a$), although this will be slightly modified below. The output of the laser exits through this cavity ($\kappa_a$ loss rate). The second cavity (mode $b$, represented in orange in Fig.~\ref{fig:physicalrealization}a), which we refer to as the ``gain'' cavity, has a frequency $\omega_b$ detuned by $\Delta$ from the transition $\left|e\right\rangle \leftrightarrow|e'\rangle$ (coupling strength $g_b$). This second cavity's output light ($\kappa_b$ loss rate) is continuously measured by a single-photon detector. We allow this single photon detector to be imperfect, with a dark-count rate $R_{\text{dark}}$ \cite{darkcountrate}. This assumption is made mainly for heuristic purposes, as shown below. While in Sec.~\ref{sec:derivationsofmasterequation} we will derive the master equation dynamics for such an imperfect detector, the master equation~(\ref{eq:model}) corresponds to the ideal-detector limit $R_{\text{dark}}=0$.

The laser drive is applied only during short ``on'' periods, during which $\Omega(t)=\Omega_0$, and is otherwise off, with $\Omega(t)=0$ (Fig.~\ref{fig:physicalrealization}c). These ``on'' periods are initiated according to a Poisson process with rate $\nu\Gamma$. Let $t_i$ denote the start time of the $i$th on period, so that $\Omega(t_i)=\Omega_0$. The laser drive is terminated when a photon is detected at the feedback detector, so that $\Omega(t_i+\Delta t_i)=0$, where $\Delta t_i$ is the elapsed time between the start of the $i$th ``on'' period and the corresponding detection event (Fig.~\ref{fig:physicalrealization}c). We require that the mean duration of an ``on'' period to be much shorter than the mean time between successive ``on'' periods
\begin{equation}
\label{eq:time}
\mathbb{E}\!\left[\Delta t_i\right]
\ll
\mathbb{E}\!\left[t_{i}-t_{i-1}\right]
=
(\nu\Gamma)^{-1}.
\end{equation}
The physical parameters for the laser model will be chosen so that Eq.~(\ref{eq:time}) holds. We note that the driving laser's coherence time can be far smaller than that of the superradiant laser it helps produce. Indeed, it need not even remain phase coherent over an entire ``on'' period. The only requirement is that its linewidth be significantly smaller than the ``gain'' cavity linewidth, $\kappa_b$.
During the ``on'' periods [$\Omega(t)=\Omega_0$], an external magnetic field is applied to shift the frequency of the atomic transition $\ket{g}\leftrightarrow\ket{e}$ by $-\Omega_0^2/\Delta$ via the Zeeman effect. This magnetic field is turned off when the driving laser is off [$\Omega(t)=0$].

The atoms are arranged in an array so that they are identically coupled to each of the modes. Specifically, the spacing between the atoms in the direction parallel to the ``output'' cavity mode $a$ (vertical direction in Fig.~\ref{fig:physicalrealization}a) should be a multiple of $\lambda_a=2\pi c/\omega_a$ (in Fig.~\ref{fig:physicalrealization}a, this multiplication factor is equal to one).
Similarly, the spacing between atoms in the direction parallel to the ``gain'' cavity mode $b$  (horizontal direction in Fig.~\ref{fig:physicalrealization}a) should be a multiple of $\lambda_b=2\pi c/\omega_b$ (in Fig.~\ref{fig:physicalrealization}a, this factor is again equal to one).
This ensures the dynamics are completely collective; that is, represented only by atomic operators that are collective (see Sec.~\ref{subsec:masterequations} below).

\subsubsection{Master equation for the physical realization}
\label{subsec:masterequations}
Here, we show the master equation of the physical realization. For simplicity, we represent the levels $g$, $e$, and $e'$ with $1$, $2$, and $3$, respectively. The non-interacting Hamiltonian of the system is ($\hbar=1$)
\begin{equation}
    H_0= \omega_a \hat{a}^\dagger \hat{a}+ \omega_b \hat{b}^\dagger \hat{b}+\omega_1 \hat{J}_{11}+\omega_2 \hat{J}_{22}+ \omega_3 \hat{J}_{33}.
\end{equation}
The master equation, in the rotating frame of the driving laser and the two cavities, is
\begin{equation}
\label{eq:a1}
    \dot{W}=-i[\hat{H},W]+{\cal L}W.
\end{equation}
Here, $W$ is the state matrix for the $N$ three-level atoms and the two cavity modes. The Hamiltonian is given by
\begin{multline}
\label{eq:a4}
\hat{H}:=\Delta \hat{J}_{33}+i\Omega(t)(\hat{J}_{13}-\text{h.c.})  + g_b(\hat{J}_{23}\hat{b}^{\dagger}+\text{h.c.}) \\ + g_a(\hat{a}^{\dagger}\hat{J}_{12}+\text{h.c.})+H_{\text{Z}}(t).
\end{multline} 
Here, $H_{\text{Z}}(t)=(\Omega^2(t)/\Delta)\hat{J}_{11}$ is the externally applied, Zeeman shift using time-dependent magnetic field. 
The superoperator for the dissipative dynamics is given by
\begin{equation}
\label{eq:L_c}
    {\cal L}W:=\kappa_a{\cal D}[\hat{a}]W+\kappa_b{\cal D}[\hat{b}]W.
\end{equation}
In the above equations, we use the notation for collective atomic operators defined as $\hat{J}_{ij}:=\sum_{a=1}^{N}\hat{\sigma}_{ij}^{a}$, with $\hat{\sigma}_{ij}^{a}:=\left|i_a\right\rangle\left\langle j_a\right|$ being the operator for atom $a$, and $i,j=1,2,3$. These satisfy the commutation relations for the U(3) group, $[\hat{J}_{i j}, \hat{J}_{k l}]=\delta_{j k} \hat{J}_{i l}-\delta_{i l} \hat{J}_{k j}$ \cite{U(3)}. We can formulate the problem using collective atomic operators because we are ignoring spontaneous emission. Here, we follow Refs.~\cite{Efiyoav,PhysRevX.12.011054,97jv-13r6} in taking the dominant form of damping to be collective, here via mode $a$; see Sec.~\ref{subsec:superradiantlaserreview}.

\subsection{Derivation of the Master equation of the model}
\label{sec:derivationsofmasterequation}
Now, we derive the master equation of the model, Eq.~(\ref{eq:model}), from the master equation of the physical realization, Eq.~(\ref{eq:a1}), subject to the conditional evolution described in Sec.~\ref{subsubsec:Implementation}. 
The conditions we assume on the physical parameters will be shown throughout the section. The technical details supporting the claims in this section can be found in Appendix \ref{thesissec:derivations}.

We will show that the gain and the loss processes of the system can be calculated independently, resulting in the first and second superoperators of the final master equation (\ref{eq:model}), respectively.
The derivation of the loss process is straightforward. 
The ``on'' periods occur only for short durations according to Eq.~(\ref{eq:time}), and thus, when calculating the dynamics of the loss process we can restrict to the durations in which the driving laser is off. Below, we will show that during ``on'' periods, the dynamics of the loss process can be neglected. We assume the ``loss'' mode $a$ is heavily damped, i.e.~$ g_a N \ll \kappa_a$. Thus, it is slaved to the atoms resulting in the loss superoperator in Eq.~(\ref{eq:model}) \cite{myPRA}, with $\Gamma:=4g_a^2/\kappa_a$ being the single atom decay rate through the ``loss'' cavity mode $a$.
The derivation of the gain process will be shown in several steps in the following subsections below.

\subsubsection{Adiabatic elimination of the third level $\ket{e'}$}
\label{subsec:adiabaticeliminatione'}
Here, we adiabatically eliminate the highly detuned third level $\ket{e'}$. The technical details of the derivation are given in Appendix~\ref{subapp:adiabatic}. We consider the near-resonant two-photon Raman transition $\ket{g}\leftrightarrow\ket{e}$ mediated by the third auxiliary level $\ket{e'}$, with detuning $\Delta$. This process, called cavity-enhanced Raman emission, occurs with the absorption of a photon from the laser drive, simultaneously creating a photon excitation in the ``gain'' mode $b$ and a collective excitation in the atoms $\ket{g}\rightarrow \ket{e}$. Cavity-enhanced Raman emission has been studied previously, both without \cite{cavityenhanced,cavityenhanced2,cavityenhaced3} and with \cite{parkinstheory,parkinsletter,parkinsdpra} superradiance.
Assuming the third level is highly detuned, in the sense  $\kappa_b,\Gamma N^2,\Omega_0\sqrt{N},g_b\sqrt{N}, g_b^2N/\kappa_b \ll \Delta $, its population is very small. This allows us to adiabatically eliminate it from the Hamiltonian (\ref{eq:a4}).
The resulting dynamics are governed by the master equation (\ref{eq:a1}) while replacing the Hamiltonian of Eq.~(\ref{eq:a4}) by the following effective Hamiltonian (see derivation in Appendix~\ref{subapp:adiabatic})
\begin{equation}
\label{eq:HH}
\hat{H}_{\mathrm{eff}}
=
\hat{H}_{\mathrm{gain}}
+
\hat{H}_{\mathrm{loss}}
+
\hat{H}_{\mathrm{detune}},
\end{equation}
with
\begin{align}
\hat{H}_{\mathrm{gain}}
&:=
i\alpha(t)(\hat{J}^{\dagger}\hat{b}^{\dagger}-\mathrm{H.c.}),
\\
\hat{H}_{\mathrm{loss}}
&:=
g_a(\hat{a}^{\dagger}\hat{J}+\mathrm{H.c.}),
\\
\hat{H}_{\mathrm{detune}}
&:=
-
\frac{\Omega^2(t)}{\Delta}\hat{J}_{11}
+ H_{\text{Z}}(t).
\end{align}
Here, $\hat{H}_{\mathrm{gain}}$ is the effective gain Hamiltonian resulting from the adiabatic elimination of the third level,  
$\hat{H}_{\mathrm{loss}}$ is the Hamiltonian for the ``loss'' process, and $\hat{H}_{\mathrm{detune}}$ is the Hamiltonian for the induced energy shift (see below).
We have defined $\alpha(t):=\Omega(t)g_b/\Delta$, $\hat{J}:=\hat{J}_{12}$, and $\alpha_0:=\Omega_0g_b/\Delta$.
 The Hamiltonian term $\hat{H}_{\mathrm{detune}}$ contains two contributions. 
The first is the light shift generated by the adiabatic elimination of the third level $\ket{e'}$. 
The second is the externally applied, time-dependent Zeeman shift. It was chosen to compensate the light shift, so that $\hat{H}_{\mathrm{detune}}(t)=0$ at all times. 
While the drive is off [$\Omega(t)=0$], only $\hat{H}_{\mathrm{loss}}$ contributes to the dynamics, as discussed above. While the drive is on [$\Omega(t)=\Omega_0$], the dynamics of the loss process involving the mode $a$ can be neglected, assuming $\Gamma\ll \Gamma_{b}$. Here, we defined $\Gamma_{b}:=4\alpha^2_0/\kappa_b$, which equals the single atom flux through the ``gain'' cavity in the limit we will consider below. 
The term $H_{\text{gain}}$ is the anti Tavis-Cummings Hamiltonian \cite{antiTavisCummings}.
It can be interpreted as the absorption of a photon from the drive, and simultaneously creating a photon excitation in the ``gain'' mode $b$ and a collective excitation in the atoms $\ket{g} \rightarrow \ket{e}$, as explained above. 
Each collective gain excitation created in the atoms $\ket{g} \rightarrow \ket{e}$ will appear as a photon detection in the detector since the mode $b$ is taken to be heavily damped (see Sec.~\ref{subsubsec:adaiabaticmodeb} below).

\subsubsection{Adiabatic elimination of mode $b$}
\label{subsubsec:adaiabaticmodeb}
Here, we derive the dynamics for the state of the atoms during an ``on'' period and the change of the state (quantum jump) after a detection of a photon in the feedback detector. 
We assume that the ``gain'' cavity mode $b$ is heavily damped, by the condition $\alpha_0 N\ll \kappa_b$, so it is slaved to the atoms, and the evolution of the atoms' state can be determined based on the feedback from the feedback detector. In the duration of the $i$th ``on'' period, prior to a click at the feedback detector, i.e.~at times $t_i\leq t< t_i+\Delta t_i$, the conditional state of the atoms evolves with non-Hermitian dynamics (see derivation in Appendix~\ref{subapp:modeb})
\begin{equation}
\label{eq:878}
    \dot{\tilde{\rho}}=-\Gamma_b({\cal A}[\hat{J}^{\dagger}]+n_\text{s}) \tilde{\rho}.
\end{equation}
Here, we remind the reader of the definition of $\Gamma_{b}:=4\alpha^2_0/\kappa_b$, as the single atom flux through the ``gain'' cavity.  Here, we have defined the saturation parameter 
\begin{equation}
\label{eq:n_s}
n_\text{s}:=R_{\text{dark}}/\Gamma_b.
\end{equation}
The saturation parameter typically appears in the standard laser master equation due to the spontaneous emission of the atoms in the gain medium  \cite{howard_atom_laser,howard1999} (see Sec.~\ref{sec:fat}). 
In our model, it appears similarly in the structure of the master equation, and with a similar physical role
(see below). 
We assume that the atomic state is normalized at the beginning of the ``on'' period, $\operatorname{Tr}[\tilde{\rho}(t_i)]=1$, so that $\operatorname{Tr}[\tilde{\rho}(t)]$ is the probability that no click, either from a genuine photon detection or from a dark count, has occurred in the feedback detector during the interval $[t_i,t]$.

Now, immediately after a click at the feedback detector which occurs within a time interval $[t_i+\Delta t_i,t_i+\Delta t_i+\text{d}t]$, the unnormalized state of the system changes to
\begin{equation}
\label{eq:rho'}
\tilde{\rho} ' (t_i+\Delta t_i+\mathrm{d}t)\text{d}t=\Gamma_b({\cal J}[\hat{J}^{\dagger}]+n_\text{s})\tilde{\rho}(t_i+\Delta t_i) \text{d}t.
\end{equation}
The trace of the density matrix in Eq.~(\ref{eq:rho'}), i.e.~$\text{Tr}[\rho ' ((t_i+\Delta t_i)^{+})]\mathrm{d}t$, is equal to the probability of a click in the detector within the time interval $[t_i+\Delta t_i,t_i+\Delta t_i+\text{d}t]$, following no click in the interval $[t_i,t_i+\Delta t_i]$. In the following, we primarily consider the ideal-detector limit, $n_{\text{s}}=0$. As shown below, this limit can be effectively realized when the laser contains sufficiently many excitations, as in the FAT laser. 
Equation~(\ref{eq:rho'}) shows that a genuine photon detection by the feedback detector, corresponding to a vanishing dark-count rate ($n_{\text{s}}=0$), corresponds to the addition of one collective excitation to the atoms. 
This can be explained intuitively. 
A single photon in the ``gain'' cavity mode leaks irreversibly from the cavity to the feedback detector on a timescale of order $\kappa_b^{-1}$. During this short time, the atom--cavity interaction is negligible: its dimensionless strength is bounded by \(
\kappa_b^{-1}\alpha_0 J_m < \kappa_b^{-1}N\alpha_0 \ll 1 .
\) Therefore, this interaction is highly unlikely to change the intracavity photon number before the photon reaches the detector.
Thus, in the bad-cavity limit, a perfect single-photon feedback detector effectively monitors the photon number in the gain cavity mode continuously. A photon in the mode $b$ and a collective excitation in the atoms created as a pair, as explained above. Thus, the feedback detector effectively monitors the gain of the number of excitations in the atoms.

\subsubsection{Update to the atoms' state from a full ``on'' period}
Now, we evaluate how the state of the atoms is changed by a full ``on'' period, i.e.,~by turning on the driving laser until a click is observed at the feedback detector.
\begin{multline}
\label{eq:a34}
{\rho((t_i+\Delta t_i)^+)}= \\ \Gamma_b({\cal J}[J^{\dagger}]+n_\text{s})\int_{0}^{\infty}\exp\left(-\Gamma_b({\cal A}[\hat{J}^{\dagger}]+n_\text{s})t\right)\text{d}t{\rho(t_i)} \\
    =({\cal J}[J^{\dagger}]+n_\text{s})(\mathcal{A}[{J^\dagger}]+n_\text{s})^{-1}\rho(t_i).
\end{multline}
We used Eqs.~(\ref{eq:878}) and (\ref{eq:rho'}) for the atoms' state evolution prior to a click at the detector, and the atoms' change of state immediately after a click is observed at the detector, respectively. The atoms cannot gain a collective excitation when they are in the fully excited state, $m=N/2$, until the loss process reduces the atoms' state to $m=N/2-1$. 
Correspondingly, the integral in Eq.~(\ref{eq:a34}) is ill defined for the fully excited atoms input state in the case $n_{\text{s}}=0$. To address this issue, we formally define the superoperator of Eq.~(\ref{eq:a34}) in the case of $n_{\text{s}}=0$ by taking the limit $n_{\text{s}}\rightarrow 0^+$, i.e. ${\cal J}[J^{\dagger}]\mathcal{A}[{J^\dagger}]^{-1}\rho:=\lim_{n_{\text{s}}\rightarrow 0^+} ({\cal J}[J^{\dagger}]+n_\text{s})(\mathcal{A}[{J^\dagger}]+n_\text{s})^{-1}\rho$.
Using this definition, we obtain ${\cal J}[J^{\dagger}]\mathcal{A}[{J^\dagger}]^{-1}\ket{N/2,N/2}\bra{N/2,N/2}=\ket{N/2,N/2}\bra{N/2,N/2}$, as desired. This definition will be used where the gain superoperator is otherwise ill-defined; see Eq.~(\ref{eq:redefine}) below. 
Evaluation of the integral in Eq.~(\ref{eq:a34}), provided that $n_{\text{s}}>0$, is straightforward, since then $\mathcal{B}=\Gamma_b({\cal A} [\hat{J}^{\dagger}]+n_{\text{s}})$ is a strictly positive superoperator, (with the set of eigenoperators $\{\ket{N/2,m}\bra{N/2,n}\}_{m,n=-N/2}^{N/2}$ of positive eigenvalues), so that the formula $\int_0^\infty\exp{(-\mathcal{B}t)\text{d}t}=\mathcal{B}^{-1}$ holds.  
The physical process described by Eq.~(\ref{eq:a34}) is unconditional, and therefore the superoperator in Eq.~(\ref{eq:a34}) is a CPTP map (defined in, e.g.~Ref.~\cite{Wiseman_2009}). In terms of population dynamics (assuming $n_{\text{s}}=0$), its action is to add one collective excitation to the atoms irrespective of its input state \cite{howard_atom_laser,howard1999}.
The average duration of the ``on'' period is approximately the inverse of the flux from the ``gain'' cavity during an ``on'' period 
\begin{equation}
\label{eq:oehdb}
\mathbb{E}[\Delta t_{i}]\approx\left(\Gamma_{b}\expectation{\hat{J}\hat{J}^{\dagger}}\right)^{-1}\approx\left(\Gamma_{b}\expectation{\hat{J}^{\dagger}\hat{J}}\right)^{-1},
\end{equation}

In each driving cycle, when the drive is on and a click is observed at the feedback detector, the reaction speed of the feedback loop to terminate the drive operation is limited by the speed of the electronics, 
$\gamma_{\text{e}}\approx 2 \pi \times 100$ MHz \cite{electronicspeed}.
According to the scheme, we require the driving laser to be terminated immediately after a click at the feedback detector. To get one collective gain excitation per cycle ($n_{\text{s}}=0$), the relevant rate to compare is the flux from the ``gain'' cavity, so that using Eq.~(\ref{eq:oehdb}), we obtain the condition 
\begin{equation}
\label{eq:electronics}
  \Gamma_b \expectation{J^\dagger J}\     \ll \gamma_{\mathrm{e}}
\end{equation}

\subsubsection{Master equation}
\label{sec:Markovdynamics}
By requiring the conditions of Eq.~(\ref{eq:time}) and (\ref{eq:electronics}), we can assume that the clicks events at the feedback detector occur at the specific times the experimentalist chooses to initiate the ``on'' periods. To obtain a Markovian master equation, we initiate the ``on'' periods at times following Poissonian statistics at a rate of $\nu \Gamma$, as described in Sec.~\ref{subsubsec:Implementation}. 
By using Eq.~(\ref{eq:a34}), we obtain the gain superoperator, and the resulting master equation
considering both the gain and the loss processes, is 
\begin{multline}
\label{eq:qfd}
\dot{\rho}=\nu\Gamma \left(({\cal J}[J^{\dagger}]+n_{\text{s}})(\mathcal{A}[\hat{J}^{\dagger}]+n_{\text{s}})^{-1}-1
\right)\rho+\Gamma\mathcal{D}[\hat{J}]\rho\\
=\Gamma\left\{\nu\mathcal{D}[\hat{J}^{\dagger}](\mathcal{A}[\hat{J}^{\dagger}]+n_{\text{s}})^{-1}\rho+\mathcal{D}[\hat{J}]\rho\right\}.
\end{multline}
Here, $\Gamma=4g_a^2/\kappa_a$. 
Equation~(\ref{eq:qfd}) generalizes the standard laser master equation (\ref{thesis:lasermasterequation}), by replacing $\hat{a}\rightarrow\hat{J}$ in the gain and loss superoperators, with the corresponding coupling constants.

The saturation parameter, $n_{\text{s}}$, which also appears in the standard laser master equation~(\ref{thesis:lasermasterequation}), plays an analogous physical role here. In the standard laser, atoms in the gain medium are pumped into the excited state at a constant rate, and each excitation is then either transferred to the cavity mode through stimulated emission or lost through spontaneous emission. Similarly, in our model, each ``on'' period is initiated at a constant rate and ends either by adding one collective excitation at enhanced superradiant rate or by losing it due to a dark-count click. Thus, spontaneous emission and dark counts play analogous roles as undesired processes that disrupt the constant gain rate. In both cases, these effects are captured by the saturation parameter, $n_{\text{s}}$, in the corresponding master equation. In addition, in both cases, when operating with enough excitations in the laser, that is, far above the threshold, these undesired processes become rare and can be neglected.
As we will see in Sec.~\ref{sec:poissonian}, constant or `flat' gain rate results in Poissonian output photon statistics, an essential property for the laser, and deviations from it cause amplitude fluctuations of the emitted beam.

The far above threshold superradiant laser model can be obtained by assuming ideal single-photon detectors, for which $R_{\text{dark}}=0$. This
gives the  final master equation for the model 
\begin{equation}
\label{eq:model1}
\dot{\rho}=\mathcal{S}\rho:=\Gamma\{\nu\mathcal{D}[\hat{J}^{\dagger}]\mathcal{A}[\hat{J}^{\dagger}]^{-1}\rho+\mathcal{D}[\hat{J}]\rho\}.
\end{equation}
Here, the superoperator $\mathcal{S}$ is defined according to Eq.~(\ref{eq:model1}).
Equation~(\ref{eq:model1}) is identical to Eq.~(\ref{eq:model}), and is the final result.
That is, we get the desired generalization of the FAT laser. Here, we use  quantum feedback control instead of gain saturation to get the desired result, in a similar way to the derivation in Refs.~\cite{howard_atom_laser,howard1999}.
Consistent with the discussion above, we formally define the gain superoperator as 
\begin{equation}
\label{eq:redefine}
 {\cal D}[\hat{J}^{\dagger}]\mathcal{A}[\hat{J}^{\dagger}]^{-1}:=\underset{n_s\rightarrow 0^+}{\text{lim} }\mathcal{D}[\hat{J}^{\dagger}]\left(\mathcal{A}[\hat{J}^{\dagger}]+n_s\right)^{-1}.   
\end{equation}
At steady state, the system's flux through the ``loss'' cavity is equal to the gain rate (gain events per unit time). This leads to   
\begin{equation}
\label{eq:ygvfujngh}
\expectation{\hat{J}^{\dagger}\hat{J}}_{\text{ss}}=\nu
\end{equation}
Dicke superradiance is characterized by enhanced intensity of light emission, so that $\nu$ is typically proportional to $N^2$ and can get up to the maximal value of $N^2/4$ [see Eq.~(\ref{eq:J_m})].
Using Eqs.~(\ref{eq:time}), (\ref{eq:oehdb}) and (\ref{eq:ygvfujngh}), we get the condition $\Gamma \ll \Gamma_b$, consistent with the assumptions above.

\subsubsection{Realizable values for the physical parameters}
Here, we show  a set of physically realizable parameter values that satisfy all the required conditions. We need to assume an additional condition, that the maximal value for the detuning $\Delta$ is 
\begin{equation}
\label{eq:Deltacondition}
    \Delta\leq \Delta_{\text{max}} \approx 2\pi \times 10^{12} \text{  Hz},
\end{equation}
ensuring it is well below the optical frequency, which is assumed to correspond to the atomic transitions and the laser output light. We remind the reader of the conditions set previously
\begin{multline}
\label{eq:gyk} 
\kappa_b,\Gamma N^2,\Omega_0\sqrt{N},g_b\sqrt{N}, g_b^2N/\kappa_b \ll \Delta, \quad N g_a \ll \kappa_a, \\
\quad \alpha_0 N\ll \kappa_b ,\quad \frac{g_a^{2}}{\kappa_a}\ll\frac{\alpha_0^{2}}{\kappa_b}, 
\quad \frac{N^{2}\alpha_0^{2}}  {\kappa_b}\ll\gamma_{\text{e}}.
\end{multline}
We aim to satisfy all inequalities (\ref{eq:gyk}) of the form $a\ll b$ by requiring $a\leq \epsilon b$, where 
$\epsilon=10^{-2}$.
In Appendix~\ref{subapp:parmaters}, we derive an optimal set of values for the physical parameters that meet our aim for all inequalities in Eqs.~(\ref{eq:Deltacondition}) and (\ref{eq:gyk}). These are
\begin{multline}
\label{eq:physicalparameters}
N=10^4,\quad g_a=2\pi\times10^2  \text{Hz}, \\\quad g_b=\Omega_0=\kappa_a=2\pi\times10^8 \ \text{ Hz}, \quad
\kappa_b=2\pi\times10^{10} \ \text{Hz}, \quad \\ \Delta =2\pi\times 10^{12} \ \text{Hz}. 
\end{multline}

\section{Steady state and Metastable States}
\label{sec:metastable}
In this section, we analyze the steady state and, more importantly, the metastable state that the laser can operate in. 
\subsection{Population dynamics and steady state}
\label{subsec:ss}
\begin{figure}
\centering
\includegraphics[width=\linewidth]{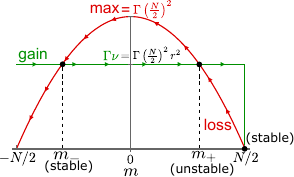}
\caption[Rates for birth and death for the birth-death Markov process]{Rates for birth (green line) and death (red line) for the birth-death Markov process Eq.~(\ref{eq:qqq}) as a function of the excitation number $m$, in the large $N$ limit. The black dots are the semiclassical fixed points: $m=m_-$ and $m=N/2$  being stable, and $m=m_+$ being unstable.}
\label{fig:stability}
\end{figure}

The population dynamics of the model are obtained from the quantum master equation (\ref{eq:model}). It is given by a classical master equation for a continuous-time birth-death Markov process \cite{Markovchain,norris_1997}
\begin{equation}
\label{eq:population}
    \dot{\rho}_m =q_{m-1,m}\rho_{m-1}+q_{m+1,m}\rho_{m+1} -(q_{m,m+1}+q_{m,m-1})\rho_m.
\end{equation}
Here, we defined $\rho_m:=\rho_{m,m}$ as the population.
The coefficients $q_{m,m+1}$ ($q_{m,m-1}$) are the transition rates for the birth (death) of the Markov process. These are given by
\begin{equation}
\label{eq:qqq}
q_{m, m+1}=\left\{\begin{array}{ll}
\nu \Gamma, & m<N / 2, \\
0, & m=N / 2,
\end{array} \quad q_{m, m-1}=\Gamma J_m^2 \right.,
\end{equation}
and are plotted in Fig.~\ref{fig:stability}. The birth rate was set in Sec.~\ref{sec:physicalrealzation} as the rate of the Poisson process which initiates the drive, adding one excitation to the system.
The coefficients $J_m$ are defined in Eq.~(\ref{eq:J_m}). 
If the experimentalist keeps track of the gain events and measures the output light with direct photodetection, the system will follow the birth-death Markov process in continuous time, with the rates of Eq.~(\ref{eq:qqq}). A birth event corresponds to the detection of a photon in the feedback detector, whereas a death event corresponds to the detection of a photon in the output light (see Sec.~\ref{sec:physicalrealzation}).

\begin{figure}
\centering
\includegraphics[width=\linewidth]{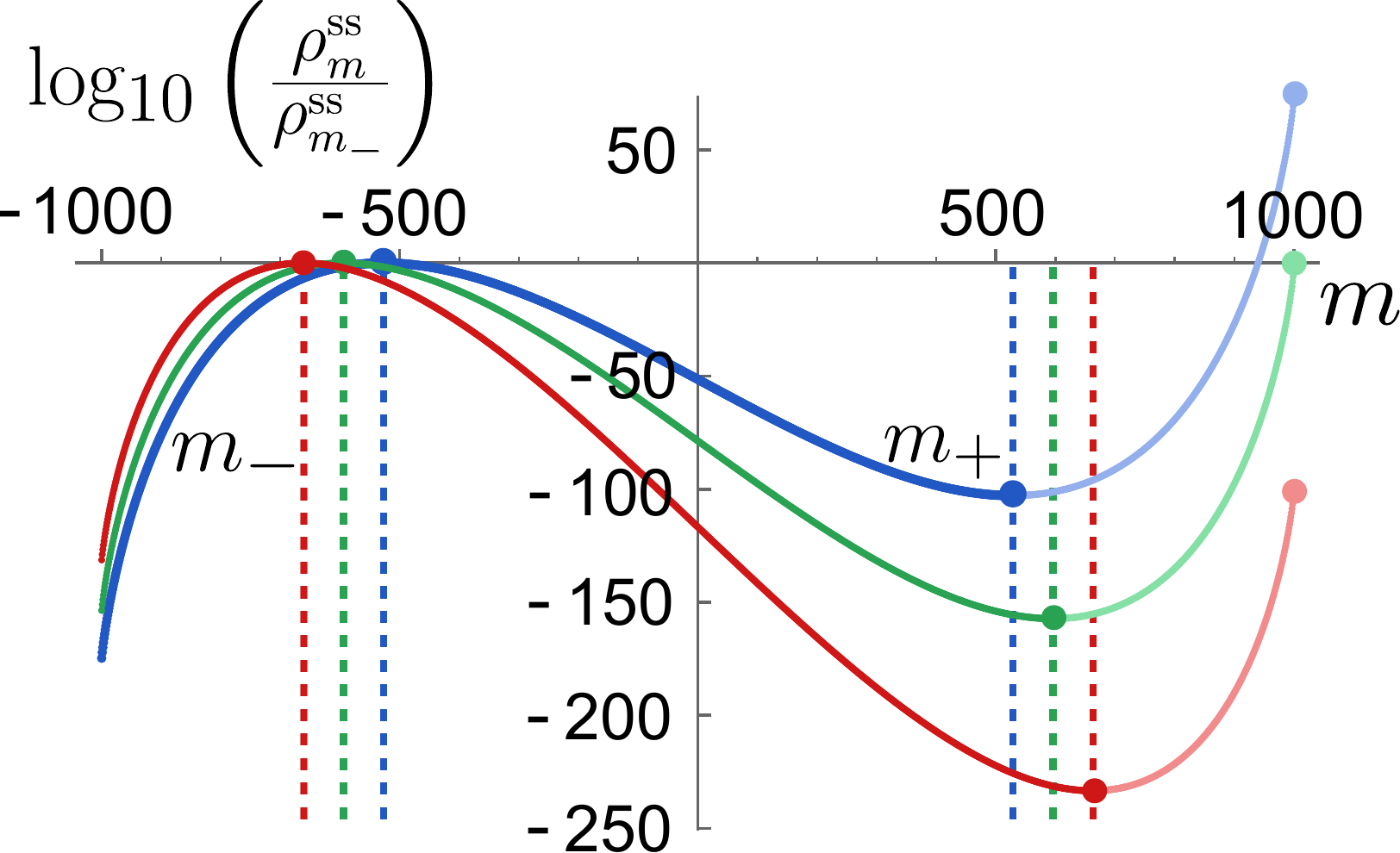}
\caption[Population of steady state comparison]{\( \log_{10} \left( \rho_m^{\text{ss}} / \rho_{m_-}^{\mathrm{ss}} \right) \) as a function of \( m \), for \( N = 2000 \) atoms, for normalized pumping rates \( r = 0.75 \) (red), \( r=0.804 \approx r_0 \) (green), and \( r=0.85 \) (blue). The local maxima at $m=N/2=1000$ are marked by dots, while the local maxima and minima, \( m_-(r) \) and \( m_+(r) \), respectively, are marked by dots and vertical dashed lines indicating their $m$-values. For each \( r \), the curve, dot, and line share the same color. The curves transition from the dark-colored to the light-colored part at \( m = m_+(r) \), indicating the truncation region of \( \rho_m^{\text{ss}} \) as defined in Eq.~(\ref{eq:truncate}). That is, in the light-colored part, \( \rho_m^{\text{ms}} = 0 \).
}
\label{fig:logss}
\end{figure}

The steady-state solution of Eq.~(\ref{eq:population}) is given by the recursion relation for a continuous-time birth–death Markov process
\cite{Markovchain}
\begin{equation}
\label{eq:ss}
    \rho_{m}^{\text{ss}} = \frac{q_{m-1,m}}{q_{m,m-1}}\rho_{m-1}^{\text{ss}}= \frac{\nu}{J_{m}^{2}} \rho_{m-1}^{\text{ss}}.
\end{equation}
From Eq.~(\ref{eq:dfgjklSDF}), we can identify the recursion relation (\ref{eq:ss}) as giving the coefficients squared for the pure state of Eq.~(\ref{eq:fgfgjkgfjuhruaer}) with $\nu=\vert \alpha \vert ^2$. Below, we will define the parameters $r$ and $m_{\pm}$ analogously to the definitions in Sec.~\ref{sec:CRSS}. 
Specifically, we define the normalized dimensionless pumping-rate coefficient \( r \) by
\begin{equation}
\label{eq:nu}
    \nu =: \frac{1}{4} N^2 r^2.
\end{equation}
For any $0<r<1$, we define $m_-$ and $m_+$ as the solutions to the equation $\nu=J_{m}^{2}$ (rounded to integers). These represent local maximum and minimum, respectively, of the solution, $\rho_m^{\text{ss}}$, to the recursion relation (\ref{eq:ss}) (see Fig.~\ref{fig:logss}). These are given approximately by
\begin{equation}
\label{eq:mpm}
    m_{\pm} \approx \pm \frac{N}{2}\sqrt{1-r^2+\frac{2}{N}+\frac{1}{N^2}}+\frac{1}{2} \approx\pm\frac{N}{2}\sqrt{1-r^{2}}.
\end{equation}
Additionally, $m=N/2$ represents another local maximum of the solution to Eq.~(\ref{eq:ss}) (see Fig.~\ref{fig:logss}).

In the asymptotic limit, $N\rightarrow\infty$, the solution of Eq.~(\ref{eq:ss}) experiences a phase transition as a function of $r$, at the critical pumping \( r = r_0 \approx 0.804 \) (see discussion below and Ref.~\cite{myarticle}). This can be observed from Fig.~\ref{fig:logss}, which illustrates the case of \( N = 2000 \) atoms. For \( r < r_0 \), \( \rho_{m_-}^{\text{ss}} \gg \rho_{N/2}^{\text{ss}} \) as $N\rightarrow\infty$, so that the population is centered near the global maximum at \( m = m_- \), and in particular, takes the shape of a Gaussian [see Eq.~(\ref{eq: gaussian}) and Fig.~\ref{fig:rec} below]. When \( r > r_0 \), \( \rho_{m_-}^{\text{ss}} \ll \rho_{N/2}^{\text{ss}} \) as $N\rightarrow\infty$, so that \( m = N/2 \) becomes the global maximum, the population is described approximately by \( \rho^{\text{ss}}_m \sim \delta_{m,N/2} \), i.e.,~all the atoms in the top state.
These features reflect the general large-\( N \) behavior \cite{myarticle}, where the critical point \( r=r_0 \) is obtained by equating the asymptotic forms for a large $N$ of \( \rho^{\text{ss}}_{m_-} \) and \( \rho^{\text{ss}}_{N/2} \) (see Sec.~\ref{subsec:metastabletime} below).
When \( r > r_0 \), the output light of the resulting system exhibits negligible coherence. This is due to the number-phase Heisenberg uncertainty relation \cite{T,mandel_wolf_1995}: a number state has maximal phase uncertainty. 

\subsection{Metastable state: semiclassical fixed point}
\label{subsec:metastable}
The above might suggest that the laser operation must be restricted to the regime $r<r_0$. However, it is possible to operate above the critical pumping threshold, $r>r_0$, with a long-lived metastable state that is practically a steady state. The motivation for operating with larger pumping is to bring the system's state population closer to $m=0$. This is needed to access the advantage of the flat area identified in Fig.~\ref{fig:coefficients}, which leads to reduced phase noise, as explained in Sec.~\ref{subsec:ourmodel}. This intuition is verified by detailed calculations in Sec.~\ref{sec:coherenceprop}.

We analyze the birth-death Markov process to obtain a simplified version of the population dynamics equation (\ref{eq:population}). 
First, we study the semiclassical dynamics, which describes how the system's excitation number $m$ tends to grow or shrink based on its current value.
Specifically, $m$ tends to grow when the birth rate exceeds the death rate, and shrink when the opposite holds (see Fig.~\ref{fig:stability}).
In the region $m_-<m<m_+$, the death rate exceeds the birth rate (see Fig.~\ref{fig:stability}), so that the Markov process tends to make $m$ shrink there. The opposite holds in the region $m_+<m$.
Recall that the semiclassical fixed points $m_+,m_-$ [defined in Eq.~(\ref{eq:mpm})] are defined as the excitation numbers where the two rates are equal. A semiclassical point is stable if a state originating nearby is attracted to it by the dynamics, and unstable if the state is repelled from it.

Figure \ref{fig:stability} shows the unstable fixed point $m_+$ and the stable fixed points $m_-$ and $m=N/2$. For a large $N$ and $0<r<1$,
the Markov process will tend to localize near either of the two stable fixed points, \( m = m_- \) and \( m = N/2 \) for a long time. Each of these is therefore associated with a long-lived metastable state, denoted \( \rho^{\text{ms}} \) and \( \rho^{\text{top}} \), respectively. We are particularly interested in \( \rho^{\text{ms}} \), where the system functions as a laser with a high degree of coherence \( \coh \). That is why we reserve the metastable abbreviation, ms, as the identifier for this state.
The semiclassical dynamics described above sets the domains where the system tends to remain as a metastable state.
The unstable fixed point $m=m_+$, which is also a local minimum of the steady state population so that the system rarely occupies it, is the transition point between the two metastable states.
If the current system's excitation number is at the region $m<m_+$, it will likely remain there for a long time, and similarly for $m>m_+$. Thus, the metastable state $\rho^{\text{ms}}$ is described by the recursion relation (\ref{eq:ss}) truncated at the transition point $m_+$, that is,
\begin{equation}
\label{eq:truncate}
\rho_{m}^{\text{ms}} \propto \rho_{m}^{\text{ss}} \ \text{for }  m \leq m_+, \quad
\rho_{m}^{\text{ms}} = 0 \ \text{for }  m > m_+.
\end{equation}
The metastable state has the same excitation statistics as the CRSS of Eq.~(\ref{eq:yofhjksjhdgkfng}) that was reviewed in Sec.~\ref{sec:CRSS}. The metastable state $\rho^{\text{top}}$ is described by the recursion relation (\ref{eq:ss}) truncated from below at the transition point $m_+$, that is,
\begin{equation}
\label{eq:truncatetop}
\rho_{m}^{\text{top}} \propto \rho_{m}^{\text{ss}} \ \text{ for }  m \geq m_+, \quad
\rho_{m}^{\text{top}} = 0 \ \text{ for }  m < m_+.
\end{equation}

The state \( \rho^{\text{ms}} \) is naturally reached by initializing the system in the ground state \( \rho = \ket{N/2, -N/2}\bra{N/2, -N/2} \), with all atoms unexcited. The system then rapidly relaxes to a transient equilibrium near \( m = m_- \) over a short timescale \( T_{\text{eq}} \). It is classified as metastable only if \( T_{\text{eq}} \ll T_{\text{ms}} \), where \( T_{\text{ms}} \) is its lifetime (see definition below). To fully exploit the laser's coherence, we assume \( T_{\text{eq}} \ll T_{\text{coh}} \ll T_{\text{ms}} \), where $T_{\text{coh}}$ is the coherence time of the laser (see Sec.~\ref{sec:coherenceprop}). 
\begin{figure}
\centering
\includegraphics[width=\linewidth]{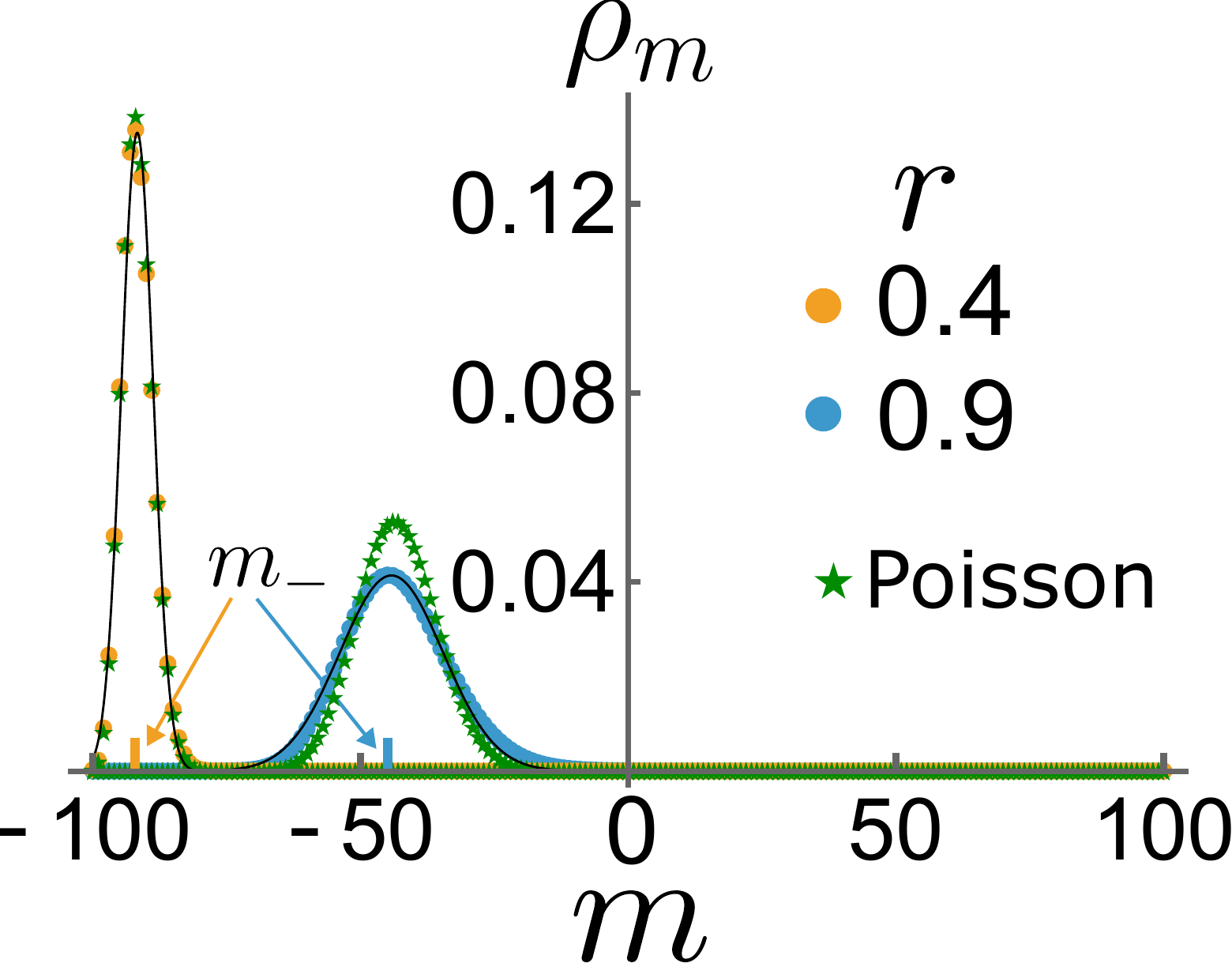}
\caption[Population of steady and metastable states]{Populations \(\rho_m^{\text{ss}}\) and \(\rho_m^{\text{ms}}\) as a function of \(m\), shown for \(N = 200\) atoms for \(r = 0.4\) (orange, steady state) and \(r = 0.9\) (blue, metastable state), respectively. The $m$-values of the peaks at $m=m_-(r)$ are indicated on the horizontal axis. The green stars indicate the corresponding Poissonian distributions with the same mean excitation number \(\mu \approx m_- + N/2\). The solid black lines show the analytical approximation of Eq.~(\ref{eq: gaussian}).}
\label{fig:rec}
\end{figure}
Truncating the state according to Eq.~(\ref{eq:truncate}) has only a negligible effect when \( r < r_0 \), since the global maximum remains at \( m = m_- \), as illustrated by the curve for \( r = 0.75 \) in Fig.~\ref{fig:logss}. In contrast, when \( r > r_0 \), truncation shifts the global maximum from \( m = N/2 \) to \( m = m_- \), as illustrated by the curve for \( r = 0.85 \) in Fig.~\ref{fig:logss}. Thus, regardless of the value of $r$,  the stable fixed point \( m = m_- \) is the global maximum of \(\rho^{\text{ms}}\).

For large $N$, this metastable state population is approximately described by a Gaussian \cite{myarticle} (see derivation in Appendix~\ref{sec:thesisderivationmsstate}) 
\begin{equation}
\label{eq: gaussian}
   \rho_m^{\text{ms}} \approx \frac{1}{\sqrt{2 \pi w^2}}\, \mathrm{e}^{-{(m-m_-)^2}/{2 w^2}},
\end{equation}
of width squared 
\begin{equation}
\label{eq:width}
    w^{2} = R[\hat{J}](m_-) \approx \frac{N r^{2}}{4 \sqrt{1 - r^{2}}}.
\end{equation}
and a center at $m_-\approx -N\sqrt{1-r^2}/2$. Here, we have defined the functional 
\begin{equation}
\label{eq:R[Q]}
    R[\hat{L}](m)
    :=
    \left(
    \left.
    \frac{\partial L_n^2}{\partial n}
    \right|_{n=m}
    \Big/
    L_m^2
    \right)^{-1}
    =
    \left(
    \bra{m}
    \frac{[\hat{L},\hat{L}^{\dagger}]}{\hat{L}^{\dagger}\hat{L}}
    \ket{m}
    \right)^{-1}.
\end{equation}
Here, $\hat{L}^\dagger\hat{L}$ is a positive-definite operator, and $[[L,L^\dagger],L^\dagger L]=0$, so that the expression in Eq.~(\ref{eq:R[Q]}) is well defined. Thus, the variance of the metastable state population distribution given in Eq.~(\ref{eq:width}) is the inverse of the relative rate of change of the coefficients $L_n^2$, evaluated at the stable fixed point $m_{\mathrm{p}}$.
Figure \ref{fig:rec} plots the analytical approximation Eq.~(\ref{eq: gaussian}).
Thus, the mean number of excitations in the atoms is approximately given by 
\begin{equation}
\label{eq:mu}
    \mu \approx m_{-}+N/2\approx \frac{N}{2}(1-\sqrt{1-r^{2}}).
\end{equation}
Using Eqs.~(\ref{eq:width}) and (\ref{eq:mu}), we have
\begin{equation}
    \frac{w^2}{\mu}=\frac{r^2}{2\sqrt{1-r^2}(1-\sqrt{1-r^2})}
\end{equation}
In the linear regime, where $r\ll1$, the superradiant laser dynamics are close to those of the FAT laser (see Sec.~\ref{subsec:ourmodel}). Thus, in this regime, the width of the Gaussian in Eq.~(\ref{eq: gaussian}) is given by $w^2\approx \mu$, which is the variance for Poissonian population of the FAT laser model \cite{howard_atom_laser,howard1999,howard_stochastic,louisell1973quantum}. 
But for beyond-SQL coherence, we want the opposite regime $0<1-r \ll1$ (see Sec.~\ref{subsec:metastabletime} below), where the metastable distribution is super-Poissonian, with $w^2 \gg \mu$.

\subsection{Metastable time}
\label{subsec:metastabletime}
To find $T_{\text{ms}}$, we study the transition from $\rho^{\text{ms}}$ to $\rho^{\text{top}}$.
In the region $m_-<m<m_+$, the death rate exceeds the birth rate (see Fig.~\ref{fig:stability}), so that the Markov process tends to shrink there. If the Markov process originates near $m_-$, it will have a mean first-passage time (see definition below and Ref.~\cite{firstpassage}) to reach the top metastable state $\rho^{\text{top}}$ that is roughly exponential in the number of states between $m_-$ and $m_+$.
It is this time that we define as the metastable time, $T_{\text{ms}}$.
According to Eq.~(\ref{eq:mpm}), this number of states decreases with increasing normalized pumping rate $r$, but increases with the number of atoms $N$ (see Fig.~\ref{fig:stability}).

\begin{figure}
\centering
\includegraphics[width=\linewidth]{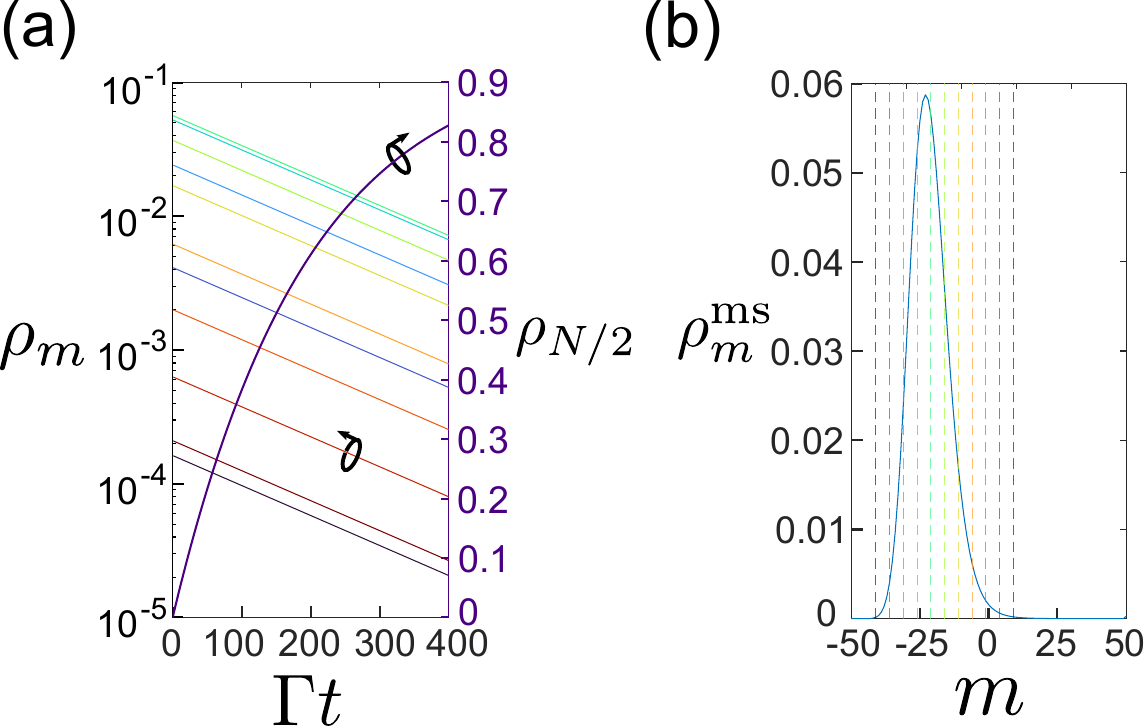}
\caption[Population dynamics]{(a) Population \( \rho_m(t) \) (left axis, log scale) as a function of time according to the dynamics of Eq.~(\ref{eq:population}), for selected values of $m$, namely \(  -41, -36, \dots, 9 \) as illustrated in (b), for \( N = 100 \) atoms and normalized pumping rate \( r = 0.9 \). The populations at $t=0$ are set by beginning the system at time $t_0=-0.3/\Gamma$ in the state \( \rho_m(t_0)=\delta_{m,-N/2} \) where all atoms are in the ground state, and evolving it for a duration of \(  0.3/\Gamma \). Because \( T_{\text{eq}} \ll \vert t_0 \vert \ll T_{\text{ms}} \), \( \rho_m(0) \approx \rho^{\text{ms}}_m \). The purple line shows the population of the maximally excited state \( \rho_{N/2} \) (right axis, linear scale). The numerical results are indistinguishable from the ones which are obtained from Eq.~(\ref{eq:expdecay}), with \( \Gamma T_{\text{ms}} = 193 \) here.
(b) The metastable-state excitation distribution at $t=0$, \( \rho_m(0) = \rho^{\text{ms}}_m \), is shown in blue. The dashed vertical lines indicate the values of \( m \) plotted in (a), with colors matching those in (a).} 
\label{fig:meta}
\end{figure}

To estimate the metastable time, \( T_{\text{ms}} \), in the limit of large $N$, analytically, we use results from Ref.~\cite{firstpassage} for the mean first passage time of a birth-death continuous-time Markov process.
The derivation is provided in Appendix~\ref{sec:thesisderivationmetastable}, and the final result is
\begin{multline}
\label{eq:T0}
  T_{\text{ms}} =  \frac{\rho_{m_{-}}^{\text{ss}}}{\rho_{m_{+}}^{\text{ss}}}\frac{2\pi}{\Gamma N\sqrt{1-r^{2}}} \\
  =\frac{2\pi\, \mathrm{e}^{2\tanh^{-1}(\sqrt{1 - r^{2}})}}{\Gamma N\sqrt{1 - r^{2}}}
\, \mathrm{e}^{2N[\tanh^{-1}(\sqrt{1 - r^{2}}) - \sqrt{1 - r^{2}}]}.
\end{multline}
This result agrees very well with numerical simulation of the population dynamics in Eq.~(\ref{eq:population}), with a typical relative error below one percent. Specifically, the resulting decay from the metastable state to the steady state is accurately described by the exponential approximation in Eq.~(\ref{eq:expdecay}), as shown in Fig.~\ref{fig:meta}.
An asymptotic expression for Eq.~(\ref{eq:T0}) in the large $N$ limit is
\begin{equation}
\label{eq:ln}
 \ln(\Gamma T_{\text{ms}}) \sim \ln\left(\frac{\rho^{\text{ss}}_{m_{-}}}{\rho^{\text{ss}}_{m_{+}}}\right)  \sim 2N[\tanh^{-1}(\sqrt{1 - r^2}) - \sqrt{1 - r^2}].
\end{equation}
We will primarily restrict our analysis to the \emph{metastable regime}, in which the laser is intended to operate, where \( \Gamma T_{\text{ms}} \) is dominated by the asymptotic expression of Eq.~(\ref{eq:ln}) and becomes exceedingly large due to its exponential scaling with \( N \).

Recall that in Sec.~\ref{subsec:ourmodel} we motivated why our laser model should operate as close as possible to $m=0$,
where the off-diagonal superradiant coefficients are flattest (see Fig.~\ref{fig:coefficients}).
This corresponds to taking $r$ as close as possible to $1$ so that the center of the population is as close as possible to $m=0$,
subject to the requirement that the system operates in the metastable regime.
We now want to consider the large $N$ limit and choose the largest pumping rate for a given $N$, however, the system should operate in the metastable regime, and ensure that $T_{\text{coh}} \ll T_{\text{ms}}$.
We choose the normalized pumping rate as a function of $N$ according to the ansatz 
\begin{equation}
\label{eq:path}
    r(N)=1-\beta N^{-q}.
\end{equation}
Here, $0<\beta,q$, so that $r(N) \rightarrow 1$ in the asymptotic limit as $N \rightarrow \infty$. 
Inserting Eq.~(\ref{eq:path}) into Eq.~(\ref{eq:ln}), we obtain
\begin{equation}
\label{T00}
    \ln(\Gamma T_{\text{ms}}) \sim {\frac{4\sqrt{2}}{3}\beta^{\frac{3}{2}}N^{1-\frac{3}{2}q}}.    
\end{equation}
From Eq.~(\ref{T00}), we clearly see that we want to choose $q < {2}/{3}$ so that the metastable time increases superpolynomially with $N$ in the asymptotic limit as $N \rightarrow \infty$. Thus, for a large enough $N$, the system operates in the metastable regime, and $T_{\text{ms}} \gg T_{\text{coh}}$, since $T_{\text{coh}}$ grows only polynomially with $N$. In Sec.~\ref{sec:Casymoptotic}, we will evaluate the coherence and linewidth for Eq.~(\ref{eq:path}), showing the coherence scales beyond the SQL, Eq.~(\ref{eq:SQL}).

Now, we present an alternative analysis to that in Sec.~\ref{subsec:ss} to explain the steady state phase transition using the above birth-death Markov process. According to its definition above, the metastable time, $T_{\text{ms}}$, is well-defined even when $r < r_0$. For heuristic purposes, we define the recovery time, 
$T_{\text{rec}}$, as the mean first passage time for the system to transition from $\rho^{\text{top}}$ to $\rho^{\text{ms}}$. This is the mean time for the system to start  operating as a laser again after it was in $\rho^{\text{top}}$. An asymptotic expression in the large $N$ limit is given by (see Appendix~\ref{sec:thesisderivationmetastable}) 
\begin{multline} 
\label{eq:recovery}
\ln(\Gamma T_{\text{rec}})\sim \ln(\rho^{\text{ss}}_{N/2}/\rho^{\text{ss}}_{m_{+}}) \\
\sim N[1 + \tanh^{-1}(\sqrt{1 - r^2}) - \sqrt{1 - r^2} + \ln(r/2)].
\end{multline}
Here, the term in the square brackets on the right hand side of Eq.~(\ref{eq:recovery}) is positive when $0<r<1$. According to Eqs.~(\ref{eq:ln}) and (\ref{eq:recovery}), the population at the unstable transition point \( m_+ \) is vanishingly small relative to the metastable peaks in the limit of large $N$, implying that the probability of the system residing between the two metastable states is negligible. This implies that transitions between the two metastable states occur as Poissonian events, with rates \( 1/T_{\text{ms}} \) and \( 1/T_{\text{rec}} \), respectively.
At the theoretical steady state, the system is a mixture of the two metastable states, obtained by averaging over the stochastic process for \( m \) on timescales much longer than \( T_{\text{ms}} \) and \( T_{\text{rec}} \).
In the asymptotic limit $N \rightarrow \infty$, when \( r < r_0 \), the system is far more likely to remain in $\rho^{\text{ms}}$, since \( T_{\text{rec}}/ T_{\text{ms}}\rightarrow 0 \), so that \( \rho^{\text{ss}} \rightarrow \rho^{\text{ms}} \) (here the arrows denotes the limit $N\rightarrow \infty$); conversely, when \( r > r_0 \), it is far more likely to remain in $\rho^{\text{top}}$, as \(   T_{\text{ms}}/T_{\text{rec}}\rightarrow0 \), yielding \( \rho^{\text{ss}}_m \rightarrow \rho^{\text{top}}_m\approx\delta_{m, N/2} \).
The critical point of the phase transition is at \( r = r_0 \approx 0.804 \), where the asymptotic exponential scalings of \( T_{\text{ms}} \) and \( T_{\text{rec}} \) with \( N \), as given in Eqs.~(\ref{eq:ln}) and (\ref{eq:recovery}), become equal. Additionally, these same equations provide the asymptotic limit of large $N$ for the expressions $\rho^{\text{ss}}_{m_{-}}/\rho^{\text{ss}}_{m_{+}}$ and $\rho^{\text{ss}}_{N/2}/\rho^{\text{ss}}_{m_{+}}$, respectively, which scale exponentially with $N$. Equating these gives the same critical point $r=r_0$, as explained in Sec.~\ref{subsec:ss}.

As stated above, finally, we derive a simplified form of the population dynamics equation~(\ref{eq:population}), applicable in the regime of large \( N \) and \( r > r_0 \), assuming the system is initialized in the metastable state \( \rho^{\text{ms}} \).
Thus, we have \( T_{\text{ms}} \ll T_{\text{rec}} \) in the limit of large $N$, so that the recovery of the system can be neglected, and \( \rho^{\text{ss}}_m \approx\rho^{\text{top}}_m \sim \delta_{m,N/2}\).
The same arguments from above apply here: the transition to $\rho^{\text{top}}$ is sharp—that is, the system is in one metastable state or the other, but not in between—and occurs at times following exponential statistics with the rate \( 1/T_{\text{ms}} \).
Consequently, the population dynamics equation (\ref{eq:population}) reduces to
\begin{equation}
\label{eq:expdecay}
    \rho_{m}(t) =\rho_{m}^{\text{ms}}\mathrm{e}^{-t/T_{\text{ms}}}+(1-\mathrm{e}^{-t/T_{\text{ms}}})\rho_{m}^{\text{ss}}.
\end{equation}
Equation~(\ref{eq:expdecay}) describes exponential decay from the metastable state \( \rho^{\text{ms}} \) to the steady state \( \rho^{\text{ss}} \), confirmed numerically in Fig.~\ref{fig:meta}.
In the metastable regime where \( \Gamma T_{\text{ms}} \) is extremely large, it suffices to consider only times \( t \ll T_{\text{ms}} \), so that the system remains in the metastable state, i.e.,~\( \rho(t) \approx \rho^{\text{ms}} \).
Equivalently, by choosing a stationary time according to \(\ T_{\text{coh}} \ll T_{\text{st}} \ll T_{\text{ms}}\), the system remains in the
metastable state with high probability for $t<T_{\text{st}}$, and the laser can be treated as stationary in these durations
(see Sec.~\ref{subsec:coherencederivation})

\section{Output photon statistics}
\label{sec:poissonian}
In this brief section, we show that our laser model exhibits Poissonian output photon statistics while operating in the metastable regime where $\Gamma T_{\text{ms}}$ is enormously large. 
We demonstrate in two equivalent ways that when the system is initialized in the metastable state of Eq.~(\ref{eq:truncate}), its population remains unchanged by the outcome of a direct photodetection of the output field.
This invariance implies that the output field has Poissonian photon statistics.

First, we derive the update in the excitation probability distribution from the birth-death process with the rates of Eq.~(\ref{eq:qqq}).
Suppose a photon is detected in the output field. The conditional probability that the excitation number after the emission is $m$, using a generalized Bayes’ theorem \cite{Wiseman_2009}, is given by 
\begin{multline} 
\label{eq:Bayes}
p'(m\vert\text{det.})=p(m+1\vert\text{det.})=
\frac{p(\text{det.}\vert m+1)p(m+1)}{p(\text{det.})} \\
=\frac{\Gamma J_{m+1}^2 \rho_{m+1}^{\text{ms}} \mathrm{d}t}{(\Gamma \nu -\Gamma \nu\rho_{m_+}^{\text{ms}})\mathrm{d}t} 
=\frac{1}{1-\rho_{m_+}^{\text{ms}}}\begin{cases}\rho_{m}^{\text{ms}},  & m\leq m_+-1  \\
0, & m \geq m_+
\end{cases}.
\end{multline}
Here, $p'(m\vert\text{det.})$ and $p(m)$ denote the probabilities for the system to be in the excitation number $m$ after and before the emission, respectively. 
The last equality in Eq.~(\ref{eq:Bayes}) follows from Eqs.~(\ref{eq:ss}) and (\ref{eq:truncate}). According to Eq.~(\ref{eq:Bayes}), the resulting state after a direct photodetection is the same state except that the last nonzero term is set to zero, i.e.~$p'(m_+\vert\text{det.})=0$, since the excitation number $m=m_++1$ is not populated and thus cannot emit a photon. In the metastable regime, using Eq.~(\ref{eq:ln}), we have $\rho_{m_+}^{\text{ms}} \sim 1/\Gamma T_{\text{ms}}$, so that $\rho_{m_+}^{\text{ms}}$ is extremely small and can be neglected. Thus, the system's state changes only negligibly after a direct photodetection at the output field, as promised.

Now, we prove the same claim in a second way, using the master equation formalism.
Define pure states having the same excitation statistics as the metastable states Eq.~(\ref{eq:truncate}). These are  the coherently radiating spin states (CRSS) of Eq.~(\ref{eq:yofhjksjhdgkfng}),
\begin{equation}
\label{eq:CRSS}
\ket{N/2, \sqrt{\nu}e^{i\phi}}_{\hat{J}} =\sum_{m=-\frac{N}{2}}^{m_+}e^{i(N/2+m)\phi}\sqrt{\rho_{m}^{\text{ms}}}\ket{{N}/{2},m}.  
\end{equation}
These states satisfy 
\begin{multline}
\label{eq:eigenstate}
    \frac{1}{e^{i\phi}\sqrt{\nu}}\hat{J}\ket{N/2,\sqrt{\nu}e^{i\phi}}_{\hat{J}} \\ =\ket{N/2,\sqrt{\nu}e^{i\phi}} _{\hat{J}}     -e^{i(N/2+m_{+})\phi}\sqrt{\rho_{m_{+}}^{\text{ms}}}\ket{N/2,m_{+}} \\ \approx \ket{N/2,\sqrt{\nu}e^{i\phi}}_{\hat{J}}.
\end{multline}
In the metastable regime, Eq.~(\ref{eq:ln}) gives
\(
\rho_{m_{+}}^{\text{ms}} \sim 1/{\Gamma T_{\text{ms}}}.
\)
The second term in the middle expression of Eq.~(\ref{eq:eigenstate}) is therefore negligible, and the CRSS states in Eq.~(\ref{eq:CRSS}) become approximate eigenstates of \( \hat{J} \) (see Sec.~\ref{sec:CRSS}). 
From the definition in Eq.~(\ref{eq:CRSS}), the metastable states of Eq.~(\ref{eq:truncate}) are uniform incoherent mixture of the CRSS of all phases
\begin{equation}
\label{eq: superposition}
\rho^{\text{ms}}=\int_{0}^{2\pi}\frac{d\phi}{2\pi}\ket{N/2, \sqrt{\nu}e^{i\phi}}_{\hat{J}} \bra{N/2, \sqrt{\nu}e^{i\phi} }.
\end{equation}
A detection of a photon in the output field corresponds to a quantum jump of the system’s state, described (up to normalization) by the transformation $\rho \rightarrow \mathcal{J}[\hat{J}] \rho$, where $\hat{J}$ is the jump operator \cite{Wiseman_2009,Markovhierarchy} associated with the loss superoperator in the master equation (\ref{eq:model}).
Thus, using Eqs.~(\ref{eq:eigenstate}) and (\ref{eq: superposition}), the metastable state remains unchanged by the outcome of a direct photodetection, proving Poissonian output photon statistics. Using input–output theory, where the output field operator is proportional to the collective atomic lowering operator \(\hat{J}\) \cite{Wiseman_2009}, the Glauber intensity correlation functions of order \(n\) satisfy 
\begin{equation}
    g^{(n)}\left(t_1, \dots, t_n\right)=\frac{\langle\hat{J}^{\dagger}\left(t_1\right) \cdots \hat{J}^{\dagger}\left(t_n\right) \hat{J}\left(t_n\right) \cdots \hat{J}\left(t_1\right)\rangle}{\prod_{j=1}^n\langle\hat{J}^{\dagger}\left(t_j\right) \hat{J}\left(t_j\right)\rangle}=1,
\end{equation}
for all positive integers \(n\) and all choices of times \(t_1, \dots, t_n\), reflecting the statistical independence of successive photodetection events. The laser's flux is \begin{equation}
\label{eq:flux}
\mathcal{N}=\Gamma \expectation{\hat{J}^\dagger(t)\hat{J}(t)}  = \Gamma \nu.
\end{equation}
The flux equals the birth rate at which excitations are added to the system, as given in Eq.~(\ref{eq:qqq}).

\section{Coherence Properties}
\label{sec:coherenceprop}
In this section, we present numerical and analytical results for the coherence properties of the laser model. First, in Sec.~\ref{subsec:coherencederivation}, we consider how to define the coherence of a metastable laser beam. In Sec.~\ref{subsec:cohanalytical}, we present the analytical results and the linewidth, and hence coherence, of the laser model. Then, we compare these analytical results with numerics, showing excellent agreement for a large $N$. In Sec.~\ref{subsec:linewidthBlochsphere}, we corroborate the analytical results for the linewidth using a geometric approach to calculate the phase noise accumulated during the loss process.
In Sec.~\ref{subsec:cohasfunctionofpumping}, we evaluate the coherence across the full range of the pumping rates. Finally, in Sec.~\ref{sec:Casymoptotic} we evaluate the linewidth and coherence in the asymptotic limit of Eq.~(\ref{eq:path}), showing achievable scaling beyond the SQL of Eq.~(\ref{eq:SQL}).

\subsection{Coherence of a metastable laser beam}
\label{subsec:coherencederivation}
Here, we consider the coherence of a general metastable laser. In Sec.~\ref{thesissec:HLtheorem} we presented the four conditions from Ref.~\cite{HL} on a laser model that are assumed in order to prove the Heisenberg limit theorem. The coherence of a laser model satisfying these four conditions is given by Eq.~(\ref{eq:C}) \cite{HL}. 
Here, we generalize condition three to include metastable laser models, such as our model, that can be considered stationary only for finite durations $T_{\text{st}}$.  
We strictly  require that $T_{\text{coh}} \ll T_{\text{st}}$, where $T_{\text{coh}}$ is the coherence time of the laser. Below, we show that the coherence $\coh$ of such a metastable laser is only negligibly smaller than the value given by Eq.~(\ref{eq:C}) for an otherwise identical, strictly stationary laser with $T_{\mathrm{st}}=\infty$.
In Sec.~\ref{sec:metastable}, we showed that our laser model when operating in the metastable regime, is stationary for durations $T_{\text{st}}$ much shorter than the metastable time, i.e.,~$T_{\text{coh}} \ll T_\text{st} \ll T_{\text{ms}}$, because it then remains in the metastable state with high probability

Recall that the coherence $\coh$ is defined as the mean number of photons in the maximally populated mode of the beam. We calculate the mean number of photons in a uniform mode of length $T_{\text{st}}$, that is defined by its bosonic annihilation operator ($c=1$)
\begin{equation}
\label{eq:b_u}
\hat{b}_{\text{u}}=\frac{1}{\sqrt{T_{\text{st}}}} \int_{0}^{T_{\text{st}}} \hat{b} (t) \, \mathrm{d} t.
\end{equation}
Here, we have $[\hat{b}_{\text{u}},\hat{b}_{\text{u}}^\dagger]=1$. We restricted to a mode of length $T_{\text{st}}$ so that the assumption of time-translation invariant statistics applies throughout the entire duration of the mode, i.e.,~for times $0 \leq t\leq T_{\text{st}}$. Consequently, within this interval, the first-order correlation function defined in Eq.~(\ref{firstorderGlauber}) depends only on the time difference, $G^{(1)}(t,t')=:G^{(1)}(t-t')$, for $0\leq t,t'\leq T_{\mathrm{st}}$.
We evaluate the mean number of photons in the mode of Eq.~(\ref{eq:b_u})
\begin{multline}
\label{eq:expectationbdagb}
\langle\hat{b}_u^\dagger\hat{b}_u\rangle 
=\frac{1}{T_{\text{st}}} \iint_{t, t^{\prime}=0}^{T_{\text{st}}} \mathrm{d} t \mathrm{d} t^{\prime} G^{(1)}(t-t') 
\\
=2\int_{0}^{T_{\text{st}}}G^{(1)}(\tau)\mathrm{d}\tau-\frac{2}{T_{\text{st}}}\int_{0}^{T_{\text{st}}}\tau G^{(1)}(\tau)\mathrm{d}\tau.
\end{multline}
Here, the second equality follows from a change of integration variables.
The family of modes defined by Eq.~(\ref{eq:b_u}) does not encompass all possible modes of the laser beam. Since $\coh$ is defined by the maximally populated mode,
$\langle \hat{b}_u^\dagger \hat{b}_u \rangle \leq \coh$. However, Ref.~\cite{HL} shows that, in the stationary limit $T_{\mathrm{st}}\rightarrow\infty$, the maximally populated mode is the uniform mode of infinite duration obtained from Eq.~(\ref{eq:b_u}) in this limit.
Thus, using Eq.~(\ref{eq:expectationbdagb}), the coherence of the metastable laser beam satisfies 
$\coh \leq 2\int_0^{\infty} d \tau \, G^{(1)}(\tau)$, where an equality in the last expression is obtained for a stationary laser beam. 
For a laser model that satisfies condition four in Sec.~\ref{thesissec:HLtheorem}, we have $G^{(1)}(t)=\mathcal{N}\exp(-\vert t \vert /T_{\text{coh}})$, for $0\leq t\leq T_{\text{st}}$, and Eq.~(\ref{eq:expectationbdagb}) reduces to 
\begin{multline}
\label{eq:uvbur}
\langle\hat{b}_{\text{u}}^\dagger\hat{b}_{\text{u}}\rangle  =
{2\mathcal{N}} T_{\text{coh}} \left(1-\frac{T_{\text{coh}}}{T_{\text{st}}}\left(1-e^{-{T_{\text{st}}/T_{\text{coh}}} }\right)\right) 
\\
={2\mathcal{N}} T_{\text{coh}} \left(1-O \left(\frac{T_{\text{coh}}}{ T_{\text{st}}}\right)\right).
\end{multline}
Thus, when $T_{\mathrm{coh}}/T_{\mathrm{st}}\ll 1$, the coherence of the metastable laser beam is only negligibly smaller than the stationary value $2\mathcal{N}T_{\mathrm{coh}}=4\mathcal{N}/\ell$ given by Eq.~(\ref{eq:C}), with equality recovered as $T_{\mathrm{st}}\rightarrow\infty$.

\subsection{Analytical and numerical results: coherence and linewidth}
\label{subsec:cohanalytical}
Here, we present analytical results for the linewidth and coherence of the laser. Then, we compare these results with numerics.
In Appendix \ref{subsecthesis:derivationoflinewidth}, we derive the following expression for the linewidth of the laser that is operating in the metastable regime
\begin{equation}
\label{eq:linewidthresult1}
    \ell
    =
    2\Gamma\int_{-N/2}^{m_+}\rho_m^{\mathrm{ms}}g(m)\,\mathrm{d}m,
\end{equation}
where we have defined
\begin{equation}
    g(m)
    :=
    \left(\pp{J_m}{m}\right)^2
    \approx
    \frac{4m^2}{N^2-4m^2}.
\end{equation}
We further simplify Eq.~(\ref{eq:linewidthresult1}) by exploiting the fact that $\rho_m^{\mathrm{ms}}$ is sharply peaked around $m_-$. Using the Gaussian approximation in Eq.~(\ref{eq: gaussian}) and expanding $g(m)$ to second order about $m_-$, we obtain
\begin{multline}
    \ell
    \approx
    2\Gamma\int_{-N/2}^{m_+}
    \frac{1}{\sqrt{2\pi w^2}}
    \mathrm{e}^{-(m-m_-)^2/(2w^2)}
    \bigg[
    g(m_-)
        \\ +g'(m_-)(m-m_-)
    +\frac{1}{2}g''(m_-)(m-m_-)^2
    \bigg]
    \,\mathrm{d}m
    \\
    \approx
    \Gamma\left[2g(m_-)+w^2g''(m_-)\right]
   \\ =
  2\Gamma g(m_-) \left[1+\frac{4-3r^2}{r^2\left(1-r^2\right)^{3/2}}\frac{1}{N}\right] \\ =2\Gamma g(m_-)\left[1+O(N^{-1})\right].
\end{multline}
To leading order in $N$, only the zeroth-order term in the expansion of $g(m)$ contributes; equivalently, the sharply peaked metastable population can be approximated by a delta function centered at $m_-$. Thus, we have
\begin{equation}
\label{eq:linewidth}
\ell=    2\Gamma
    \left.
    \left(\frac{\partial J_m}{\partial m}\right)^2
    \right|_{m=m_-}= \Gamma \frac{\left(\frac{\partial J^2_m}{\partial m} \right)^2}{2J^2_m}\Bigg|_{m=m_-} \approx 2\Gamma \frac{1-r^2}{r^2}.
\end{equation}
The final approximation in Eq.~(\ref{eq:linewidth}) follows by substituting an approximate value of \(m_-\approx-N\sqrt{1-r^2}/2\) [see Eq.~(\ref{eq:mpm})]. Here, the gain and loss processes generate equal phase noise and therefore contribute equally to the linewidth (see Appendix~\ref{subsecthesis:derivationoflinewidth}).

To derive the coherence, we assume that the laser operates in the metastable regime and use Eq.~(\ref{eq:C}), whose applicability to the metastable regime was established in Sec.~\ref{subsec:coherencederivation}. Thus, using Eqs.~(\ref{eq:C}), (\ref{eq:flux}), and (\ref{eq:linewidth}), we obtain 
\begin{equation}
\label{eq:coherence}
\coh =8\left[R[\hat{J}](m_-)\right]^2 \approx \frac{N^2 r^4}{2(1-r^2)}.
\end{equation}
Here, the functional $R[\hat{J}](m)$ is defined in Eq.~(\ref{eq:R[Q]}). Thus, the coherence is proportional to the square of the inverse relative rate of change of the coefficients $J_m^2$, evaluated at the stable fixed point $m_{\mathrm{-}}$. This confirms the intuition presented in Sec.~\ref{subsec:ourmodel}: metastable states centered in regions where $J_m^2$ varies more slowly exhibit higher coherence. The results of Eqs.~(\ref{eq:linewidth}) and (\ref{eq:coherence}) are calculated using the assumption that the laser remains in the metastable state at all times, and thus, are reliable estimates only for pumping rates for which the laser operates in the metastable regime (see Secs.~\ref{subsec:coherencederivation} and \ref{sec:metastable}).
According to Eqs.~(\ref{eq:width}) and (\ref{eq:coherence}), the superradiant laser satisfies $\coh=8V(\hat{n})^2$, and therefore satisfies the bound in Eq.~(\ref{eq:irfhtcmgohhvnhkjfhfgfh897653958}).

\begin{figure}
    \centering
    \includegraphics[width=\linewidth]{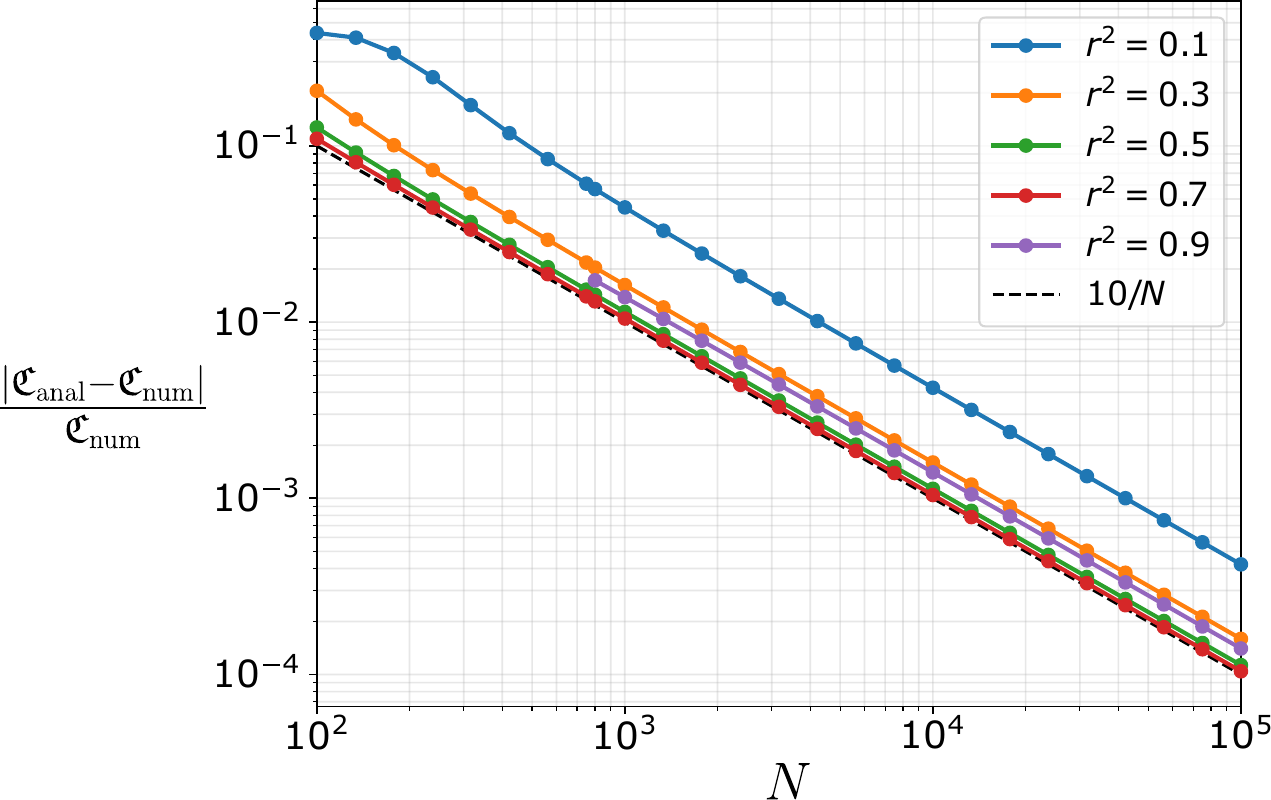}
\caption{Comparison of the analytical coherence $\mathfrak{C}_{\mathrm{anal}}$, Eq.~(\ref{eq:coherence}), with
the numerically computed coherence $\mathfrak{C}_{\mathrm{num}}$ (Appendix \ref{subsec:numericalmethod}). The relative error
$|\mathfrak{C}_{\mathrm{anal}}-\mathfrak{C}_{\mathrm{num}}|/\mathfrak{C}_{\mathrm{num}}$
is plotted as a function of $N$ on logarithmic axes for $r^2 = 0.1$, 0.3, 0.5, 0.7 and 0.9 (solid lines with markers). For $r^2 = 0.9$, the data start at $N = 800$ to ensure that $T_{\mathrm{ms}} \gg T_{\text {coh }}$. The black dashed line shows
$10/N$ for reference.}
    \label{fig:analyticsvsnumerics}
\end{figure}

The analytical expression for the coherence, Eq.~(\ref{eq:coherence}), is compared with numerical results in Fig.~\ref{fig:analyticsvsnumerics}. The numerical data in all figures in this section are obtained using the method described in Appendix~\ref{subsec:numericalmethod}. 
For all pumping rates shown, the relative error decreases approximately as $N^{-1}$ at large $N$, supporting the asymptotic validity of the analytical expression.

\subsection{Phase noise accumulation on the Bloch sphere}
\label{subsec:linewidthBlochsphere}
\begin{figure}
    \centering
\includegraphics[width=\linewidth]{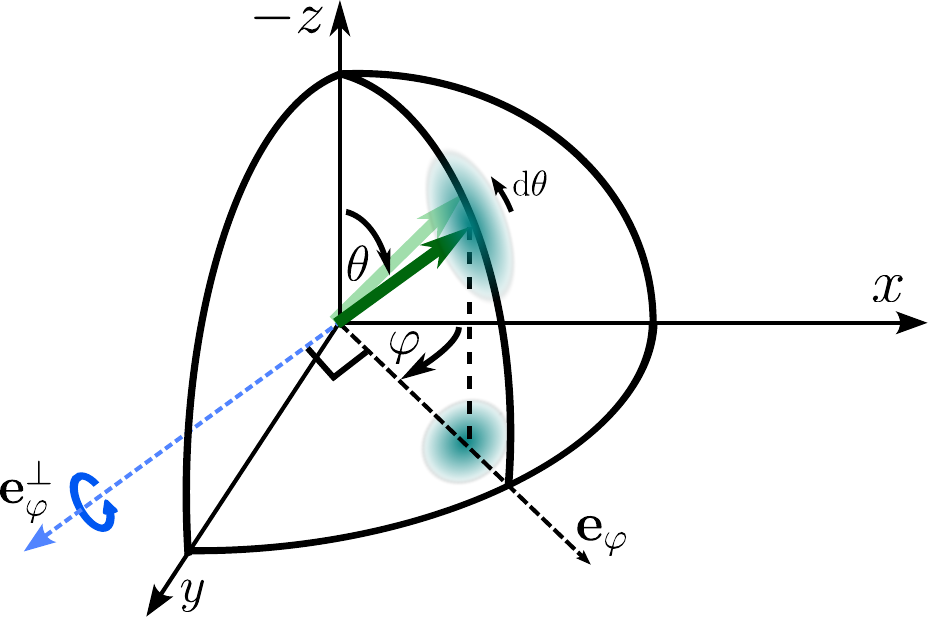}
\caption[Loss process illustrated on the inverted Bloch sphere]{Loss process illustrated on the inverted Bloch sphere (ground state at top). 
The mean pseudo-spin of Eq.~(\ref{eq:mkoinhl}) is indicated by the dark green arrow, with pseudo-spin fluctuations shown by a blue-gray uncertainty ellipse; its projection onto the $x$--$y$ plane is circular, a manifestation of a laser beam with Poissonian output photon statistics \cite{myarticle}. The loss process of Eq.~(\ref{eq:lossprocess}) tends to rotate the state around the dashed-blue axis according to Eq.~(\ref{eq:rotmeanspin}), so that the coordinate $\theta$ tends to decrease according to Eq.~(\ref{eq:thetadecrease}). The light green arrow indicates the mean pseudo spin after the infinitesimal rotation.}
    \label{fig:rotphasespace}
\end{figure}
Here, we illustrate how phase noise accumulates during the loss process using geometric phase space methods on the Bloch sphere. We show that the phase noise accumulation rate due to the loss equals half the linewidth given by Eq.~(\ref{eq:linewidth}).
We assume that our laser model evolves according to a pure phase diffusion of the CRSS, and that its state is initialized in the CRSS state 
\begin{equation}
\label{eq:mkoinhl}
\rho(t_0)=\ket{N/2,\alpha}_{\hat{J}}\bra{N/2,\alpha}.
\end{equation}
with
\begin{equation}
    \alpha:=\frac{N}{2}\sin\theta e^{-i\varphi}.
\end{equation}
Here, we introduced the parameters $\theta$ and $\varphi$. The coordinate $\theta$ is chosen according to 
\begin{equation}
\label{eq:oterjihffghjk}
    \sin \theta:=r,
\end{equation}
where $r$ is the normalized pumping rate coefficient as usual.
The coordinate $\varphi$ corresponds to the phase of the laser. 
The Bloch-vector expectation values for the state of Eq.~(\ref{eq:mkoinhl}) are given by 
\begin{multline}
\label{eq:pseudospin}
    \expectation{\hat J_x}=N/2 \sin{\theta}\cos{\varphi}, \quad  \expectation{\hat J_y}= N/2 \sin{\theta}\sin{\varphi}, \\ \expectation{\hat J_z}=- N/2 \cos{\theta}.
\end{multline}
Figure~\ref{fig:rotphasespace} illustrates the mean pseudo-spin components of Eq.~(\ref{eq:pseudospin}) and their fluctuations on the Bloch sphere for the state of Eq.~(\ref{eq:mkoinhl}). 
Define unit vectors in the transverse plane (see Fig.~\ref{fig:rotphasespace})
\begin{equation}
    \mathbf{e}_\varphi= \cos\varphi \mathbf{x}+ \sin \varphi \mathbf{y}, \quad \mathbf{e}_\varphi^{\perp}= -\sin \varphi \mathbf{x} +\cos \varphi \mathbf{y}.
\end{equation}
Here, $\mathbf{e}_\varphi$ points along the projection of the Bloch-vector expectation value onto the $xy$ plane, while $\mathbf{e}_{\varphi}^{\perp}$ denotes the corresponding orthogonal direction within that plane.  
Define the pseudo spin operators in the direction of the unit vectors
\begin{equation}
   \hat J_\varphi=  {\mathbf{\hat{J}}} \cdot \mathbf{e}_\varphi=   
   \cos\varphi \hat J_ x+ \sin \varphi \hat J_y=\frac{1}{2}(e^{-i\varphi}\hat{J}^{\dagger}+e^{i\varphi}\hat{J}), 
   \end{equation}
\begin{equation}
\label{eq:Jvarphiperp}
   \hat{J}_\varphi^{\perp}=  {\mathbf{\hat{J}}}\cdot \mathbf{e}_\varphi^{\perp}= -\sin\varphi \hat J_x+ \cos \varphi \hat J_y=\frac{-i}{2}(e^{-i\varphi}\hat{J}^{\dagger}-e^{i\varphi}\hat{J}).
\end{equation}
Here, $\mathbf{\hat J}:=\hat J_x \mathbf{x}+\hat J_y \mathbf{y}+\hat J_z \mathbf{z}$.
Define the phase variance \cite{spinsqueezingreview} 
\begin{equation}
\Delta\varphi^{2}:=\frac{(\Delta \hat J_\varphi^{\perp})^2}{\expectation{\hat{J}_\varphi}^{2}}.
\end{equation}
This quantity is the variance of small fluctuations of the laser phase about its mean value $\varphi$.

The loss process for the master equation of the model of Eq.~(\ref{eq:model}) is given by
\begin{equation}
\label{eq:lossprocess}
\dot{\rho}=\Gamma\mathcal{D}[\hat J] \rho.
\end{equation}
Below, we calculate the phase noise that accrues during the evolution of the loss process of Eq.~(\ref{eq:lossprocess}) on the state of Eq.~(\ref{eq:mkoinhl}) from time $t_0$ to the time $t_0+\mathrm{d}t$, where $\mathrm{d}t$ is an infinitesimal time step.
Now, recall that the state of Eq.~(\ref{eq:mkoinhl}) is the steady state solution of Eq.~(\ref{thesiseq:DDsuperradiance}) when taking $\alpha=-2i\Omega/\Gamma$. Taking $\dot {\rho}=0$ in Eq.~(\ref{thesiseq:DDsuperradiance}) we get the following expression for the loss process given in Eq.~(\ref{eq:lossprocess}) acting on the state $\rho(t_0)$:
\begin{equation}
\label{eq:rotmeanspin}
    \Gamma\mathcal{D}[\hat J]\rho(t_0)=-i\Gamma\frac{N}{2}\sin{\theta}[\hat{J}_\varphi^{\perp},  \rho(t_0)].
\end{equation}
The right hand side of Eq.~(\ref{eq:rotmeanspin})  corresponds to a unitary evolution of an infinitesimal rotation on the Bloch sphere at the time $t_0$ (see Fig.~\ref{fig:rotphasespace}), since $\hat J_\varphi^{\perp}$ is composed of the generators for the rotation group SO(3). Using Eq.~(\ref{eq:rotmeanspin}), this rotation tends to decrease the coordinate $\theta$ (see Fig.~\ref{fig:rotphasespace}) according to 
\begin{equation}
\label{eq:thetadecrease}
    \dd{\theta}{t}\bigg\vert_{t=t_0}=-\Gamma\frac{N}{2}\sin \theta.
\end{equation}
We note that Eqs.~(\ref{eq:rotmeanspin}) and (\ref{eq:thetadecrease}) are valid only at the time $t_0$ and not after that. We calculate, using Eqs.~(\ref{eq:mkoinhl}), (\ref{eq:Jvarphiperp}), and (\ref{eq:CRSS})
\cite{myarticle}
\begin{equation}
    (\Delta \hat J_\varphi^{\perp})^2=  
\frac{1}{4}\expectation{[J,J^\dagger]}=-\frac{1}{2}\expectation{J_z}=
        \frac{N}{4} \cos{\theta}.
\end{equation}
We now calculate the phase noise that is gained during the loss process of Eq.~(\ref{eq:lossprocess}) on the state of Eq.~(\ref{eq:mkoinhl})
\begin{multline}
\label{eq:phasnoise123}
\dd{\Delta\varphi^{2}}{t}\bigg\vert_{t=t_0}=(\Delta\hat{J}_{\varphi}^{\perp})^{2}\frac{\mathrm{d}}{\mathrm{d}t}\bigg\vert_{t=t_0}\left(\frac{1}{\expectation{\hat{J}_{\varphi}}^{2}}\right) \\
= \frac{-2}{N} \frac{\cos^2{\theta}}{\sin^3{\theta}}\dd{\theta}{t}\bigg\vert_{t=t_0}=\Gamma\frac{\cos^2\theta}{\sin^2\theta}=\Gamma\frac{1-r^2}{r^2}.
\end{multline}
Here, the expression $(\Delta\hat{J}_{\varphi}^{\perp})^{2}$ remains unchanged after the infinitesimal rotation since it corresponds to the variance of $\mathbf{\hat J}$ in the direction parallel to the axis of rotation (see Fig.~\ref{fig:rotphasespace}).
In the transition from the third expression to the fourth one we used Eq.~(\ref{eq:thetadecrease}). In the transition from the fourth expression to the last one, we used Eq.~(\ref{eq:oterjihffghjk}).

The loss-induced phase diffusion rate, Eq.~(\ref{eq:phasnoise123}), is half the linewidth as calculated in Eq.~({\ref{eq:linewidth}}). However, as noted above, in Appendix~\ref{subsecthesis:derivationoflinewidth}, we show that the gain and loss processes generate equal amounts of phase noise. Thus, the total phase-noise accumulation rate under the master equation (\ref{eq:model}) equals the linewidth in Eq.~({\ref{eq:linewidth}}), as expected.

\subsection{Coherence as a function of the pumping rate}
\label{subsec:cohasfunctionofpumping}
\begin{figure}
\centering
\includegraphics[width=\linewidth]{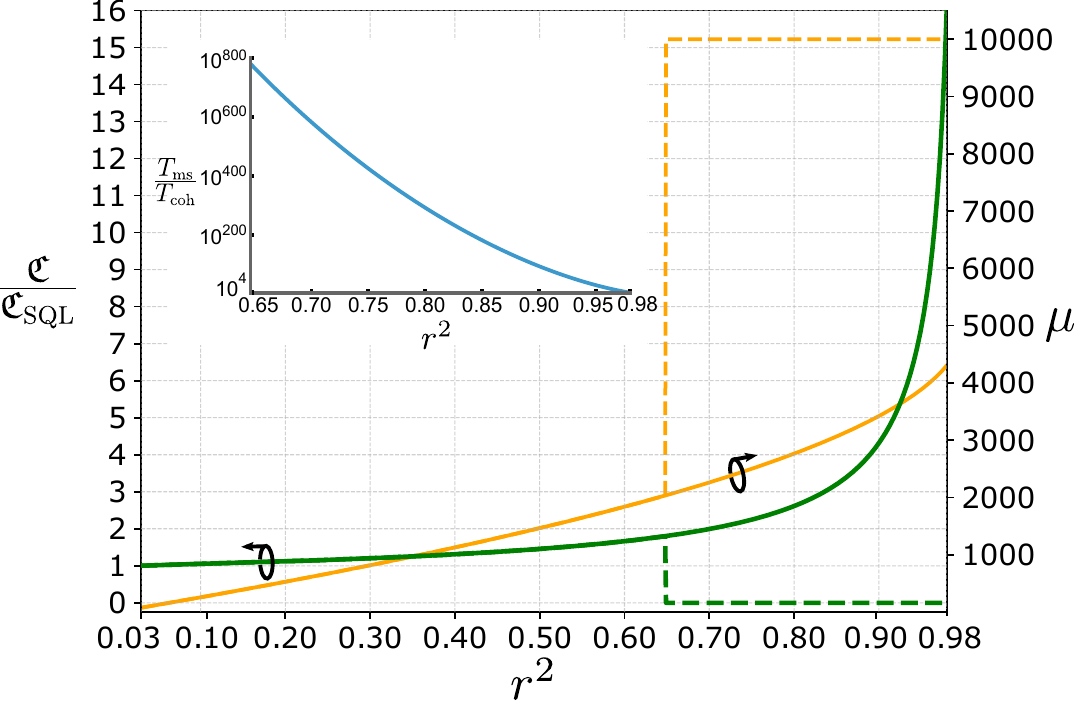}
\caption[$\coh/\coh_{\text{SQL}}$ as a function of the normalized pumping rate $r^2 = 4\nu/N^2$ for $N=10^4$ atoms]{Normalized coherence, $\coh/\coh_{\text{SQL}}$ as a function of the normalized pumping rate $r^2 = 4\nu/N^2$ for $N=10^4$ atoms, shown by the green curve (left axis). The orange curve shows the mean number of excitations, $\mu$ (right axis). For pumping rates above the critical value, $r^2 > r_0^2 \approx 0.65$, the solid lines represent the metastable-state solutions, and the dashed lines represent the steady-state solutions. The numerical and analytical results are indistinguishable for all curves shown in the plot. Inset: The ratio of the metastable time to the coherence time, $T_{\text{ms}}/T_{\text{coh}}$, as a function of $r^2$ (analytical solution).}
    \label{fig:cohmu}
\end{figure}
Here, we consider the coherence, $\coh$, across the full range of pumping rates. In Fig.~\ref{fig:cohmu}, we plot ${\coh}/{\coh_{\text{SQL}}}$ as a function of the normalized pumping rate $r^2={4\nu}/{ N^2}$ for a finite but large number of atoms, $N=10^4$,  where $\coh_{\text{SQL}}$ is the SQL for laser coherence of Eq.~(\ref{eq:SQL}). 

For small values of $r^2$, the superradiant coefficients $J_m^2$ are approximately linear in the excitation number $m+N/2$ (see Fig.~\ref{fig:coefficients}); 
in this linear regime, Eq.~(\ref{eq:model}) maps onto the FAT master equation~(\ref{eq:FAT}), where the bosonic annihilation operator $\hat a$ is realized by the collective atomic operator with the appropriate normalization $\hat{a}\mapsto\hat J/\sqrt{N}$, and with the parameter substitutions $\kappa \mapsto N\Gamma$, and $P/\kappa \mapsto \nu/N = Nr^2/4$.
When $\mu \gg 1$, the solution to this master equation is pure phase diffusion for the CRSS, Eq.~(\ref{eq:CRSS}); which are, in this limit, coherent spin states (CSS) \cite{mandel_wolf_1995,spinsqueezingreview}. That is, the dynamics are described by all the atoms with the same population and share a common phase that undergoes pure phase diffusion. Since the population of each atom is much smaller than one, the atoms can be considered as classical dipoles when interacting with radiation. However, unlike the usual low-excitation regime, where coherent laser light drives atoms into classical dipoles, here, collective excitations are added to the atoms one by one at a constant rate, and the atoms themselves generate classical light.

The radiation dipole $\hat{J}$ is $N$ times larger than that of a single atom, and correspondingly, the atoms emit light at an enhanced emission rate $N^2$ larger than that of a single atom with the same population (per atom). In the low pumping regime $r^2\ll1$, we have $\mu\approx Nr^2/4$, a flux of $\mathcal{N} \approx \Gamma N\mu$, and a linewidth $\ell \approx \Gamma N/2\mu \approx 2\Gamma/r^2 $.
In Fig.~\ref{fig:cohmu}, we omit the plot for $r^2 \leq 0.03$, for which $\mu \lesssim 75 $, as this is insufficient to produce pure phase diffusion.
It is worth noting that including spontaneous emission during the loss process—that is, the decay $\ket{e}\rightarrow\ket{g}$ into free-space modes, would not change the results significantly. In this case, the atomic state remains a CSS and undergoes pure phase diffusion, and only a small fraction of the coherent emission is radiated into free-space modes. Most of the light is instead collected by the loss cavity, whose output exceeds the free-space emission by a factor that scales as $N$ because of the collective superradiant enhancement discussed above.

As the pumping rate coefficient $r^2$  increases, the dependence of the superradiant coefficients $J_m^2$ in the excitation number $m+N/2$ becomes sublinear (i.e., in Fig.~\ref{fig:coefficients}, the slope of $J_m^2$ as function of $m$ decreases). Correspondingly, the coherence in Fig.~\ref{fig:cohmu} gradually exceeds the SQL, with a deviation that grows with $r^2$. Within the stable (pre-metastable) regime, this enhancement reaches a maximum of $\coh/\coh_{\text{SQL}} \approx 1.8$ at $r^2=r_0^2\approx 0.65$. This ratio value persists in the large-$N$ limit (see Fig.~\ref{fig:asyma} below).

For $r^2>r_0^2\approx 0.65$ the system's steady state experiences a phase transition, being close to the top state $m=N/2$ with $\mu=N$ (see Sec.~\ref{sec:metastable}), which generates light with negligible coherence (see Sec.~\ref{subsec:ss}). 
In contrast, for the desired long-lived metastable state, $\coh/\coh_{\text{SQL}}$ continues to grow with $\mu$. As $r^2$ approaches 1, the slope of $\coh/\coh_{\text{SQL}}$ as function of $r^2$ grows rapidly. 
The inset of Fig.~\ref{fig:cohmu} shows $T_{\text{ms}}/T_{\text{coh}}$ for the same configuration, with $T_{\text{ms}}$ being much larger than the coherence time, $T_{\text{coh}}$, i.e, the metastable regime, for all values of $r$ in the plot. Specifically, at the last point on the $r^2$ axis, $r^2=0.98$, we still have $T_{\text{ms}}/T_{\text{coh}}\approx 2.3\times 10^{4}\gg 1$. We omit the plot for $r^2>0.98$ because when $T_{\text{ms}}$ is no longer much greater than $T_{\text{coh}}$ we cannot use the approach of Sec.~\ref{subsec:coherencederivation}. At this last point, $\coh/\coh_{\text{SQL}}=16$.

\subsection{Coherence beyond the SQL at the asymptotic limit $N\rightarrow \infty$}
\label{sec:Casymoptotic}
\begin{figure}
\centering    
\includegraphics[width=\linewidth]{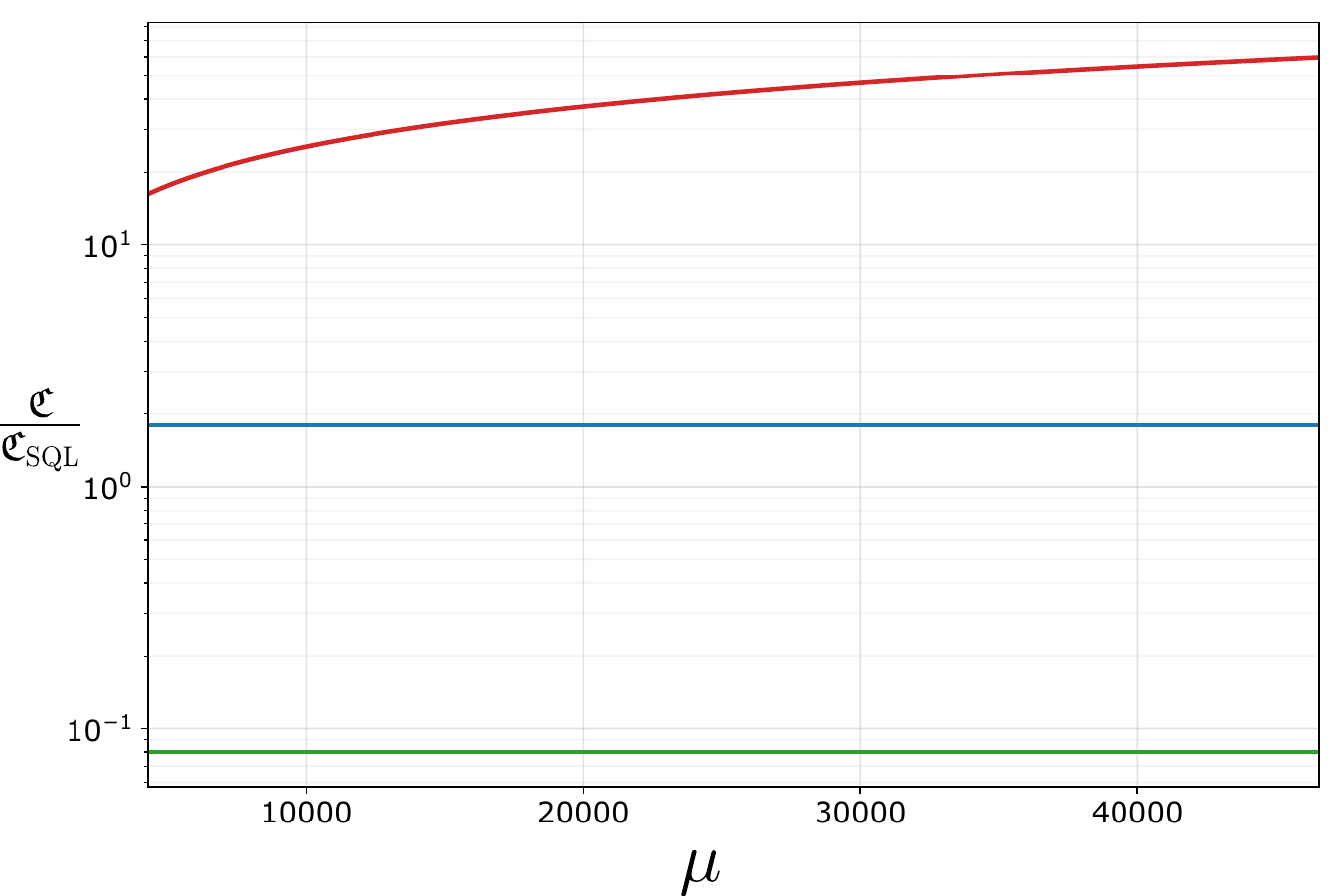}
\caption[$\coh/\coh_{\text{SQL}}$ as a function of the number of excitations, $\mu$]{$\coh/\coh_{\text{SQL}}$ as a function of the number of excitations, $\mu$. The red curve represents the solution for the pumping rate trajectory given by $\beta=3$, $q=0.62$ in Eq.~(\ref{eq:path}). Numerical and analytical results are indistinguishable in the plot. The blue curve is the maximal achievable $\coh/\coh_{\text{SQL}}$ at steady state, obtained with a normalized pumping rate coefficient just below the critical value, $r=0.8<r_0\approx 0.804$.
The green curve represents the model of the standard superradiant laser, showing the maximal achievable value of $\coh/\coh_{\text{SQL}}$ for the model, given in Eq.~(\ref{eq:coh/mu^2}).}
\label{fig:asyma}
\end{figure}
In this section, we will show that in the macroscopic limit for $N$, our model can surpass the SQL by a factor that grows without limit in $\mu$. More specifically, we find the maximal scaling of the coherence as a function of the number of excitations $\mu$. 
Recall that in Sec.~\ref{subsec:generalmodelsphasenoise} we motivated why our laser model should operate as close as possible to the excitation number $m=0$, where the off-diagonal superradiant coefficients are flattest (see Fig.~\ref{fig:coefficients}). The same conclusion follows directly from the expression for the coherence in Eq.~(\ref{eq:coherence}).
This corresponds to taking the normalized pumping rate $r$ as close as possible to $1$ so that the center of the population is as close as possible to $m=0$,
subject to the requirement that the system operates in the metastable regime.
In Sec.~\ref{subsec:metastabletime}, we considered the large $N$ limit, and examined the ansatz of Eq.~(\ref{eq:path}) for an asymptotic scaling of the pumping rate as function of $N$. If we set $q<2/3$ in Eq.~(\ref{eq:path}), we showed that the system operates in the metastable regime in which $T_{\text{ms}} \gg T_{\text{coh}}$, if $N$ is large enough.
We substitute Eq.~(\ref{eq:path}) into Eqs.~(\ref{eq:linewidth}) and (\ref{eq:coherence}), obtaining the following asymptotic expressions for the linewidth and coherence:
\begin{align}
\label{eq:linewidthasymptotic}
\ell &\approx 4 \Gamma\beta N^{-q}=\Gamma \Omega(N^{-{2}/{3}}), \\
\label{eq:cohasymptotic}
\coh &\approx \frac{N^{2+q}}{4\beta}=O\left(N^{\frac{8}{3}}\right).
\end{align}
Here, the last terms in Eqs.~(\ref{eq:linewidthasymptotic}) and (\ref{eq:cohasymptotic}) denote the minimal and maximal scaling for the linewidth and coherence, respectively, where we used the Bachmann-Landau notation \cite{BigO}.
Moreover, any power in $N$ up to $8/3$ is possible for the coherence scaling. Since $\mu\approx N/2$ in the asymptotic limit $N \rightarrow \infty$, the same scaling law applies as a function of the number of excitations $\mu$, i.e.~$\coh=O(\mu^{8/3})$.
This result is unprecedented for a laser in the optical regime, as all current models are limited by the SQL (\ref{eq:SQL}), $\coh=O(\mu^2)$.

Figure \ref{fig:asyma} plots $\coh/\coh_{\text{SQL}}$ as a function of the mean number of excitations $\mu$.
The red curve shows the results for the pumping-rate trajectory defined by the parameters $\beta=3$ and $q=0.62$ in Eq.~(\ref{eq:path}), for which the ratio of the metastable time to the coherence time remains approximately $T_{\text{ms}}/T_{\text{coh}} \approx 1.6 \times 10^4$ (similar to the ratio at $r^2=0.98$ in Fig.~\ref{fig:cohmu}) over the full range of $\mu$ shown.
In addition, the maximal achievable value of $\coh/\coh_{\text{SQL}}$ in the steady state ($r^2=0.64$) is shown in blue. For comparison, we also show the maximal achievable value of $\coh/\coh_{\text{SQL}}$ for the standard superradiant laser model, Eq.~(\ref{eq:coh/mu^2}) , which is two to three orders of magnitude smaller than that of our model, for the number of atoms shown in the plot.

\begin{figure}
\centering
\includegraphics[width=\linewidth]{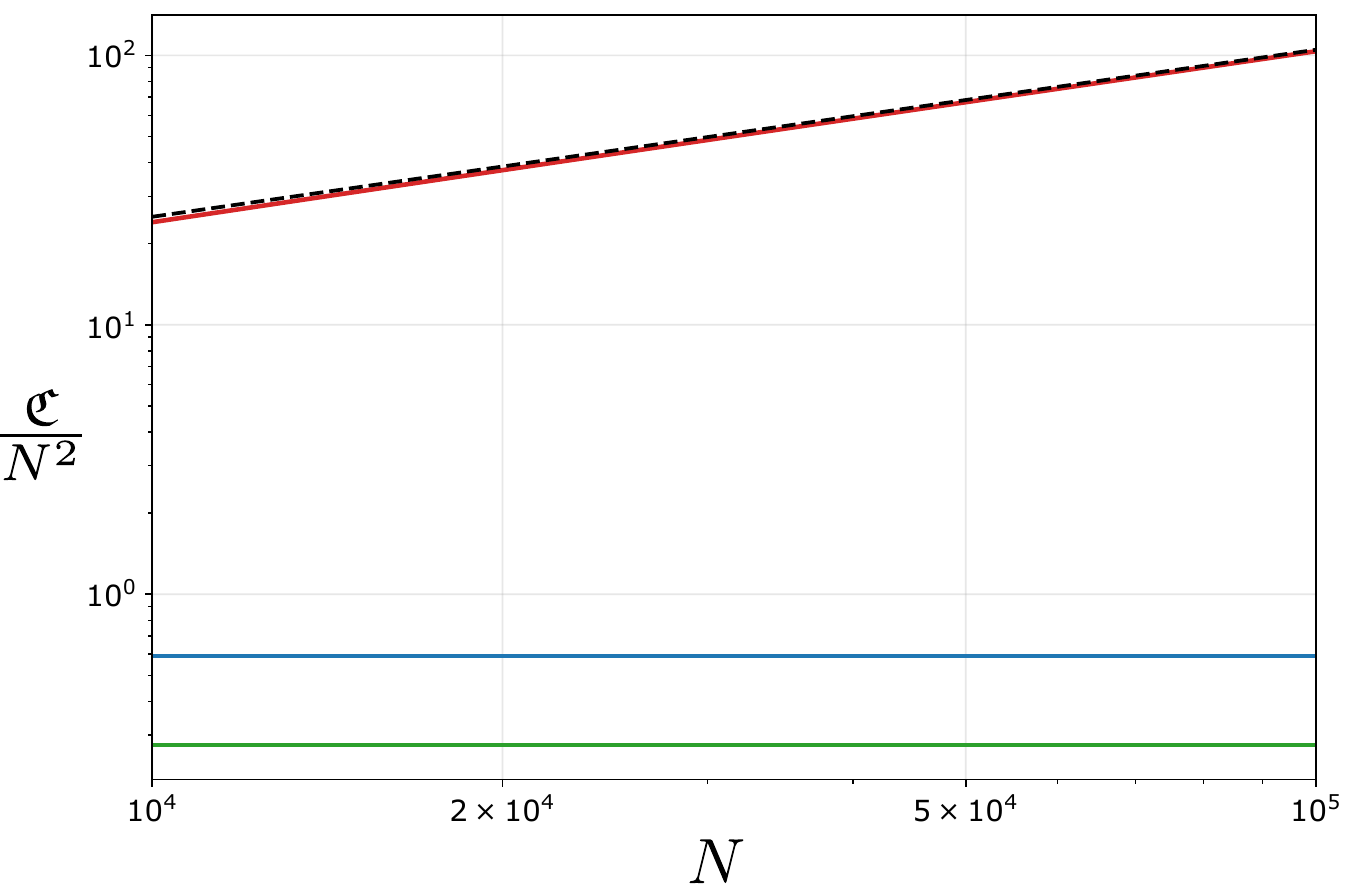}
\caption[${\coh}/{N^2}$ as a function of the number of atoms $N$]{${\coh}/{N^2}$ as a function of the number of atoms $N$. The red curve represents the solution for the pumping rate trajectory given by $\beta=3$, $q=0.62$ in Eq.~(\ref{eq:path}) (the same data as in Fig.~\ref{fig:asyma} are used). The respective asymptotic expression of Eq.~(\ref{eq:cohasymptotic}) is plotted in a dashed black line. The blue curve is the maximal achievable $\coh/N^2$ at steady state, obtained with a normalized pumping rate coefficient just below the critical value, $r=0.8<r_0\approx 0.804$. Numerical and analytical results are indistinguishable in the plot. The green line represents the model of the standard superradiant laser, showing the maximal achievable value of ${\coh}/{N^2}$ for the model, given in Eq.~(\ref{eq:coh/N^2}).}
\label{fig:asymb}
\end{figure}

\begin{figure}
\centering
\includegraphics[width=\linewidth]{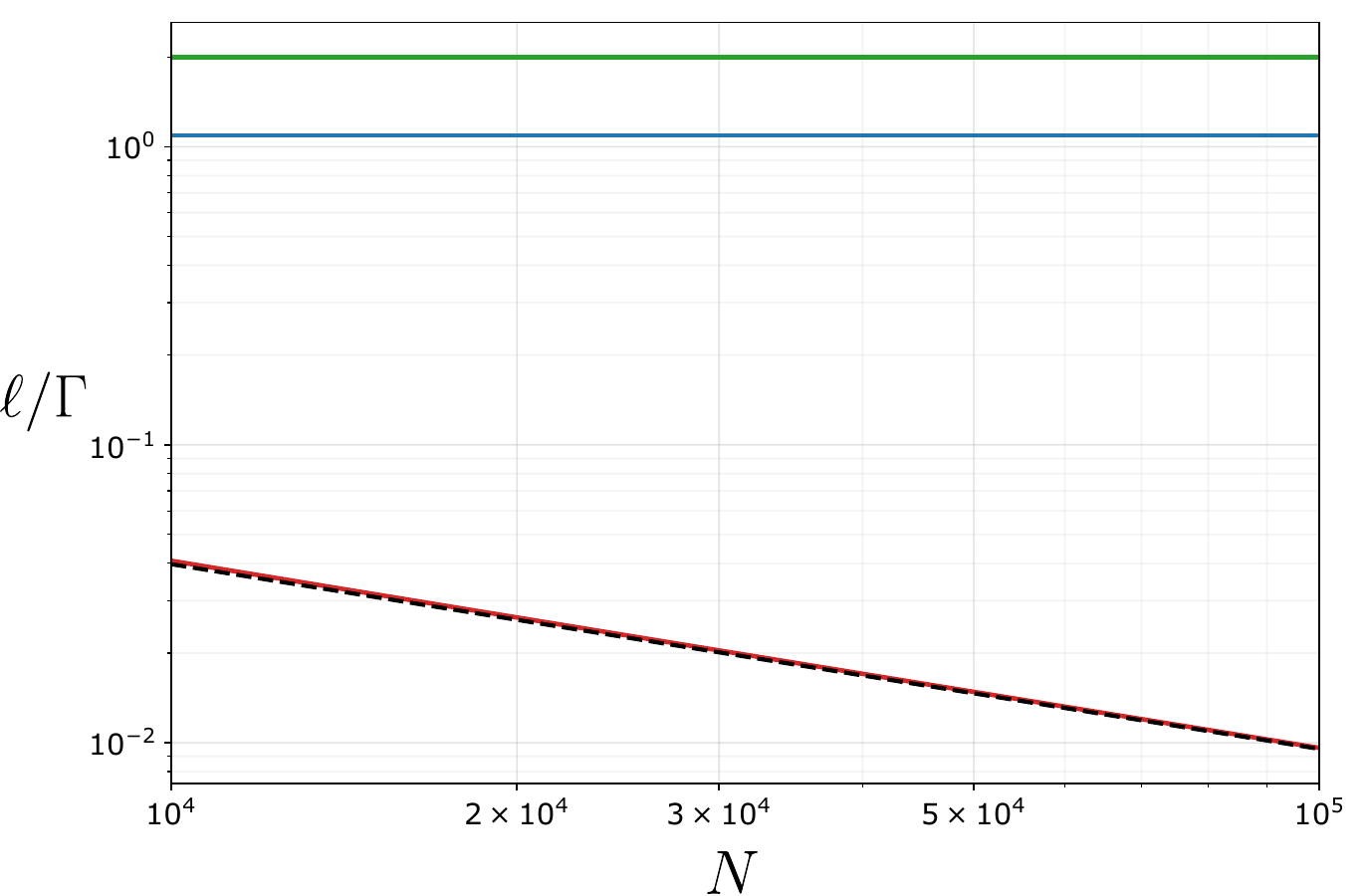}
\caption[$\ell/\Gamma$ as a function of the number of atoms $N$]{$\ell/\Gamma$ as a function of the number of atoms $N$. The red curve represents the solution for the pumping-rate trajectory given by $\beta=3$, $q=0.62$ in Eq.~(\ref{eq:path}). The respective asymptotic expression of Eq.~(\ref{eq:linewidthasymptotic}) is plotted in a dashed black line. The blue curve is the minimal achievable $\ell/\Gamma$ at steady state, obtained with a normalized pumping rate coefficient just below the critical value, $r=0.8<r_0\approx 0.804$. Numerical and analytical results are indistinguishable in the plot. The green line represents the model of the standard superradiant laser, showing the linewidth obtained at peak flux, i.e.~$\nu=N/2$ in Eq.~(\ref{eq:line}). }
\label{fig:asymc}
\end{figure}

To present the coherence as a function of the atom number $N$, rather than the mean number of excitations $\mu$, Fig.~\ref{fig:asymb} replots the data from Fig.~\ref{fig:asyma} as $\coh/N^2$ versus $N$.
Figure \ref{fig:asymc} plots $\ell/\Gamma$ as function of the number of atoms $N$, for the same pumping-rate trajectory defined by the parameters $\beta=3$ and $q=0.62$ in Eq.~(\ref{eq:path}).
The linewidth data in Fig.~\ref{fig:asymc} were obtained from the coherence calculated using the method described in Appendix~\ref{subsec:numericalmethod}, together with Eqs.~(\ref{eq:C}) and (\ref{eq:flux}).
With only $N=10^5$ atoms, our model can achieve a linewidth more than two orders of magnitude narrower than that of the standard superradiant laser model with the same transition rate $\Gamma$.

\section{Generalized FAT laser models}
\label{sec:generalFAT}
Finally, before concluding, we provide a qualitative analysis of the generalized FAT laser models defined by Eq.~(\ref{eq:generalmodel}), building on our analysis of the superradiant laser model in the preceding sections. In Sec.~\ref{sec:metastablegeneral}, we discuss the metastable states in which these models can operate, and in Sec.~\ref{subsec:generalcoherenceporperties}, we analyze the output photon statistics and coherence properties.

\begin{figure}
\centering
\includegraphics[width=\linewidth]{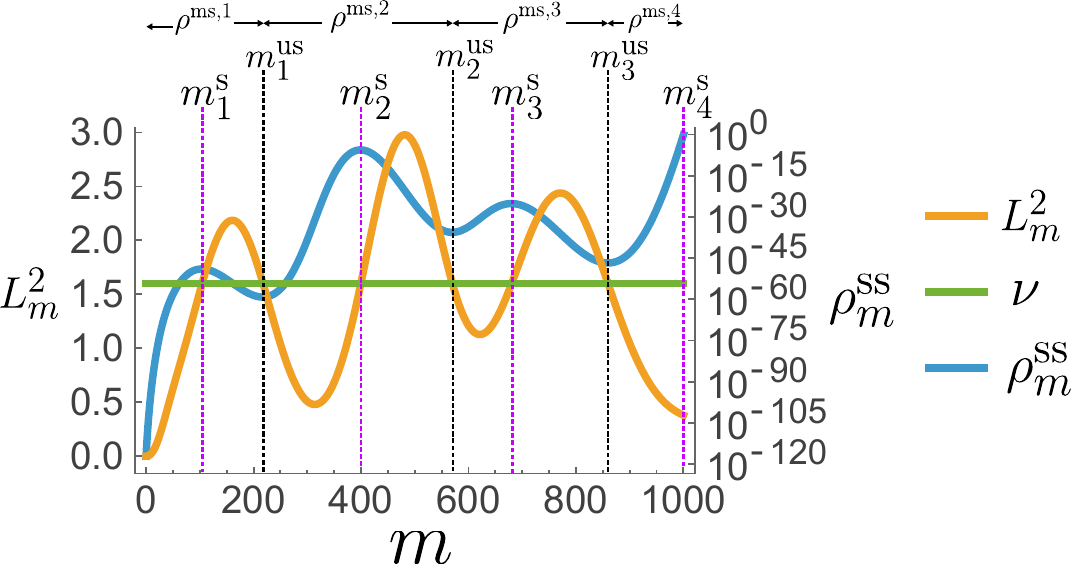}
\caption{A representative example of a generalized FAT laser model, of dimension $D=1001$, exhibiting multiple metastable states. 
The orange curve shows the squared off-diagonal coefficients $L_m^2$, and the green line shows the dimensionless pumping parameter $\nu$, both with respect to the left axis. The blue curve shows the corresponding steady state populations $\rho_m^{\mathrm{ss}}$ obtained from Eq.~(\ref{eq:ssgeneral}) for the plotted $L_m^2$ and $\nu$, on the logarithmic right axis. The intersections $L_m^2=\nu$ define alternating stable and unstable fixed points, denoted by $m_i^{\mathrm{s}}$ and $m_i^{\mathrm{us}}$, respectively. The unstable fixed points partition the excitation-number range into metastable intervals, indicated by the arrows, with corresponding metastable states $\rho^{\mathrm{ms},i}$ localized around $m_i^{\mathrm{s}}$. The metastable state $\rho^{\mathrm{ms},4}$ does not support laser operation.}
    \label{fig:generalmetastable}
\end{figure}
\subsection{Metastable states}
\label{sec:metastablegeneral}
The population dynamics of the quantum master equation~(\ref{eq:generalmodel}) is analogous to that of the superradiant model, Eq.~(\ref{eq:population}). It is given by the classical master equation of a continuous-time birth--death Markov process \cite{Markovchain,norris_1997}
\begin{equation}
\label{eq:populationL}
    \dot{\rho}_m
    =
    \tilde{q}_{m-1,m}\rho_{m-1}
    +
    \tilde{q}_{m+1,m}\rho_{m+1}
    -
    \left(
    \tilde{q}_{m,m+1}
    +
    \tilde{q}_{m,m-1}
    \right)\rho_m .
\end{equation}
We remind the reader that the populations $\rho_m$ are defined for $0\leq m \leq D-1$.
The birth and death rates, $\tilde{q}_{m,m+1}$ and $\tilde{q}_{m,m-1}$, respectively, are given by
\begin{equation}
\label{eq:qqqL}
    \tilde{q}_{m,m+1}
    =
    \begin{cases}
        \nu\Gamma, & m<D-1,\\
        0, & m=D-1,
    \end{cases}
    \qquad
    \tilde{q}_{m,m-1}
    =
    \Gamma L_m^2 .
\end{equation}
The steady-state solution of Eq.~(\ref{eq:populationL}) is given by the recursion relation for a continuous-time birth–death Markov process \cite{Markovchain}
\begin{equation}
\label{eq:ssgeneral}
    \rho_m^{\mathrm{ss}}
    =
    \frac{\nu}{L_m^2}\rho_{m-1}^{\mathrm{ss}},
\end{equation}
for $1 \leq m \leq D-1$. 
Thus, the population dynamics and the steady state differ from those of the superradiant model only through the replacement $J_m^2\mapsto L_m^2$. Accordingly, Eqs.~(\ref{eq:qqq}) and (\ref{eq:ss})  are mapped onto Eqs.~(\ref{eq:qqqL}), and (\ref{eq:ssgeneral}), respectively.

We next analyze the semiclassical dynamics, following Sec.~\ref{sec:metastable}, where stable and unstable fixed points were defined. Unlike the superradiant model, the general dynamics generated by the rates of Eq.~(\ref{eq:qqqL}) may exhibit multiple stable and unstable fixed points. We remind the reader that $L_m^2$ varies slowly with $m$. Since the Hilbert space contains no basis states with $m<0$, we must have $L_0^2=0$.
To illustrate the general case for the dynamics of Eq.~(\ref{eq:populationL}), we consider a representative example that is shown in Fig.~\ref{fig:generalmetastable}. The figure plots the coefficients $L_m^2$, the pumping parameter $\nu$, and the corresponding steady-state populations $\rho_m^{\mathrm{ss}}$ given by Eq.~(\ref{eq:ssgeneral}), as functions of $m$.
The interior fixed points are determined by the intersections $L_m^2=\nu$. We denote the stable and unstable fixed points by $m_i^{\mathrm{s}}$ and $m_i^{\mathrm{us}}$, respectively. Crossings with positive slope are stable, whereas crossings with negative slope are unstable. We identify the upper boundary, $m=D-1$, as a stable fixed point when $L_{D-1}^{2}<\nu$, since the transition rate into it then exceeds that out of it.

Since $L_0^2=0<\nu$, the first fixed point must lie in the interior and be stable. The fixed points then alternate between stable and unstable as $m$ increases,
\begin{equation}
0<m_1^{\mathrm{s}}
<
m_1^{\mathrm{us}}
<
m_2^{\mathrm{s}}
<
m_2^{\mathrm{us}}
<
\ldots< m_{i_{\text{max}}}^\text{s}.
\end{equation}
The first and the last fixed points are always stable.

Each stable fixed point $m_i^{\mathrm{s}}$ is associated with a metastable interval $\mathcal{I}_i$, defined as
\begin{equation}
m_i^{\text{s}} \in\mathcal{I}_i
= \begin{cases}
 [
0,
m_1^{\mathrm{us}}
], & i=1, \\ 
[
m_{i-1}^{\mathrm{us}},
m_i^{\mathrm{us}}], & 1<i<i_{\text{max}},\\
[m_{i_{\text{max}}-1}^{\text{us}},D-1], &i=i_{\text{max}}.
\end{cases}
\end{equation}
The population distribution of the $i$th metastable state is obtained by restricting the steady-state population in Eq.~(\ref{eq:ssgeneral}) to the interval $\mathcal{I}_i$:
\begin{equation}
\label{eq:metastablegeneral}
\rho_{m}^{\mathrm{ms},i}
\propto
\begin{cases}
\rho_m^{\mathrm{ss}}, & m\in\mathcal{I}_i,\\
0, & m\notin\mathcal{I}_i.
\end{cases}
\end{equation}
For each metastable state, $\rho^{\mathrm{ms},i}$, we define the mean first-passage times (see Sec.~\ref{sec:metastable})  for transition to the adjacent metastable states on its left and right as $T_{\mathrm{ms},i}^{\leftarrow}$ and $T_{\mathrm{ms},i}^{\rightarrow}$, respectively. The metastable time $T_{\mathrm{ms},i}$ for $\rho^{\text{ms},i}$ is therefore determined by
\begin{equation}
\frac{1}{T_{\mathrm{ms},i}}
=
\frac{1}{T_{\mathrm{ms},i}^{\leftarrow}}
+
\frac{1}{T_{\mathrm{ms},i}^{\rightarrow}}.
\end{equation}
If no adjacent metastable state exists in either direction, the corresponding transition time is taken to be infinite. The corresponding metastable transition times can be calculated based on the method used for the superradiant laser model, detailed in Appendix~\ref{sec:thesisderivationmetastable}.
If $L_{D-1}^{2}<\nu$, the metastable state  $\rho_{m}^{\mathrm{ms},i_{\text{max}}}$ does not support operation of a laser since its population variance is comparable to unity (see Sec.~\ref{sec:metastable}).
In the representative example shown in Fig.~\ref{fig:generalmetastable}, this state corresponds to $\rho^{\mathrm{ms},4}$.

\subsection{Output photon statistics and coherence properties}
\label{subsec:generalcoherenceporperties}
Here, we assume the generalized FAT laser model is initialized in one of the metastable states, $\rho^{\mathrm{ms},i}$, that supports laser operation. Additionally, we assume that $T_{\mathrm{ms},i}$ is enormously large and, as in Sec.~\ref{sec:metastable}, refer to this as the metastable regime. We show that, in this regime, the laser model exhibits ideal Poissonian output photon statistics, and we find the linewidth and the coherence of the laser.

The Poissonian output photon statistics can be demonstrated as in the superradiant case in Sec.~\ref{sec:poissonian}; here, we follow the second method presented in Sec.~\ref{sec:poissonian}.
The metastable state in Eq.~(\ref{eq:metastablegeneral}) can be expressed as a uniform incoherent mixture of approximate eigenstates of the lowering operator $\hat{L}$ over all phases
\begin{equation}
\label{eq:generalmodelsuperposition}
\rho^{\mathrm{ms},i}
=
\int_{0}^{2\pi}
\frac{\mathrm{d}\phi}{2\pi}
\ket{\rho^{\mathrm{ms},i},\sqrt{\nu}e^{i\phi}}_{\hat{L}}
\bra{\rho^{\mathrm{ms},i},\sqrt{\nu}e^{i\phi}}.
\end{equation}
Here, we define the approximate eigenstates of $\hat{L}$ according to
\begin{equation}
\label{eq:generaleig12}
\ket{\rho^{\mathrm{ms},i},\sqrt{\nu}e^{i\phi}}_{\hat{L}}
:=
\sum_{m\in\mathcal{I}_i}
\mathrm{e}^{im\phi}
\sqrt{\rho_m^{\mathrm{ms},i}}
\ket{m}.
\end{equation}
These states satisfy, 
\begin{multline}
\label{eq:generaleigenstatesL}
\frac{1}{\sqrt{\nu}e^{i\phi}}\hat{L}\ket{\rho^{\mathrm{ms},i},\sqrt{\nu}e^{i\phi}}_{\hat{L}}= \ket{\rho^{\mathrm{ms},i},\sqrt{\nu}e^{i\phi}}_{\hat{L}} \\+\frac{L_{m_{i-1}^{\text{us}}}}{\sqrt{\nu}}\sqrt{\rho_{m_{i-1}^{\text{us}}}^{\mathrm{ms},i}}\mathrm{e}^{i(m_{i-1}^{\text{us}}-1)\phi}\ket{m_{i-1}^{\text{us}}-1}-\mathrm{e}^{im_{i  }^{\text{us}}\phi}\sqrt{\rho_{m_{i}^{\text{us}}}^{\mathrm{ms},i}}\ket{m_{i}^{\text{us}}} \\
\approx\ket{\rho^{\mathrm{ms},i},\sqrt{\nu}e^{i\phi}}_{\hat{L}}.
\end{multline}
Equation~(\ref{eq:generaleigenstatesL}) applies directly to the interior metastable states. For \(i=1\), the first boundary term is absent because \(L_0=0\). For \(i=i_{\max}\), we set \(m_{i_{\max}}^{\mathrm{us}}\equiv D-1\) in the second boundary term for notational convenience, although this boundary is not an unstable fixed point. In the metastable regime, the two terms neglected in the final approximation in Eq.~(\ref{eq:generaleigenstatesL}) are expected to be negligibly small, as in Sec.~\ref{sec:metastable}, so that the states of Eq.~(\ref{eq:generaleig12}) become approximate eigenstates of $\hat{L}$.
Applying the same arguments of Sec.~\ref{sec:poissonian} to Eqs.~(\ref{eq:generalmodelsuperposition}) and (\ref{eq:generaleigenstatesL}) proves that the output photon statistics of the laser are Poissonian.
The flux of the laser operating in the metastable state is given by
\begin{equation}
\label{eq:generalflux}
    \mathcal{N}=\Gamma \expectation{\hat L^\dagger \hat L}=\Gamma\sum_{m\in\mathcal{I}_i} \rho_m^{\text{ms},i} L_m^2 \approx \Gamma \int_{m\in\mathcal{I}_i} \rho_m^{\text{ms},i} L_m^2 \mathrm{d}m.
\end{equation}
Here, in the last step, we approximated the sum by an integral. The result is intuitively expressed as the population $\rho_m^{\mathrm{ms},i}$ gives the relative time the system spends at excitation level $m$, while $\Gamma L_m^2$ is the flux from the level $m$.

In Appendix~\ref{subsecthesis:derivationoflinewidth}, we derive the linewidth for the superradiant laser of Eq.~(\ref{eq:model}). The same derivation applies to the generalized FAT laser of Eq.~(\ref{eq:generalmodel}), initialized in a metastable state $\rho^{\mathrm{ms},i}$ that supports laser operation. Under the same smoothness and time-scale assumptions, we replace the
metastable interval $[-N/2,m_+]$ with $I_i$ and make the substitutions
$\hat{J}\mapsto\hat{L}$, $\rho^{\mathrm{ms}}\mapsto\rho^{\mathrm{ms},i}$, and
\begin{equation}
\ket{N/2,\sqrt{\nu}e^{i\phi}}_{\hat{J}}
\mapsto
\ket{\rho^{\mathrm{ms},i},\sqrt{\nu}e^{i\phi}}_{\hat{L}}.
\end{equation}
The required phase-mixture representation and approximate eigenstate relation are given by Eqs.~(\ref{eq:generalmodelsuperposition}) and (\ref{eq:generaleigenstatesL}), respectively, while Eqs.~(\ref{eq:ssgeneral}) and (\ref{eq:metastablegeneral}) imply $\nu\rho_{m-1}^{\mathrm{ms},i}=L_m^2\rho_m^{\mathrm{ms},i}$ in the interior of $I_i$. Neglecting boundary contributions at both ends of $I_i$, the derivation yields
\begin{equation}
\label{eq:generallinewidth}
    \ell= 2\Gamma\int_{m\in\mathcal{I}_i} \rho_m^{\text{ms},i}\left(\pp{L_m}{m}\right)^2 \text{d}m.
\end{equation}
The gain and loss processes contribute equally to the linewidth.

Equation~(\ref{eq:generallinewidth}) admits a simple interpretation. The population \(\rho_m^{\mathrm{ms},i}\) determines the fraction of time that the system spends near excitation level \(m\), whereas \(\left(\partial L_m/\partial m\right)^2\) quantifies the local sensitivity of the phase gain near that level. This can be understood by considering the isometric gain superoperator (see Secs.~\ref{thesissec:HLmodel} and \ref{subsec:HL}), with flat off-diagonal coefficients, which uniformly shifts both populations and coherences without degrading the latter and therefore generates no phase noise.  When the coupling is locally flat, $\partial L_m/\partial m=0$, the corresponding gain process acts locally in the same manner as the isometric gain superoperator. More generally, Eq.~(\ref{eq:generallinewidth}) shows that the local sensitivity to the phase noise generated by each of the gain and loss processes increases with $\lvert\partial L_m/\partial m\rvert$, confirming the intuition presented in Sec.~\ref{subsec:ourmodel}.

To find the coherence, we assume the laser operates in the metastable regime and we use the approach of Sec.~\ref{subsec:coherencederivation}. Using Eqs.~(\ref{eq:C}), (\ref{eq:generalflux}), and (\ref{eq:generallinewidth}), we obtain
\begin{equation}
  \label{eq:generalC}
   \coh= \frac{2\int_{m\in\mathcal{I}_i}\rho_{m}^{\text{ms},i}L_{m}^{2}\mathrm{d}m}{\int_{m\in\mathcal{I}_i}\rho_{m}^{\text{ms},i}\left(\pp{L_{m}}{m}\right)^{2}\text{d}m}.
\end{equation}
This result explains why the Heisenberg-limited laser models of Eqs.~(\ref{thesiseq:HLmodel}) and (\ref{eq:HL}) have a large coherence for a given mean excitation number $\mu$. As seen in Fig.~\ref{fig:HLpop}, the population of the HL models becomes very small near the edges, precisely where $(\partial L_m/\partial m)^2$ becomes significantly large.
The generalized FAT laser models should obey the Heisenberg limit for laser coherence because they satisfy the four conditions of the Heisenberg limit theorem (see Secs.~\ref{thesissec:HLtheorem} and \ref{subsec:coherencederivation}).

\section{Conclusions}
\label{sec:conclusions}
To conclude, we provide a summary in Sec.~\ref{subsec:summary}, discuss the results in Sec.~\ref{subsec:discussion}, and outline future work in Sec.~\ref{sec:futurework}.
\subsection{Summary}
\label{subsec:summary}
In this paper, we proposed the first laser model operating in the optical regime that can surpass the SQL for laser coherence in terms of scaling. The model is based on Dicke superradiance and is a generalization of the far above threshold standard laser model. That is, its master equation (\ref{eq:model}) is obtained by the replacement $\hat{a}\mapsto\hat{J}$ in the superoperators of the standard FAT master equation (\ref{eq:FAT}).

Equivalently, the master equation of our model is obtained from that of the standard superradiant laser model by replacing its gain process, which consists of noncollective pumping, with a collective pumping process characterized by a `flat' gain rate. This `flat' gain rate mimics the effect of gain saturation that occurs in the FAT laser as a consequence of its nonlinear dynamics. To realize this collective gain process, we propose cavity-enhanced Raman emission into an auxiliary, heavily damped ``gain'' mode that couples identically to all atoms. By monitoring the output of this mode and using the resulting signal to control the driving laser via feedback, one can obtain the desired `flat' gain rate.

We showed that the `flat' gain in our model gives rise to ideal Poissonian output photon statistics. Furthermore, we showed that the optimal laser operation occurs in a metastable regime in which the metastable lifetime is extraordinarily long and far exceeds the coherence time. Within this regime, the coherence can approach the limiting scaling $\coh\propto\mu^{8/3}$, thereby surpassing the SQL for laser coherence. Correspondingly, the linewidth can approach the limiting scaling $\ell\propto\Gamma N^{-2/3}$, where $\Gamma$ is the Purcell-enhanced emission rate into the cavity.

Our analysis shows that generalized FAT laser models exhibit characteristics similar to those of our superradiant laser model. The central results are general expressions for their linewidth and coherence; in particular, the linewidth is governed by the population-weighted average local sensitivity proportional to $\left(\partial L_m/\partial m\right)^2$, with the gain and loss processes contributing equally to the phase noise.

\subsection{Discussion}
\label{subsec:discussion}
As stated above, our laser model is the first proposed optical platform with coherence scaling beyond the SQL. Previous proposals for in-principle surpassing of the SQL for laser coherence have been restricted to the microwave domain and rely on superconducting circuit-QED systems \cite{HL,circutQED}. By contrast, high-coherence optical lasers have a substantially broader range of applications \cite{RevModPhys.78.1279,RevModPhys.87.637,nextligo}. To our knowledge, our model is also the first laser proposal in which pumping is implemented using quantum feedback control at the level of individual excitations, such that excitations are added one at a time.

Compared with the standard superradiant laser model, our model offers three advantages: its linewidth is not bounded from below by the Purcell-enhanced emission rate into the cavity, $\Gamma$; its output exhibits ideal Poissonian photon statistics; and its collective pumping process avoids atomic heating. The latter point is the main reason why the standard superradiant laser model has not been realized experimentally yet \cite{rugged,superradiantlasetexpirement2,Hollandfullycollective,PhysRevXcrossover}.

\subsection{Future work} 
\label{sec:futurework}
Here, we discuss possible directions for future work.
First, the proposed physical realization could be implemented in a waveguide rather than in cavities. This would enhance the desired atom--field interaction relative to undesired spontaneous emission into free space modes.
A waveguide implementation can be obtained using three-level atoms in a ladder configuration, with approximately equally spaced energy levels, so that the $1\leftrightarrow 2$ and $2\leftrightarrow 3$ transition frequencies are approximately equal (these can slightly differ because of the large detuning $\Delta$ in our model). These $1\leftrightarrow 2$ and $2\leftrightarrow 3$ transitions should be coupled to different polarization modes of the waveguide. The atoms should be fixed in place along the waveguide with adjacent atoms separated by one resonant wavelength. Consequently, the gain and loss modes couple identically to the collective atomic degrees of freedom.

Second, the Poissonian pumping protocol adopted here was chosen for analytical tractability rather than optimal performance. Regular pumping, as well as more sophisticated feedback protocols that use the measured durations of the ``on'' periods as inputs to optimal-control methods \cite{optimalcontrol1,optimalcontrolfiltering}, could extend the metastable time. This could enable operation closer to $m=0$, thereby increasing the coherence, narrowing the linewidth, and possibly improving the scaling of both quantities. More generally, quantum control could engineer laser beams with prescribed, possibly time-dependent, output properties.

Finally, exploring new physical realizations of generalized FAT laser models is a direction for future work. Two approaches to obtaining the `flat' gain rate required by these models have already been identified: gain saturation in the standard FAT laser model and quantum feedback control in our superradiant laser model. Further realizations may be possible on various optical and microwave platforms and may offer a promising route to laser coherence beyond the SQL.

\begin{acknowledgments}
The authors acknowledge the support of the Griffith University eResearch Service and Specialised Platforms Team. This research was supported by the Australian Government through the Australian Research Council’s Discovery Projects funding scheme (Project DP220101602) and by an Australian Government Research Training Program (RTP) Scholarship. 
\end{acknowledgments}

\appendix

\section{Physical realization- details of derivations}
\label{thesissec:derivations}
Here, we provide details of several derivations in Sec.~\ref{sec:physicalrealzation}.
\subsection{Adiabatic elimination of the third level}
\label{subapp:adiabatic}

Here, we consider the dynamics of Eq.~(\ref{eq:a1}) and adiabatically eliminate the third auxiliary level from the Hamiltonian of Eq.~(\ref{eq:a4}), obtaining the Hamiltonian of Eq.~(\ref{eq:HH}). While the driving laser is switched off \([\Omega(t)=0]\), the interaction mediated by the third level is absent, and the adiabatic elimination therefore becomes trivial.
In the remainder of this subsection, we restrict the calculation to times within an ``on'' period, for which \(\Omega(t)=\Omega_0\). Specifically, we consider the \(i\)th ``on'' period, so that \(t_i \le t \le t_i+\Delta t_i\).
We calculate the evolution of the system from time \(t\) to \(t+\delta t\), where \(\delta t\) is a small time interval.
We first specify the time step \(\delta t\) used to evolve the system in Eq.~(\ref{eq:a1}). That is, to ensure that we can perform independently the adiabatic elimination of the third level restricting to the first three terms in the Hamiltonian of Eq.~(\ref{eq:a4}).  We require
\begin{equation}
\label{eq:deltat}
\delta t \ll \kappa_b^{-1},(\Gamma N^2)^{-1},(\Omega_0^2N/\Delta )^{-1}.
\end{equation}
When taking the conditions of Eq.~(\ref{eq:deltat}) into account, the other terms in the dynamics of Eq.~(\ref{eq:a1}) that do not involve the third level [these include the last two terms in the Hamiltonian of Eq.~(\ref{eq:a4}) and the dissipative dynamics of Eq.~(\ref{eq:L_c})] will be added up to the resulting master equation after the adiabatic elimination of the third level (see below).

We move to a rotating frame \begin{equation}
\label{eq:b5}
{W}_{\text{rot}}(t):=e^{i\Delta \hat{J}_{33}t}W(t) e^{-i\Delta \hat{J}_{33}t}.
\end{equation}
The state ${W}_{\text{rot}}$ evolves unitarily with the Hamiltonian
\begin{equation}
\label{eq:n10}
\hat{H}_{\text{rot}}:=i\Omega_0\left(\hat{J}_{13}e^{-i\Delta t}-\text{h.c.}\right)+g_b\left(\hat{J}_{23}e^{-i\Delta t}\hat{b}^{\dagger}+\text{h.c.}\right),
\end{equation}
We perform a Magnus expansion \cite{Magnus1} for the state ${{W}_{\text{rot}}(t)}$ up to second order
\begin{equation}
\label{eq:b10}
{W}_{\text{rot}}(t+\delta t)\approx \hat{U}_{\text{rot}}(t,\delta t) {{W}_{\text{rot}}(t)}\hat{U}_{\text{rot}}^\dagger(t,\delta t),\end{equation}
with the unitary operator
\begin{equation}
   \hat {U}_{\text{rot}}(t,\delta t)  :=\exp[\hat{\Sigma}_{\text{rot}}^{1}(t,\delta t)+\hat{\Sigma}_{\text{rot}}^{2}(t,\delta t)],
\end{equation}
where we defined
\begin{equation}
\label{eq:b12}
  \hat \Sigma_{\text{rot}}^{1}(t,\delta t):=-i\int_{t}^{t+\delta t}\hat{H}_{\text{rot}}\left(t_{1}\right)\text{d}t_{1},
\end{equation}
\begin{equation}
\label{eq:b13}
 \hat  \Sigma_{\text{rot}}^{2}(t,\delta t):= -\frac{1}{2}\int_{t}^{t+\delta t}\text{d}t_{1}\int_{t}^{t_{1}}\text{d}t_{2}\left[\hat{H}_{\text{rot}}\left(t_{1}\right),\hat{H}_{\text{rot}}\left(t_{2}\right)\right].
\end{equation}
The Magnus expansion up to the second order, Eq.~(\ref{eq:b10}), preserves unitarity \cite{Magnus1}.
Using Eq.~(\ref{eq:b10}) and moving back from the rotating frame to the initial frame by using Eq.~(\ref{eq:b5}), we obtain
\begin{equation}
    W(t+\delta t)=\hat U(t,\delta t)W(t) \hat U^\dagger(t,\delta t),
    \end{equation}
\begin{equation}
\label{eq:gpm}
  \hat  U(t,\delta t):=e^{-i\Delta \hat{J}_{33}\delta t}\exp[\hat{\Sigma}_{1}(t,\delta t)+\hat{\Sigma}_{2}(t,\delta t)],
    \end{equation}
where we defined
\begin{equation}
\label{eq:b15}
    \hat{\Sigma}_i(t,\delta t):=e^{-i\Delta \hat{J}_{33}t}{\Sigma}_{\text{rot}}^{i}(t,\delta t)e^{i\Delta \hat{J}_{33} t},
\end{equation}
for $i=1,2$.

The master equation (\ref{eq:a1}), and in particular the Hamiltonian of Eq.~(\ref{eq:n10}), is composed only of atomic operators which are collective, $\hat{J}_{ij}$. These satisfy the commutation relations for the U(3) group, $[\hat{J}_{i j}, \hat{J}_{k l}]=\delta_{j k} \hat{J}_{i l}-\delta_{i l} \hat{J}_{k j}$ \cite{U(3)}.
Consequently, the Hamiltonian of the system is symmetric to permutations of the atoms, and so $[\Pi^\alpha_{\text{sym}},\hat H_{\text{rot}}]=0$, where $\Pi^{\alpha}_{\text{sym}}$ is the permutational symmetry operator \cite{SU(4)FrancoNori}. Thus, when all atoms start in the ground state, the state of the atoms at any subsequent time belongs to the Hilbert subspace spanned by the fully symmetric states with respect to permutations of the atoms, denoted by $\ket{n_1,n_2,n_3}_{\text{s}}$ \cite{collectivepumpingprl,collectivepumpingpra}, where $n_i$ represents the number of atoms in level $i$ and $n_1+n_2+n_3=N$. In this notation, the initial state of the atoms is given by $\ket{N,0,0}_{\text{s}}$. 
The $N+1$ Dicke states, which are the basis of the Hilbert space for our system, are denoted in this notation as $\ket{N/2,m}=\ket{N/2-m,N/2+m,0}_{\text{s}}$.

To perform the adiabatic elimination, we have to set conditions on the physical parameters in Eq.~(\ref{eq:n10}). To do that, we estimate the operator norm for the operators $\hat{J}_{ij}$ in the appropriate Hilbert space where the third level is depleted. The operator norm is defined as $\|\hat{A}\|_{\mathrm{op}}=\sup _{\langle{\psi}\vert{\psi}\rangle=1} \sqrt{\bra{\psi}\hat{A}^\dagger\hat{A} \ket{\psi}}$.
This estimation will be done below. We introduce six atomic annihilation and creation operators, $\hat{z}_i$ and $\hat{z}_i^\dagger$ $(i=1,2,3)$, which lower and raise the occupation numbers $n_i$ as $\hat{z}_1 \ket{n_1,n_2,n_3}=\sqrt{n_1}\ket{n_1-1,n_2,n_3}$ and $\hat{z}_1^\dagger \ket{n_1,n_2,n_3}=\sqrt{n_1+1}\ket{n_1+1,n_2,n_3}$, etc., and satisfy bosonic commutation relations, i.e.~$[\hat{z}_i,\hat{z}^\dagger_j]=\delta_{i,j}$. Moreover, we are led to Schwinger’s representation of the Lie algebra \cite{schwinger1952angular} of the group U(3) by expressing our nine atomic polarization and population operators as $\hat{J}_{ij}=\hat{z}_i^\dagger\hat{z}_j$ \cite{rep}, i.e.,~$\hat{J}_{12}\ket{n_1,n_2,n_3}=\sqrt{(n_1+1)n_2}\ket{n_1+1,n_2-1,n_3}$, etc. 
Using the above expressions, and since the third level is depleted ($n_3=0$), we obtain,
$\|\hat{J}_{13}\|_{\mathrm{op}}=\|\hat{J}_{23}\|_{\mathrm{op}}=\sqrt{N}$, $\|\hat{J}_{12}\|_{\mathrm{op}}\approx N/2$, and $\|\hat{J}_{11}\|_{\mathrm{op}}=\|\hat{J}_{22}\|_{\mathrm{op}}=N$.

Using Eq.~(\ref{eq:b12}), we calculate the first order coefficient in Eq.~(\ref{eq:b15})
\begin{multline}
\label{eq:b17}
  \hat{\Sigma}_{1}(t,\delta t)=\frac{2}{\Delta}\sin\left(\frac{\Delta\delta t}{2}\right) \\ \times\left\{ \Omega_{0}\left(\hat{J}_{13}e^{-i\Delta\delta t/2}-\text{h.c.}\right)-ig_{b}\left(\hat{J}_{23}e^{-i\Delta\delta t/2}\hat{b}^{\dagger}+\text{h.c.}\right)\right\}.
\end{multline}
We can notice that the expression in equation (\ref{eq:b17}) is oscillating, and it is vanishing i.e.~$\hat{\Sigma}_{1}(t,\delta t)=0$, for $\delta t=(\Delta/2\pi)^{-1}n$, when $n$ is an integer. Thus, we can choose $\delta t=(\Delta/2\pi)^{-1}$ and evolve the system in small discrete time steps of $\delta t$. Substituting this into conditions (\ref{eq:deltat}), and requiring the amplitude of oscillation of the expression in Eq.~(\ref{eq:b17}) to be much smaller than one, we get a new set of conditions 
\begin{equation}
    \kappa_b,\Gamma N^2,\Omega_0\sqrt{N},g_b\sqrt{N} \ll \Delta .
\end{equation}  
To derive the above conditions, we replaced the operators in Eq.~(\ref{eq:b17}) by their operator norm, which was calculated above. 
Using Eq.~(\ref{eq:b13}), we calculate the second order coefficient in Eq.~(\ref{eq:b15})
\begin{multline}
\label{eq:b18}
  \hat{\Sigma}_{2}(t,\delta t)=\left(-i\delta t+ \frac{i}{\Delta}\sin\left(\Delta\delta t\right) \right) \\
 \times \bigg\{-\frac{\Omega_{0}^{2}}{\Delta}\hat{J}_{11}-i\frac{\Omega_{0}g_{b}}{\Delta}\hat{J}_{12}\hat{b}+i\frac{\Omega_{0}g_{b}}{\Delta}\hat{J}_{21}\hat{b}^{\dagger} \\ 
 + \left[\frac{g_{b}^{2}}{\Delta}\hat{J}_{32}\hat{J}_{23} 
  +\frac{\Omega_{0}^{2}}{\Delta}\hat{J}_{33}+\frac{g_{b}^{2}}{\Delta}\left(\hat{J}_{33}-\hat{J}_{22}\right)b^{\dagger}\hat{b}\right]\bigg\}.
\end{multline}
Since we evolve the system in time steps of $\delta t=(\Delta/2\pi)^{-1}$, the sinusoidal term, which is the second term in the round brackets, vanishes.  The last term in the expression in the square brackets  will be shown to contribute only negligibly to the resulting equations (\ref{eq:psi0}) and (\ref{eq:psi1}) in Appendix~\ref{subapp:modeb} below by assuming 
\begin{equation}
    g_b^2 N \ll \kappa_b \Delta.
\end{equation}

The adiabatic elimination is performed by projecting the unitary dynamics of Eq.~(\ref{eq:gpm}) into the Hilbert space of the two atomic levels 1 and 2, without changing significantly the dynamics. Formally, we define $\hat\Pi_3$ as the projection operator into the third level, and define the orthogonal projection operator as $\bar{\Pi}_3=1-\hat\Pi_3$. Thus, using Eqs.~(\ref{eq:gpm}), (\ref{eq:b17}), and (\ref{eq:b18}), in conjunction with the accompanying explanations provided within the text, i.e., the particular expressions we neglected in the relevant equations, the unitary evolution operator from time $t$ to time $t+\delta t$ after the adiabatic elimination is given by 
\begin{multline}
\label{eq:ldfjgjdflgj}
 \hat {U}_{\text{a}}(t,\delta t)\approx\bar{\Pi}_3 \hat U(t,\delta t) \bar{\Pi}_3 = \\ \bar{\Pi}_3 - i \delta t\left\{ i\alpha(t)(\hat{J}^\dagger\hat{b}^{\dagger}-\text{h.c.}) -
\frac{\Omega^2(t)}{\Delta}\hat{J}_{11} \right\}
\end{multline} 
Here, we remind the reader of the definitions $\alpha(t):=\Omega(t)g_b/\Delta$, and $\hat{J}:=\hat{J}_{12}$.

Now, as promised above, we need to include the interaction terms in the dynamics of the master equation (\ref{eq:a1}) that had been neglected before performing the adiabatic elimination by using the conditions of Eq.~(\ref{eq:deltat}).
Thus, the final master equation is given by
\begin{equation}
\label{eq:rsdfsdfsdf}
    \dot{W}=-i[\hat{H}_{\text{eff}},W]+{\cal L}W.
\end{equation}
Here, $W$ is the state matrix for the $N$ two-level atoms and the two cavity modes. The effective Hamiltonian is given by
\begin{multline}
\label{eq:ungdjfsldnfsdf}
\hat{H}_{\text{eff}}=i\alpha(t)(\hat{J}^\dagger\hat{b}^{\dagger}-\text{h.c.})+g_a(\hat{a}^{\dagger}\hat{J}_{12}+\text{h.c.}) \\ -
\frac{\Omega^2(t)}{\Delta}\hat{J}_{11}
+ H_{\text{Z}}(t).
\end{multline} 
Here, $H_{\text{Z}}(t)=(\Omega^2(t)/\Delta)\hat{J}_{11}$ is an externally applied Zeeman shift using time-dependent magnetic field. The two last terms in the Hamiltonian of Eq.~(\ref{eq:ungdjfsldnfsdf}) cancel, as explained in Sec.~\ref{subsec:adiabaticeliminatione'}.
The Hamiltonian of Eq.~(\ref{eq:ungdjfsldnfsdf}) is identical to that of Eq.~(\ref{eq:HH}).
The superoperator for the dissipative dynamics in Eq.~(\ref{eq:rsdfsdfsdf}) is given by
\begin{equation}
    {\cal L}W:=\kappa_a{\cal D}[\hat{a}]W+\kappa_b{\cal D}[\hat{b}]W.
\end{equation}

\subsection{Adiabatic elimination of the mode $b$}
\label{subapp:modeb}
Here, we derive the dynamics for the state of the atoms during an ``on'' period and the quantum jump after the detection of a photon in the feedback detector. Without quantum feedback control, using Eq.~(\ref{eq:rsdfsdfsdf}), the system's dynamics are governed by the master equation 
\begin{equation}
\label{}
    \dot{W}=\alpha(t)[\hat{J}^\dagger\hat{b}^{\dagger}-\text{h.c.},W]+\kappa_b \mathcal{D}[\hat{b}]W.
\end{equation}
Here, we neglected the dynamics with the mode $a$ as explained in Sec.~\ref{subsec:adiabaticeliminatione'}.

We assume that the mode $b$ is heavily damped, by the condition
\begin{equation}
\label{eq:assumptiondamped}
    \alpha_0 N \ll \kappa_b.
\end{equation}
Thus, the mode $b$ becomes slaved to the atoms (see below). In the duration of the $i$th on period, and prior to a detection of a photon in the feedback detector, the state of the system, denoted by $\ket{\tilde{\psi}(t)}$, evolves with the non-Hermitian Hamiltonian 
\begin{equation}
\label{eq:Hamiltoniannh}
\hat {H}_{\text{nh}}=i\alpha_0(\hat{J}^\dagger\hat{b}^{\dagger}-\text{h.c.}) -\frac{i}{2}\kappa_b \hat{b}^\dagger\hat{b}.
\end{equation}
We assume $\braket{\tilde{\psi}(t_i)}{\tilde{\psi}(t_i)}=1$, so that $\braket{\tilde{\psi}(t)}{\tilde{\psi}(t)}$
is then the probability that no genuine photon has reached the feedback detector during the time interval $[t_i,t]$. In the final step of the calculation, we will modify this no-count probability to include the effect of dark counts. We expand the state of the system in orders of $\epsilon=\alpha_0 N/\kappa_b\ll 1$
\begin{equation}
\label{eq:state}
    \ket{\tilde{\psi}(t)}=\ket{\tilde{\psi}_{0}(t)}\ket 0_{b}+\ket{\tilde{\psi}_{1}(t)}\ket 1_{b}+O(\epsilon^2).
\end{equation}
Here, $\ket{i}_{b}$ is the state of $i$ photons in the mode $b$, and $\ket{\tilde{\psi}_i(t)}$ is a state of the system in the Hilbert space of the atoms. The Schr\"{o}dinger equation for the conditional state, under the non-Hermitian Hamiltonian of Eq.~(\ref{eq:Hamiltoniannh}) and the expansion of Eq.~(\ref{eq:state}), is
\begin{equation}
\label{eq:psi0}
    \frac{\text{d}\ket{\tilde{\psi}_{0}}}{\text{d}t}=-\alpha_0\hat{J}\ket{\tilde{\psi}_{1}}+\kappa_bO(\epsilon^{3})=\kappa_bO(\epsilon^{2}),
\end{equation}
\begin{equation}
\label{eq:psi1}
  \frac{\text{d}\ket{\tilde{\psi}_{1}}}{\text{d}t}=\alpha_0\hat{J}^{\dagger}\ket{\tilde{\psi}_{0}}-\frac{1}{2}\kappa_b\ket{\tilde{\psi}_{1}}+\kappa_bO(\epsilon^{2})=\kappa_b O(\epsilon).
\end{equation}
Examining Eq.~(\ref{eq:psi1}), we notice that the state $\ket{\tilde{\psi}_1}$ can be slaved to $\ket{\tilde{\psi}_0}$, by the assumption of Eq.~(\ref{eq:assumptiondamped}). Thus, by setting ${\text{d}\ket{\tilde{\psi}_1}}/{\text{d}t}=0$, we obtain
\begin{equation}
\label{eq:mjn}
    \ket{\tilde{\psi}_{1}}=\frac{2\alpha_0}{\kappa_b}\hat{J}^{\dagger}\ket{\tilde{\psi}_{0}}+O(\epsilon^2)=O(\epsilon).
\end{equation}
Substituting Eq.~(\ref{eq:mjn}) into Eq.~(\ref{eq:psi0}), we obtain
\begin{equation}
\label{eq:tgy}
\frac{\text{d}\ket{\tilde{\psi}_{0}}}{\text{d}t}=-\frac{2\alpha_0^{2}}{\kappa_b}\hat{J}\hat{J}^{\dagger}\ket{\tilde{\psi}_{0}}+\kappa_b O(\epsilon^3).
\end{equation}

After tracing out mode $b$, the unnormalized atomic density operator, conditioned on the absence of any click, including dark counts, at the feedback detector is
\begin{equation}
\label{eq:condstate}
\begin{aligned}
\tilde{\rho}(t)
&:= e^{-R_{\mathrm{dark}}(t-t_i)}
\operatorname{Tr}_b\!\left[
|\tilde{\psi}(t)\rangle\langle\tilde{\psi}(t)|
\right]\\
&= e^{-R_{\mathrm{dark}}(t-t_i)}
\left[
|\tilde{\psi}_0(t)\rangle\langle\tilde{\psi}_0(t)|
+ O(\epsilon^2)
\right].
\end{aligned}
\end{equation}
The exponential factor accounts for the probability that no dark count occurs during $[t_i,t]$.
Using Eq.~(\ref{eq:tgy}) and (\ref{eq:n_s}), we obtain the conditional evolution for this unnormalized atomic density operator conditioned on the absence of any click
\begin{equation}
\label{eq:1asr3r}
    \dot{\tilde{\rho}}=-\Gamma_b({\cal A}[\hat{J}^{\dagger}]+n_\text{s}) \tilde{\rho}+\kappa_b O(\epsilon^3).
\end{equation}
Here, we remind the reader of the definition of $\Gamma_{b}:=4\alpha^2_0/\kappa_b$, as the single atom flux through the ``gain'' cavity, and $n_{\text{s}}$ is defined in Eq.~(\ref{eq:n_s}). We also remind the reader of the definition of the superoperator $\mathcal{A}[\hat{A}] \hat{B}:=\frac{1}{2}\{\hat{A}^{\dagger} \hat{A}, \hat{B}\}$. 
Equation (\ref{eq:1asr3r}) is the desired result for the conditional evolution of the atoms' state prior to any click at the detector. 

Now, immediately after a genuine photon detection at the detector, which occurs within the time interval $[t_i+\Delta t_i,t_i+\Delta t_i+\text{d}t]$, the unnormalized state of the system changes to
\begin{equation}
\label{eq:poil}
 W'(t_i+\Delta t_i+\mathrm{d}t)\text{d}t=\kappa_b{\cal J}[\hat{b}]\ket{\tilde{\psi}(t_i+\Delta t_i)}\bra{\tilde{\psi}(t_i+\Delta t_i)} \text{d}t.
\end{equation}
The trace of this density matrix is equal to the probability of a genuine detection of a photon within the time interval from $t$ to $t+\text{d}t$.
We remind the reader of the definition of the superoperator $\mathcal{J} [\hat{A}] \hat{B}:=\hat{A} \hat{B} \hat{A}^{\dagger}$.
By tracing out the mode $b$ from Eq.~(\ref{eq:poil}) and using the expansion (\ref{eq:state}), we obtain
\begin{multline}
\label{eq:ofem}
\text{Tr}_{b}\left[W'(t_i+\Delta t_i+\mathrm{d}t)\right]\text{d}t \\=\kappa_b \ket{\tilde{\psi}_1(t_i+\Delta t_i)}\bra{\tilde{\psi}_1(t_i+\Delta t_i)}\text{d}t +\kappa_b  O\left(\epsilon^3\right)\mathrm{d}t.
\end{multline}
Using Eqs.~(\ref{eq:mjn}), (\ref{eq:condstate}), and (\ref{eq:ofem}), we obtain the atomic density operator conditioned on a click in the feedback detector, where the click may originate either from a genuine photon detection or from a dark count
\begin{multline}
\label{eq:ooruhsd}
\tilde{\rho} ' (t_i+\Delta t_i+\mathrm{d}t)\text{d}t \\
=\Gamma_b({\cal J}[\hat{J}^{\dagger}]+n_\text{s})\tilde{\rho}(t_i+\Delta t_i)\mathrm{d}t+\kappa_b O\left(\epsilon^3\right)\text{d}t.
\end{multline}
Equation (\ref{eq:ooruhsd}) is the desired result for the atoms' state matrix, immediately after a click at the feedback detector which occurs within a time interval $[t_i+\Delta t_i,t_i+\Delta t_i+\text{d}t]$, after tracing out the mode $b$.

\subsection{Optimal values for the physical parameters}
\label{subapp:parmaters}
Here, we describe how to identify an optimal set of experimentally realizable values for the physical parameters to realize experimentally the master equation (\ref{eq:model}). We remind the reader that the parameters appearing in this master equation are $\Gamma=4g_a^2/\kappa_a$ and the pumping rate $\nu$, with $\nu$ assumed to be of order $N^2$.

To this end, we use an optimization procedure, described below, to determine suitable parameter values. All the conditions that need to be satisfied in order to derive the master equation are given in Eqs.~(\ref{eq:Deltacondition}) and (\ref{eq:gyk}).
We enforce all inequalities of the form $a\ll b$ in Eq.~(\ref{eq:gyk})
by requiring $a\leq\epsilon b$, with $\epsilon=10^{-2}$.
By satisfying the above conditions optimally, we mean choosing the physical parameters so that the required inequalities are satisfied as tightly as possible. That is, the relevant parameter ratios are kept close to their allowed bounds, rather than being separated by unnecessarily large margins. This yields physical parameter values that are closer to one another and therefore more likely to be experimentally realizable.
To achieve this, we use the method of quadratic programming \cite{quadraticprog}.
Denote for each of the physical parameters an associated parameter: $\Omega_0=\epsilon^{\Omega^*_{0}}$, $\Delta=\epsilon^{\Delta^*}$, etc. We transform conditions (\ref{eq:gyk}) by taking a logarithm base $\epsilon$ on each side of the inequalities, and setting $2 \pi \times \text{Hz}=1$ (arbitrary units), thereby obtaining a system of linear inequalities
\begin{equation}
\label{eq:b36}
    \kappa_b^*-\Delta^*-1\geq0,
\end{equation}
\begin{equation}
    -\Delta^* +\frac{N^*}{2}+\Omega^*_0 -1\geq 0, 
\end{equation}
\begin{equation}
    -\Delta^* +g_b^*+\frac{N^*}{2}-1\geq 0,
\end{equation}
\begin{equation}
    -\Delta^* +2g_b^*-\kappa_b^*+N^*-1\geq 0,
\end{equation}
\begin{equation}
    g_a^*-\kappa_a^*+N^*-1\geq 0,
\end{equation}
\begin{equation}
    2 \Delta^* -2 g_b^*+2 g_a^*+\kappa_b^*-\kappa_a^*-2 \Omega^*_0 -1\geq 0,
\end{equation}
\begin{equation}
\label{eq:b45}
    \Delta^*-\Delta^*_{\text{max}} \geq 0 ,
\end{equation}
with $\Delta^*_{\text{max}}=\log _{\epsilon}\left( 10^{12}\right)=-6$, and
\begin{equation}
    2 g_a^*+2N^*-\kappa_a^*-\Delta^*-1\geq0,
\end{equation}
\begin{equation}
    g_b^*+\Omega^*
    _0-\Delta^*+N^*-\kappa_b^*-1\geq0,
\end{equation}
\begin{equation}
\label{eq:b46}
    -2 \Delta^* -\gamma^*_e+2 g_b^*-\kappa_b^*+2 N^*+2 \Omega_0^* -1\geq 0,
\end{equation}
with $\gamma^*_e=\log _{\epsilon}\left(10^{8}\right)=-4$. To achieve the goal stated above, we minimize a cost function $f_{\epsilon,N^*,\Delta^*_{\text{max}},\gamma^*_e}(g_a^*, g_b^*, \Omega^*_0, \kappa_a^*, \kappa_b^*, \Delta^*)$ that is the sum of all quadratic terms, which are the squares of the left-hand side of the inequalities (\ref{eq:b36})-(\ref{eq:b46}). Minimizing a cost function while satisfying a set of inequalities has a well-known solution method called quadratic programming \cite{quadraticprog}. Using the method, and by setting $N=10^4$ and $\epsilon=10^{-2}$, we get the parameters
\begin{multline}
\tag{\ref{eq:physicalparameters}}
N=10^4,\quad g_a=2\pi\times10^2  \text{Hz}, \\\quad g_b=\Omega_0=\kappa_a=2\pi\times10^8 \ \text{ Hz}, \quad
\kappa_b=2\pi\times10^{10} \ \text{Hz}, \quad \\ \Delta =2\pi\times 10^{12} \ \text{Hz}.
\end{multline}

\section{Analytical derivations: population and metastable time}
\subsection{Analytical derivation of the population}
\label{sec:thesisderivationmsstate}
Here, we derive the metastable state population, Eq.~(\ref{eq: gaussian}), in the limit of large $N$. 
We define the function
\begin{equation}
\label{eq:f}
f(m):=\ln\left(\frac{\rho^{\text{ss}}_{m}}{\rho^{\text{ss}}_{-N/2}}\right)=\sum_{q=-N/2+1}^{m}\ln\left(\frac{\nu} {J_{q}^{2}}\right).
\end{equation}
The last equality is obtained using Eq.~(\ref{eq:ss}).
We expand \(f(m)\) to quadratic order around the local extrema at \(m_\pm\); around the local maximum at $m=m_{-}$
\begin{equation}
\label{eq:m-}
    f(m)\approx f(m_{-})-\frac{1}{2w^2}\left(m-m_{-}\right)^{2}+\ldots,
\end{equation}
and around the local minimum at $m=m_+$
\begin{equation}
\label{eq:m+}
    f(m)\approx f(m_{+})+\frac{1}{2w^2}\left(m-m_{+}\right)^{2} +\ldots.
\end{equation}
Here, we obtain
\begin{multline}
\label{eq:width1}
    w^{2} = -\left( \pp{^2f(m)}{m^2}\Bigg|_{m=m_-}\right)^{-1} = \left( \pp{^2f(m)}{m^2}\Bigg|_{m=m_+}\right)^{-1} \\ = R[\hat{J}](m_-) \approx \frac{N r^{2}}{4 \sqrt{1 - r^{2}}} .
\end{multline}
The functional in Eq.~(\ref{eq:width1}) is defined in Eq.~(\ref{eq:R[Q]}).
The normalized pumping coefficient 
$r$ is given in Eq.~(\ref{eq:nu}). 
The higher-order terms in Eqs.~(\ref{eq:m-}) and (\ref{eq:m+}) can be shown to be small in the large $N$ limit \cite{myarticle}.
Using Eqs.~(\ref{eq:f}) and (\ref{eq:m-}), we obtain the metastable state population [defined in Eq.~(\ref{eq:truncate})] 
\begin{equation}
    \rho_m^{\text{ms}} \approx \frac{1}{\sqrt{2 \pi w^2}}\, \mathrm{e}^{-{(m-m_-)^2}/{2 w^2}},
\end{equation}
with width $w$, and centered at $m_-$.

\subsection{Analytical derivation of the metastable time and recovery time}
\label{sec:thesisderivationmetastable}
Here, we derive Eqs.~(\ref{eq:T0}) and (\ref{eq:recovery}) in the limit of large \( N \). 
We derive it using the birth-death Markov process in continuous time with the rates (\ref{eq:qqq}).
The derivation is based on results from Ref.~\cite{firstpassage} for the mean first passage time (see definition below) of a birth-death continuous-time Markov process.
We define \( E(t) \) as the system’s excitation number at time \( t \), so that \( E(t) = m \) indicates excitation level \( m \).
The probability distribution of \( E(t) \) is therefore given by \( P(E(t) = m) = \rho_m(t) \), where \( \rho_m(t) \) evolves according to the population dynamics equation~(\ref{eq:population}).
We also define the random variable \( \tau_m \) as the first-passage time for the excitation number for a transition from \( m-1 \) to \( m \), that is,
\begin{equation}
\label{tau_m}
    \tau_{m}:=\min \{ t>0:E(t)=m \mid E(0)=m-1 \},
\end{equation}
with a mean \cite{firstpassage}
\begin{equation}
\label{eq:tau}
   \mathbb{E}[{{\tau}_{m}}]=\frac{1}{\nu \Gamma \rho^{\text{ss}}_{m-1}}\sum_{n=-\frac{N}{2}}^{m-1}\rho^{\text{ss}}_{n}.
\end{equation}
Recall that the metastable time is defined as the mean first-passage time for the excitation number to transition from \( m=m_- \) to the top state \( m=N/2 \).
 Using the definition in Eq.~(\ref{tau_m}), and Eq.~(\ref{eq:tau}), we calculate   
\begin{align}
\label{eq:0}
T_{\text{ms}} &:=\sum_{m=m_{-}}^{\frac{N}{2}}\mathbb{E}[{\tau}_{m}]=\frac{1}{\nu\Gamma}\sum_{m=m_{-}}^{N/2}\frac{1}{\rho_{m-1}^{\text{ss}}}\sum_{n=-\frac{N}{2}}^{m-1}\rho_{n}^{\text{ss}} \\
&\approx\frac{1}{\nu\Gamma}\int_{-\infty}^{\infty}\mathrm{exp}\left[-f(m_{+})-(m-m_{+})^{2}/2w^{2}\right]\mathrm{d}m \notag \\ \label{eq:orkhbjjgfty}
&\times\int_{-\infty}^{\infty}\mathrm{exp}\left[f(m_{-})-(m'-m_{-})^{2}/2w^{2}\right]\mathrm{d}m'\\
&\approx\frac{\rho_{m_{-}}^{\text{ss}}}{\rho_{m_{+}}^{\text{ss}}}\frac{2\pi}{\Gamma N\sqrt{1-r^{2}}}.
 \end{align}
In the last term of Eq.~(\ref{eq:0}), each of the two summands---\( 1/\rho^{\text{ss}}_{m-1} \) and \( \rho^{\text{ss}}_n \)---can be approximated by Gaussians of width \( w \), centered at \( m = m_{+}+1 \) and \( n = m_{-} \), respectively, using Eqs.~(\ref{eq:m+}) and (\ref{eq:m-}). Thus, the transition from Eq.~(\ref{eq:0}) to Eq.~(\ref{eq:orkhbjjgfty}) is made by this approximation and by replacing  the sums with integrals and extending their limits to \( (-\infty, \infty) \), without significantly affecting the result.
To evaluate $\rho_{m_-}^{\text{ss}}/\rho_{m_+}^{\text{ss}}$, we use Eq.~(\ref{eq:f}), and so calculate
\begin{multline}
\label{eq:pp}
f(m_{-})-f(m_{+}) = -\sum_{q={m_-+1}}^{m_+}\ln\left(\frac{\nu} {J_{q}^{2}}\right) \\  \approx - \int_{{m_-}}^{m_+} \ln\left(\frac{\nu} {J_{q}^{2}}\right) \mathrm{d}q
\\=(2N+2)\text{tanh}^{-1}\left(\frac{\sqrt{(N+1)^{2}-{4\nu}}} {N+1}\right) \\-2\sqrt{(N+1)^{2}-{4\nu}} \\
 \approx 2N\left(\tanh^{-1}(\sqrt{1-r^{2}}) 
 - \sqrt{1-r^{2}}\right) \\ +2\tanh^{-1}(\sqrt{1-r^{2}}).
\end{multline}
In the final step, we approximated the expression in the asymptotic limit of large \( N \).
Using Eqs.~(\ref{eq:orkhbjjgfty}) and (\ref{eq:pp}), we obtain the final result for the metastable time
\begin{multline}
\label{eq:945gje489u}
 T_{\text{ms}} = \frac{\rho_{m_{-}}^{\text{ss}}}{\rho_{m_{+}}^{\text{ss}}}\frac{2\pi}{\Gamma N\sqrt{1-r^{2}}} \\
 =\frac{2\pi\, \mathrm{e}^{2\tanh^{-1}(\sqrt{1 - r^{2}})}}{\Gamma N\sqrt{1 - r^{2}}}
\, \mathrm{e}^{2N[\tanh^{-1}(\sqrt{1 - r^{2}}) - \sqrt{1 - r^{2}}]}.
\end{multline}
Equation~(\ref{eq:945gje489u}) is identical to Eq.~(\ref{eq:T0}) from the main text.

Recall that in Sec.~\ref{subsec:metastabletime}, we defined the recovery time as the mean first passage time for the system to transition from $\rho^{\text{top}}$ back to $\rho^{\text{ms}}$. It can be calculated similarly to the derivation of $T_{\text{ms}}$ above, with the final result in the asymptotic limit of large $N$
\begin{multline} 
\ln(\Gamma T_{\text{rec}})\sim \ln(\rho^{\text{ss}}_{N/2}/\rho^{\text{ss}}_{m_{+}}) \\
\sim N[1 + \tanh^{-1}(\sqrt{1 - r^2}) - \sqrt{1 - r^2} + \ln(r/2)].
\end{multline}

\section{Analytical derivation of the linewidth}
\label{subsecthesis:derivationoflinewidth}
Here, we derive the linewidth of the superradiant laser model of Eq.~(\ref{eq:model}). The same method applies to the linewidth of the generalized FAT laser models of Eq.~(\ref{eq:generalmodel}), as explained in Sec.~\ref{subsec:generalcoherenceporperties}.

The first-order normalized Glauber coherence function can be evaluated using input-output theory, where the electric field operator for the output of the laser is proportional to the lowering operator $\hat{J}$, thus yielding 
\begin{equation}
\label{eq:y123}
    g^{(1)}(\tau)=\frac{\langle \hat{J}^{\dagger}(t+\tau)\hat{J}(t)\rangle _{\text{ms}}}{\langle \hat{J}^{\dagger}(t)\hat{J}(t)\rangle _{\text{ms}}}.
\end{equation}
Here, $\expectation{\cdot}_{\text{ms}}$ denotes the expectation value calculated for the metastable state, $\rho^{\text{ms}}$, and we restrict to the metastable regime in which $T_{\text{coh}}\ll T_{\text{ms}}$.
Using Eq.~(\ref{eq:model1}) and the quantum regression theorem, Eq.~(\ref{eq:y123}) reduces to  
\begin{equation}
\label{eq:g}
   g^{(1)}(\tau)=\frac{ \text{Tr}\left[\hat{J}^{\dagger}e^{\tau{\mathcal{S}}}\left(\hat{J}\rho^{\text{ms}}\right)\right]}{\langle \hat{J}^{\dagger}(t)\hat{J}(t)\rangle _{\text{ms}}}.
\end{equation}
Here, the superoperator $\mathcal{S}$ is defined in Eq.~(\ref{eq:model1}). Since $g^{(1)}(\tau)$ is invariant under phase shifts, Eq.~(\ref{eq: superposition}) implies that in Eq.~(\ref{eq:g}) one can take $\rho^{\text{ms}}=\left|N/2, \sqrt{\nu}\right\rangle_{\hat{J}} \left\langle  N/2,\sqrt{\nu} \right|$,
where $\left|N/2,\sqrt{\nu}\right\rangle_{\hat{J}}$ are the CRSS of Eq.~(\ref{eq:yofhjksjhdgkfng}). Thus, Eq.~(\ref{eq:g}) becomes
\begin{equation}
\label{eq:miuv}
    g^{(1)}(\tau)=\frac{1}{\sqrt{\nu}}\text{Tr}[\hat{J}^{\dagger}\rho(\tau)],
\end{equation}
where $\rho(t)$ obeys the master equation (\ref{eq:model}), and 
\begin{equation}
\label{eq:s56df345789fg}
\rho(0)=\left|N/2,\sqrt{\nu}\right\rangle_{\hat{J}} \left\langle N/2, \sqrt{\nu} \right|.
\end{equation}
To obtain Eq.~(\ref{eq:miuv}) from Eq.~(\ref{eq:g}), we used the eigenstate relation of Eq.~(\ref{eq:eigenstate}). 
Using Eq.~(\ref{eq:miuv}), we calculate
\begin{equation}
\label{eq:g1g1}
    g^{(1)}(\tau)=\frac{1}{\sqrt{\nu}}\sum_{m=-N/2}^{m_+-1}J_{m+1}\rho_{m,m+1}(\tau).
\end{equation}

Equation (\ref{eq:g1g1}) shows that $g^{(1)}(\tau)$ depends only on the first off-diagonal coherences. Thus, we consider the dynamics of the first off-diagonal coherences associated with the master equation (\ref{eq:model}). Defining $x_{m}:=\rho_{m,m+1}$, these satisfy the set of equations
\begin{multline}
\label{eq:94785638fjhghfh}
\dot{x}_{
m}=\Gamma\bigg[\nu\frac{2J_{m}J_{m+1}}{J_{m}^{2}+J_{m+1}^{2}}x_{m-1}-\nu x_{m}+J_{m+1}J_{m+2}x_{m+1}\\
-\frac{1}{2}\left(J_{m}^{2}+J_{m+1}^{2}\right)x_{m}\bigg]=:\Gamma\left(M_{1}\mathbf{x}\right)_{m}.
\end{multline}
Here, $\mathbf{x}$ is a vector of length $N$, with components $x_{m}$, where $-N/2\leq m\leq N/2-1$. We define the matrix $M_1$ according to Eq.~(\ref{eq:94785638fjhghfh}).
We determine the eigenvalue $\lambda$ of $M_1$ with the largest real part. This eigenvalue is real and negative, and we denote the corresponding eigenvector by $\mathbf{y}$. Below, we show that the initial coherence vector $\mathbf{x}(0)$ is well approximated by $\mathbf{y}$, giving a single coherence decay rate $-\Gamma\lambda=1/T_{\mathrm{coh}}$.

It will be useful to examine the population dynamics of the master equation (\ref{eq:model}) as it will be shown below.
The population dynamics are given in Eq.~(\ref{eq:population}) in Sec.~\ref{sec:metastable}, and revisited here:
\begin{multline}
\label{eq:jkhr348hgjsdfdf}
    \dot{p}_{m}=\Gamma\left[\nu p_{m-1}-\nu p_{m}+J_{m+1}^{2}p_{m+1}-J_{m}^{2}p_{m}\right] \\
    =:\Gamma(M_{0}\mathbf{p})_{m}.
\end{multline}
Here, $\mathbf{p}$ is a vector of dimension $N+1$ with the components of the populations $p_m:=\rho_m$ where $-N/2 \leq m\leq N/2$. The matrix $M_0$ is defined according to Eq.~(\ref{eq:jkhr348hgjsdfdf}).

To compare two matrices $A$ and $B$ that are not necessarily of the same dimension, we restrict both matrices to their common set of indices. On this common domain, we define
\begin{equation}
    C := A-B ,
\end{equation}
or, equivalently,
\begin{equation}
    C_{mn} := A_{mn}-B_{mn},
\end{equation}
for all pairs of indices $(m,n)$ for which both matrix elements are defined.
The same convention is used for vectors and for matrix--vector products. 
That is, when comparing two vectors of different lengths, we restrict them to their common set of indices. 
Similarly, for a matrix $A$ and a vector $\mathbf{v}$ that are not defined on the same full index set, the matrix--vector product is understood to be evaluated only on the common domain on which the relevant matrix elements and vector components are defined. 
Thus,
\begin{equation}
    \left(A\mathbf{v}\right)_m
    =
    \sum_n A_{mn} v_n ,
\end{equation}
where the sum is restricted to values of $n$ for which both $A_{mn}$ and $v_n$ are defined.
Since our system is restricted to the metastable interval $-N/2 \leq m\leq m_+$, we further reduce all the matrices to act on this Hilbert subspace if needed. 
We calculate using Eqs.~(\ref{eq:94785638fjhghfh}), (\ref{eq:jkhr348hgjsdfdf}), the difference between $M_1$ and $M_0$ 
\begin{multline}
\label{eq:jgirty94503wrer254}
    \left(\delta M \mathbf{x}\right)_{m}:=((M_1-M_0)\mathbf{x})_m \\
    =-\nu\left(\frac{(J_{m+1}-J_{m})^{2}}{J_{m}^{2}+J_{m+1}^{2}}\right)x_{m-1}\\
    -J_{m+1}\left(J_{m+1}-J_{m+2}\right)x_{m+1}-\frac{1}{2}\left(J_{m+1}^{2}-J_{m}^{2}\right)x_{m}.
\end{multline}
Comparing Eqs.~(\ref{eq:jkhr348hgjsdfdf}) and (\ref{eq:jgirty94503wrer254}), we find that the ratio of the largest matrix-element magnitudes of $\delta M$ and $M_0$ is at most of order $\epsilon:=\max_m |J_{m+1}-J_m|/J_m\ll1$, where the maximum is taken over the populated interior. This motivates treating $\delta M$ as a small perturbation of $M_0$.

Define the vector $\mathbf{y}^{0}$ by setting its $m$th component equal to the metastable population,
\begin{equation}
\label{eq:skjhg3938347}
    y^{0}_m = \rho^{\mathrm{ms}}_m .
\end{equation}
In the metastable approximation, this vector is treated as a zero-eigenvalue eigenvector of $M_0$. Strictly speaking, since the state is metastable rather than stationary, the corresponding eigenvalue is negative but very small in the metastable regime. We therefore neglect it here and write
\begin{equation}
\label{eq:sjklndtoigj3045345345}
    M_0 \mathbf{y}^{0}=0 .
\end{equation}
Since $M_1=M_0+\delta M$, where $\delta M$ is only a small perturbation, we expect the zero eigenvalue of $M_0$ to be shifted to a small magnitude real negative eigenvalue $\lambda$ of $M_1$. Correspondingly, the associated eigenvector $\mathbf{y}=\mathbf{y}^0+\delta\mathbf{y}$ of $M_1$ is expected to remain close to $\mathbf{y}^{0}$. We can assume that every eigenvalue of $M_0$ has a corresponding eigenvalue of $M_1$ that is close to it in the complex plane, and vice versa. All the eigenvalues of the matrices $M_0$ and $M_1$ must have real non-positive parts. Since zero is the eigenvalue of $M_0$ with the largest real part, we expect the corresponding eigenvalue $\lambda$ of $M_1$ to be the one whose negative real part has the smallest magnitude.

For the initial state of Eq.~(\ref{eq:s56df345789fg}), the smoothness of the metastable population distribution of the CRSS implies $x_m(0)=\sqrt{\rho_m^{\mathrm{ms}}\rho_{m+1}^{\mathrm{ms}}}\approx\rho_m^{\mathrm{ms}}=y_m^0$. Together with the perturbative approximation $\mathbf{y}\approx\mathbf{y}^0$, this implies that the initial coherence vector $\mathbf{x}(0)$ is approximately proportional to the slow eigenvector $\mathbf{y}$ of $M_1$. Its evolution is therefore approximately $\mathbf{x}(\tau)\approx e^{\Gamma\lambda\tau}\mathbf{x}(0)$, which, upon substitution into Eq.~(\ref{eq:g1g1}), gives $g^{(1)}(\tau)\simeq e^{\Gamma\lambda\tau}$ so that $1/T_{\mathrm{coh}}=-\Gamma\lambda$, as stated above.

We evaluate $\lambda$ perturbatively from the eigenvalue equation $M_1 \mathbf{y}=\lambda \mathbf{y}$,
\begin{equation}
    (M_0+\delta M) (\mathbf{y}^0+\delta \mathbf{y})=\lambda(\mathbf{y}^0+\delta\mathbf{y}).
\end{equation}
Keeping terms first order in the perturbation and using Eq.~(\ref{eq:sjklndtoigj3045345345}), we obtain
\begin{equation}
\label{eq:okaertj03450345234}
    M_0 \delta \mathbf{y}+ \delta M \mathbf{y}^0 =\lambda\mathbf{y}^0.
\end{equation}
We define the vector $\mathbf{l}^{\rm T}=(1,1,\dots,1)$, which is a left eigenvector of $M_0$ with eigenvalue zero, i.e.~$\mathbf{l}^{\rm T}M_0=0$. This relation expresses conservation of the total probability under the population dynamics of Eq.~(\ref{eq:jkhr348hgjsdfdf}). Additionally, we have $\mathbf{l}^T\mathbf{y}^0=1$, due to the normalization of the metastable state by its definition.
Using these relations, and multiplying Eq.~(\ref{eq:okaertj03450345234}) by $\mathbf{l}^\text{T}$ from the left gives
\begin{equation}
\label{eq:ksdhgksdjk;rgjq349053}
   \lambda= \mathbf{l}^\text{T}\delta M \mathbf{y}^0=\sum_n (\delta M \mathbf{y}^0)_n .
\end{equation}

We evaluate Eq.~(\ref{eq:ksdhgksdjk;rgjq349053}) using Eqs.~(\ref{eq:jgirty94503wrer254}) and (\ref{eq:skjhg3938347}) 
\begin{align}
\notag
\lambda&= -\sum_{m}\bigg[\nu\frac{(J_{m+1}-J_{m})^{2}}{J_{m}^{2}+J_{m+1}^{2}}\rho_{m-1}^{{\rm ms}} \\ 
&+J_{m+1}\left(J_{m+1}-J_{m+2}\right)\rho_{m+1}^{{\rm ms}}+\frac{1}{2}\left(J_{m+1}^{2}-J_{m}^{2}\right)\rho_{m}^{{\rm ms}}\bigg]  \label{eq:098767890}\\
&=-\sum_{m}\left[\nu\frac{(J_{m+1}-J_{m})^{2}}{J_{m}^{2}+J_{m+1}^{2}}\rho_{m-1}^{{\rm ms}}+\frac{1}{2}(J_{m+1}-J_{m})^{2}\rho_{m}^{{\rm ms}}\right]\label{eq:09876789dfgre0}\\
&\approx-\sum_{m}\left[\frac{(J_{m+1}-J_{m})^{2}}{2}\frac{\nu \rho_{m-1}^{{\rm ms}}}{J_{m}^{2}}+\frac{1}{2}(J_{m+1}-J_{m})^{2}\rho_{m}^{{\rm ms}}\right]\label{eq:82374kjsdh}\\
&=-\sum_{m}\left[\frac{1}{2}(J_{m+1}-J_{m})^{2}\rho_{m}^{{\rm ms}}+\frac{1}{2}(J_{m+1}-J_{m})^{2}\rho_{m}^{{\rm ms}}\right] \label{eq:ioaweurtpioasdfg}\\
&=-\sum_{m}(J_{m+1}-J_{m})^{2}\rho_{m}^{{\rm ms}}  \approx 
-\int_{-N/2}^{m_+} \rho_m^{\text{ms}}\left(\pp{J_m}{m}\right)^2 \text{d}m. \label{eq:jqo,v976}
\end{align}
In the step between Eq.~(\ref{eq:098767890}) to Eq.~(\ref{eq:09876789dfgre0}) we shifted the summation index in the term proportional to $\rho_m^\text{ms}$ and combined the resulting term. In the step between Eq.~(\ref{eq:09876789dfgre0}) to Eq.~(\ref{eq:82374kjsdh}) we approximated $J_m^2+J_{m+1}^2\approx2 J_m^2$, which is valid since $J_m^2$ varies slowly with $m$. In the step between Eq.~(\ref{eq:82374kjsdh}) to Eq.~(\ref{eq:ioaweurtpioasdfg}), we used Eqs.~(\ref{eq:ss}) and (\ref{eq:truncate}). Finally, we replace the discrete sum by its continuum approximation. 
As seen explicitly in Eq.~(\ref{eq:ioaweurtpioasdfg}), the first and second terms arise from the gain and loss processes, respectively, showing that each contributes one half of the total phase noise.
Throughout these manipulations, we neglect edge terms at the boundaries of the metastable interval, since $\rho_m^{\rm ms}$ is localized away from these boundaries.
The linewidth equals twice the reciprocal of the coherence time, and thus we obtain the final result
\begin{equation}
\label{eq:linewidthresult}
    \ell= -2\Gamma \lambda=2\Gamma\int_{-N/2}^{m_+} \rho_m^{\text{ms}}\left(\pp{J_m}{m}\right)^2 \text{d}m
\end{equation}
In Sec.~\ref{sec:coherenceprop}, we further simplify this expression using the sharply peaked population approximation.

 \section{Numerical Method to calculate the coherence}
\label{subsec:numericalmethod}
Here, we briefly describe the numerical method used to compute the coherence, \(\mathfrak C\), following Ref.~\cite{HL}. While we derive here coherence of the superradiant laser model of Eq.~(\ref{eq:model}), the same procedure applies to any generalized FAT laser model of Eq.~(\ref{eq:generalmodel}). In particular, this is how we generated the data in Fig.~\ref{fig:HL} for the generalized FAT laser model that achieves the HL for laser coherence, Eq.~(\ref{eq:HL}). 
We assume that the system operates in the metastable regime, where $T_{\text{ms}}$ is much longer than the coherence time, so that $\rho^{\text{ms}}$ can be treated as stationary (see Sec.~\ref{subsec:coherencederivation}).
For $\tau\geq0$, the quantum regression theorem gives the first-order Glauber coherence function defined in Eq.~(\ref{firstorderGlauber}) as
\begin{equation}
\label{eq:khaf9345kfgjdl90}
    G^{(1)}(\tau)
    =
    \Gamma
    \operatorname{Tr}
    \!\left[
        \hat{J}^\dagger
        e^{\mathcal{S}\tau}
        \left(\hat{J}\rho^{\text{ms}}\right)
    \right],
\end{equation}
where $\mathcal{S}$ is the Liouvillian defined in Eq.~(\ref{eq:model1}).
A similar expression, Eq.~(\ref{eq:g}), is given in Appendix~\ref{subsecthesis:derivationoflinewidth}.
In Sec.~\ref{subsec:coherencederivation}, we showed that the coherence for a stationary laser beam is given by \cite{HL}
\begin{equation}
\label{eq:9858949gjkdfjfg934}
    \coh
    =
    2\int_0^\infty G^{(1)}(\tau)\,\mathrm{d}\tau.
\end{equation}

Substituting Eq.~(\ref{eq:khaf9345kfgjdl90}) into Eq.~(\ref{eq:9858949gjkdfjfg934}) and evaluating the time integral gives
\begin{equation}
\label{eq:uiert8fjgk5t}
    \coh
    =
    -2\Gamma
    \operatorname{Tr}
    \!\left[
        \hat{J}^\dagger
        \operatorname{inv}[\mathcal{S}]
        \left(\hat{J}\rho^{\text{ms}}\right)
    \right].
\end{equation}
The formal integration uses the identity
$\int_0^\infty e^{\mathcal{S}\tau}\,\mathrm{d}\tau=-\mathcal{S}^{-1}$,
where $\mathcal{S}^{-1}$ denotes the inverse superoperator of $\mathcal{S}$.
This identity holds when all eigenvalues of $\mathcal{S}$ have strictly negative real parts.
Within the metastable approximation, however, $\mathcal{S}\rho^{\text{ms}}=0$, so $\mathcal{S}$ is not invertible.
We therefore compute $\operatorname{inv}[\mathcal{S}]$ in Eq.~(\ref{eq:uiert8fjgk5t}) by inverting $\mathcal{S}$ on its decaying subspace, with the stationary mode excluded (see Ref.~\cite{HL} for technical details).
Equation~(\ref{eq:uiert8fjgk5t}) already expresses the coherence in a form that can be evaluated numerically using linear algebra.
To further simplify the calculation, we follow Ref.~\cite{HL} and vectorize each operator, mapping it to a vector in Liouville space.
In this representation, left and right multiplication by operators are expressed using Kronecker products, allowing us to construct a simple matrix representation of $\mathcal{S}$ acting on the vectorized operators. Moreover, 
the trace in Eq.~(\ref{eq:uiert8fjgk5t}) becomes an inner product between vectors.
The coherence can then be evaluated using standard matrix operations.

\section{Semiclassical dynamics of the standard superradiant laser}
\label{app:superrdiantlasersemi}
Here, we present a geometrical analysis of the semiclassical dynamics of the standard superradiant laser. To our knowledge, this geometrical interpretation has not been discussed in previous studies. The model is reviewed in Sec.~\ref{subsec:superradiantlaserreview}, and its dynamics are described by the master equation~(\ref{eq:basicsuperr}).

\begin{figure}
\centering
\includegraphics[width=0.9\linewidth]{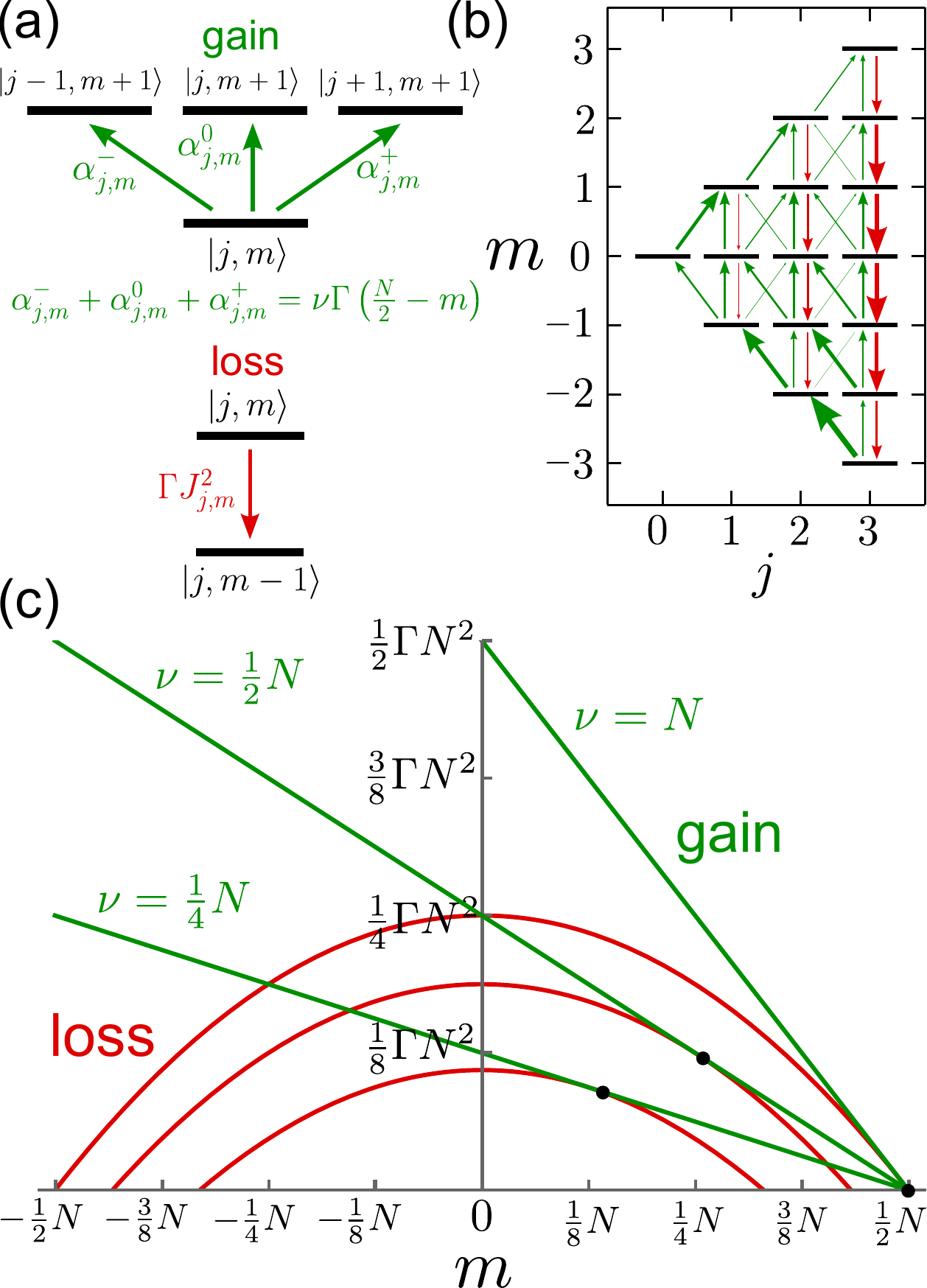}
\caption[Population dynamics of the standard superradiant laser in the Dicke basis.]{
Population dynamics of the standard superradiant laser in the Dicke basis.
(a) Transition rates from a Dicke state $\ket{j,m}$ in Eq.~(\ref{eq:standard-SR-2D-Markov}). Noncollective pumping increases $m$ by one and can transfer population to the sectors $j-1$, $j$, and $j+1$, whereas collective loss preserves $j$ and lowers $m$ by one.
(b) The corresponding two-dimensional jump process on the Dicke lattice, shown for $N=6$ atoms and $\nu=N/2$. Green arrows denote pumping transitions and red arrows denote collective-loss transitions; the arrow thickness indicates the relative transition rate.
(c) Drift in the $m$ direction. The green lines show the total pumping rate $D_m^{+}(m,\nu)$ given in Eq.~(\ref{eq:sum-pumping-rates}). At fixed $\nu$, this rate depends only on $m$, and different green lines correspond to different values of $\nu$, as indicated in the plot. The red curves show the approximate collective-loss rate $D_m^{-}(j,m)$ given in Eq.~(\ref{eq:lossrate}), which depends on both $m$ and $j$; different red curves correspond to different values of $j$. The black dots mark tangencies between the green lines and red curves, corresponding to fixed points of the coupled semiclassical drift (see text).}
\label{fig:rates}
\end{figure}

Since Eq.~(\ref{eq:basicsuperr}) is invariant under permutations of the atoms, the population dynamics can be represented in terms of the Dicke states $\ket{j,m}$, with the degeneracy label (see Sec.~\ref{subsec:Dicke}) suppressed \cite{SU(4)FrancoNori,HollandSU(4)}. We denote by $P_{j,m}$ the population associated with $\ket{j,m}$, summed over this degeneracy. The populations obey the classical master equation for a two-dimensional continuous-time Markov process:
\begin{multline} 
\label{eq:standard-SR-2D-Markov}
\dot{P}_{j,m}
={}
\alpha_{j+1,m-1}^{-} P_{j+1,m-1}
+\alpha_{j,m-1}^{0} P_{j,m-1} \\
+\alpha_{j-1,m-1}^{+} P_{j-1,m-1}
+\Gamma J_{j,m+1}^{2} P_{j,m+1} \\
-\left(
\alpha_{j,m}^{-} 
+\alpha_{j,m}^{0}
+\alpha_{j,m}^{+}
+\Gamma J_{j,m}^{2}
\right)P_{j,m}.
\end{multline}
Equation~(\ref{eq:standard-SR-2D-Markov}) has the standard gain--loss structure of a classical Markov master equation. The corresponding transition rates are shown in Fig.~\ref{fig:rates}a, with explicit expressions given in Ref.~\cite{SU(4)FrancoNori}. Noncollective pumping increases $m$ by one and changes $j$ by $-1$, $0$, or $+1$. The total pumping rate is \cite{PhysRevA2010}
\begin{equation}
\label{eq:sum-pumping-rates}
D_m^+(m,\nu):=\alpha_{j,m}^{-}+\alpha_{j,m}^{0}+\alpha_{j,m}^{+}
=\nu\Gamma\left(\frac{N}{2}-m\right).
\end{equation}
When the atomic ensemble is in the Dicke state $\ket{j,m}$, $N/2-m$ atoms are in the ground state $\ket{g}$. Since each ground-state atom is pumped at rate $\nu\Gamma$, the total pumping rate is $\nu\Gamma(N/2-m)$, independent of $j$ at fixed $m$. 
By contrast, collective loss preserves $j$ and lowers $m$ by one, as shown in Fig.~\ref{fig:rates}a. Using Eq.~(\ref{eq:J_{j,m}}), the loss rate can be approximated for a large $j$ as 
\begin{equation}
\label{eq:lossrate}
D_{m}^{-}(j,m): =    \Gamma J_{j,m}^{2}\approx \Gamma(j^{2}-m^{2}).
\end{equation}
Fig.~\ref{fig:rates}b plots the transition rates of the Markov process for $N=6$ atoms. However, $N=6$ atoms does not represent the large $N$ limit, which will be analyzed below.

We now consider the leading semiclassical drift in the $m$ direction for large $N$, treating $m$ and $j$ as continuous variables. In Fig.~\ref{fig:rates}c, the green lines show the pumping rate $D_m^+(m)$, Eq.~(\ref{eq:sum-pumping-rates}), for different values of $\nu$, while the red curves show the approximate collective-loss rate $D_m^-(m,j)$, Eq.~(\ref{eq:lossrate}), for different values of $j$. The net drift in $m$ is
\begin{equation}
\label{eq:standard-SR-m-drift}
\begin{aligned}
D_m(j,m,\nu)
&:=D_m^+(m,\nu)-D_m^-(j,m)\\
&\approx \Gamma\left[\nu\left(\frac{N}{2}-m\right)-j^2+m^2\right].
\end{aligned}
\end{equation}
At fixed $j$, the zeros of $D_m$ correspond to intersections of a green line and a red curve. A stationary point of the coupled dynamics additionally requires zero drift in $j$.
Since collective loss preserves $j$, the drift in this direction arises entirely from noncollective pumping. Expanding the transition rates given in Ref.~\cite{SU(4)FrancoNori} to leading order gives
\begin{equation}
\label{eq:standard-SR-j-drift}
\begin{aligned}
D_j(j,m,\nu)
&:=\alpha_{j,m}^{+}-\alpha_{j,m}^{-}\\
&\simeq\frac{\nu\Gamma}{2j}\left(Nm-j^2-m^2\right).
\end{aligned}
\end{equation}

For $0<\nu<N$, solving $D_m=D_j=0$ using the leading-order expressions in Eqs.~(\ref{eq:standard-SR-m-drift}) and (\ref{eq:standard-SR-j-drift}) gives the interior fixed point
\begin{equation}
\label{eq:standard-SR-tangency}
m_*=\frac{\nu}{2},
\qquad
j_*=\frac{1}{2}\sqrt{\nu(2N-\nu)}.
\end{equation}
Assuming that the steady state is localized around the fixed point $(j_*,m_*)$ given in Eq.~(\ref{eq:standard-SR-tangency}), the corresponding mean excitation number and photon flux are, to leading order,
\begin{equation}
\label{eq:standard-SR-geometric-results}
\begin{aligned}
\mu&=\frac{N}{2}+m_*=\frac{N+\nu}{2},\\
\mathcal{N}&=\Gamma(j_*^2-m_*^2)
=\frac{\Gamma}{2}\nu(N-\nu).
\end{aligned}
\end{equation}
These expressions recover the semiclassical results of Ref.~\cite{PhysRevA2010}, given in Eqs.~(\ref{eq:mmm6}) and (\ref{eq:fl}).
Within the leading-order approximation, solving $D_m=0$ and $\partial_m D_m=0$ also yields the interior fixed point given in Eq.~(\ref{eq:standard-SR-tangency}). These conditions require equal pumping and collective-loss rates and equal slopes with respect to $m$, at fixed $j$ and $\nu$, corresponding to a tangency between a green line and a red curve in Fig.~\ref{fig:rates}c.

\bibliography{sample}






\end{document}